\documentclass[11pt,a4paper]{article} 
\pdfoutput=1      
\usepackage{jcappub}
\usepackage{bm}
\usepackage{times}

\newcommand{\NC}{\mathbb{C}}  
\newcommand{\dea}{{\delta n}} 
\renewcommand{\AA}{\mathcal{A}}
\newcommand{\BB}{\mathcal{B}} 
\newcommand{\CC}{\mathcal{C}} 
\newcommand{\VV}{\mathcal{U}} 
\newcommand{\ax}{\alpha_{\chi}}
\newcommand{\px}{\varphi_{\chi}}
\newcommand{\vx}{v_{\chi}}      
\newcommand{\pv}{\Psi}          
\newcommand{\CCG}{\mathcal{G}}  
\newcommand{\LL}{\mathcal{L}}   
\newcommand{\dz}{\delta z}      
\newcommand{\dzg}{\dz_\chi}
\newcommand{\VP}{V_\parallel}   
\newcommand{\pvP}{\pv_\parallel}
\newcommand{\CCP}{\CC_\parallel}
\newcommand{\TCP}{C_\parallel}
\newcommand{\BBP}{\BB_\parallel}
\newcommand{\PP}{\mathcal{P}}        
\newcommand{\ddD}{\delta\mathcal{D}} 
\newcommand{\hDD}{\hat\DD}
\newcommand{\ttt}{\theta}          
\newcommand{\pp}{\phi}             
\newcommand{\eett}{\breve{\ttt}}   
\newcommand{\eepp}{\breve{\pp}}    
\newcommand{\up}[1]{{\rm #1}}
\newcommand{\bdv}[1]{{\bf #1}}
\newcommand{\bdi}[1]{\hbox{\boldmath{$#1$}}}
\newcommand{\beeq}{\begin{equation}}
\newcommand{\eneq}{\end{equation}}
\newcommand{\bear}{\begin{eqnarray}}
\newcommand{\enar}{\end{eqnarray}}
\newcommand{\nnn}{\nonumber \\}
\newcommand{\RA}{\rightarrow}
\newcommand{\OO}{\mathcal{O}}
\newcommand{\nhat}{\bdi{\hat n}}
\newcommand{\HH}{\mathcal{H}}  
\newcommand{\rbar}{\bar r}     
\newcommand{\DD}{\mathbb{D}}   
\newcommand{\dL}{\mathcal{D}_L}
\newcommand{\dA}{\mathcal{D}_A}
\newcommand{\dnu}{\delta\nu}   
\newcommand{\dhnu}{\widehat{\Delta\nu}} 
\newcommand{\drr}{\delta r}     
\newcommand{\dtt}{\delta\ttt}   
\newcommand{\dpp}{\delta\pp}    
\newcommand{\ee}{\breve{e}}     

\begin{document}

\begin{titlepage}

\setcounter{page}{1} \baselineskip=15.5pt \thispagestyle{empty}

\bigskip

\vspace{1cm}
\begin{center}
{\fontsize{20}{28}\selectfont \bfseries Gauge-Invariant Formalism 
of Cosmological Weak Lensing}
\end{center}

\vspace{0.2cm}

\begin{center}
{\fontsize{13}{30}\selectfont Jaiyul Yoo,$^{a,b}$, Nastassia Grimm$^a$,
Ermis Mitsou$^a$,\\ Adam Amara$^c$, Alexandre Refregier$^c$}
\end{center}

\begin{center}
\vskip 8pt
\textsl{$^a$ Center for Theoretical Astrophysics and Cosmology,
Institute for Computational Science}\\
\textsl{University of Z\"urich, Winterthurerstrasse 190,
CH-8057, Z\"urich, Switzerland}

\vskip 7pt

\textsl{$^b$Physics Institute, University of Z\"urich,
Winterthurerstrasse 190, CH-8057, Z\"urich, Switzerland}

\vskip 7pt

\textsl{$^c$Institute for Particle Physics and Astrophysics,
Department of Physics, ETH Z\"urich, 
Wolfgang-Pauli-Strasse 27, 8093 Z\"urich, Switzerland}

\vskip 7pt

\today

\end{center}

\note{jyoo@physik.uzh.ch, ngrimm@physik.uzh.ch, ermitsou@physik.uzh.ch,
alexandre.refregier@phys.ethz.ch, adam.amara@phys.ethz.ch}

\vspace{1.2cm}
\hrule \vspace{0.3cm}
\noindent {\sffamily \bfseries Abstract} \\[0.1cm]
We present the gauge-invariant formalism of cosmological weak lensing,
accounting for all the relativistic effects due to the 
scalar, vector, and tensor perturbations at the linear order.
While the light propagation is fully described by the geodesic
equation, the relation of the photon wavevector to the physical quantities
requires the specification of the frames, where they are defined.
By constructing the local tetrad bases at the observer
and the source positions, we clarify the relation of the weak lensing
observables such as the convergence, the shear, and the rotation
to the physical size and shape defined in the source rest-frame
and the observed angle and redshift measured in the observer rest-frame.
Compared to the standard lensing formalism, additional relativistic effects
contribute to all the lensing observables.
We explicitly verify the gauge-invariance of the lensing observables 
and compare our results to previous work. In particular, we demonstrate that 
even in the presence of the vector and tensor perturbations,
the physical rotation of the lensing observables
vanishes at the linear order, while the tetrad basis rotates along 
the light propagation compared to a FRW coordinate. Though the latter
is often used as a probe of primordial gravitational waves, the rotation
of the tetrad basis is indeed {\it not} a physical observable. We further
clarify its relation to the E-B decomposition in weak lensing.
Our formalism provides a transparent and comprehensive
perspective of cosmological weak lensing. 
\vskip 10pt
\hrule

\vspace{0.6cm}
\end{titlepage}

\noindent\hrulefill

\tableofcontents

\noindent\hrulefill

\setcounter{page}{1}
\pagenumbering{arabic}

\section{Introduction}
Significant and impressive progress has been made in observational cosmology 
in recent decades. However, the standard cosmology is still full of mystery, 
demanding further explanations both in terms of theoretical understanding and observational breakthrough. In particular, the late-time cosmic acceleration due to dark energy poses great challenges to theorists and observers alike. 
To tackle this pressing problem, numerous large-scale surveys have been undertaken to map the matter distribution of the Universe over the Hubble volume out to high redshifts. The current generation of these surveys include the Kilo Degree Survey (KiDS; \cite{KIDS15}), which is a wide-field program being carried out using the VLT Survey Telescope, 
the Dark Energy Survey (DES; \cite{DESC05}), which is nominally a five year survey that will cover 5000 square degrees using the Blanco 4~m telescope,
and the Subaru Hyper Suprime-Cam 
survey,\footnote{http://hsc.mtk.nao.ac.jp/ssp} which will cover 1400
square degrees.
These experiments represent a substantial improvement over previous measurements and are a significant stepping-stone to the next generation of experiments, known as stage-IV. These future programs include the Large Synoptic Survey Telescope (LSST; \cite{LSST04}), an 8-m dedicated ground based facility, Euclid \cite{EUCLID11} and the Wide Field Infrared Survey Telescope (WFIRST; \cite{WFIRST12}), which are two planned space based missions. Together these experiments will effectively cover the full observable extra-galactic sky and measure the shapes of roughly a billion galaxies.

One of the main goals of these cosmological experiments is to measure gravitational lensing. Gravitational lensing deals with light propagation in the Universe (see, e.g.,
\cite{SCEHFA92,MELLI99,BASC01,REFRE03,HEAVE03,MUVAET08} for recent reviews). 
As light travels towards us from distant objects, its path is perturbed due to intervening mass. In the case when these perturbations are small, the path is typically affected by several structures along the line-of-sight. This regime is known as weak lensing or cosmic shear and leads to distortions in the observed properties of the distant sources. Since first detection in the late 1990s 
\cite{BAREEL00,WITYET00,VAMEET00},
 the field of weak gravitational lensing has matured into one of the core cosmological probes. Current results include the KiDS weak lensing analysis of a 450 deg$^2$ survey \cite{KIDS17}, which reported a 5\% precision on the 
parameter $S_8$, a combination of $\sigma_8$ and $\Omega_m$,
where $\Omega_m$ is the cosmic matter density parameter and~$\sigma_8$ is
the amplitude of matter density fluctuations today. 
More recently the DES analysis of the first year data (Y1) \cite{DES17}, which covers an area of 1321 deg$^2$, achieved 3\% precision on $S_8$. This steady improvement in precision will increase and should accelerate as we enter the era of stage IV experiments. However, the ambitious goals enabled by these impressive observational facilities can only be achieved, if the theoretical predictions are at the same level of precision as those set by observations.

Sachs \cite{SACHS61,KRSA66}  was the first to develop the basic foundation 
for the propagation of the gravitational waves, and building on this framework
the early work on weak lensing was performed by Gunn \cite{GUNN67}.
With observational progress in 1990s, cosmological weak lensing received large attention, and many researchers \cite{SCWE88,MIRAL91a,MIRAL91b,BLSAET91,KAISE92,KASQ93,SESCEH94}
established the standard weak lensing formalism, which is further completed in the following years  \cite{JASE97,KABAET98,HU99,HUWH01,CRNAET02}.
However, despite these theoretical and observational developments in the
past decades, there has not been a complete gauge-invariant description of
cosmological weak lensing.
Here we derive the gauge-invariant formalism of cosmological weak lensing.
The key ingredient for such a task is to identify the missing physics
in the standard weak lensing formalism, which makes the predictions 
for the lensing observables gauge-dependent. In recent years,  great attention
has been paid to the relativistic description of galaxy clustering
\cite{YOFIZA09,YOO10,BODU11,CHLE11,JESCHI12,YOZA14} 
(see \cite{YOO14a} for review), by which the gauge ambiguities in the standard 
theoretical predictions are highlighted.
The dominant contribution to galaxy clustering is the matter density 
fluctuation, and it is well known that the matter density power spectrum
differs significantly near the horizon among different choices of gauge 
condition. It was quickly understood and argued \cite{YOFIZA09} that 
the standard model for galaxy clustering was incomplete and
our (correct) theoretical descriptions for any cosmological observables should 
be gauge-invariant; The key reason for such inadequacy in the standard model
was that ``unobservable quantities'' are used to build theoretical predictions.
For example, the observed redshift~$z$ is a physical observable, 
but our description
of the radial position (or sometimes called ``true redshift''~$\bar z$
without the redshift-space distortion) is gauge-dependent. 
In gravitational lensing, the observed
angular position~$\nhat$ of the source is physical, 
but our description of the (unlensed) 
``true angular position''~$\bdi{\hat s}$ is again gauge-dependent.
The standard weak lensing formalism built in terms of such unobservable
quantities is inevitably incomplete.
In fact, it is well known that
the lensing convergence~$\kappa$ in Eq.~\eqref{eq:kappa} 
is gauge-dependent and hence it cannot be directly associated
with the physical lensing observable we measure. It is shown
\cite{BONVI08,SCJE12a} that the lensing 
convergence we measure is the (gauge-invariant) fluctuation~$\ddD$
in Eq.~\eqref{eq:ddD} in the luminosity distance (or the angular diameter 
distance), which includes not only the standard lensing convergence~$\kappa$,
but also other relativistic contributions.
Here we present a complete and coherent description of the
gauge-invariant lensing formalism.

Establishing the proper descriptions of the observable quantities such as the
observed angle and redshift is the first step, and this requires the 
specification of the observer, in particular, the observer frame, in which 
the metric is Minkowski~$\eta_{ab}$. The second step is to establish the
proper descriptions of the physical quantities at the source position
that will be measured by the observer, and this step also requires the
specification of the source and its rest-frame, where the physical quantities
such as the size and the shape of galaxies are defined. 
For both cases, the tetrad basis provides such a link needed to connect 
the light propagation in the Friedmann-Robertson-Walker (FRW) universe to 
the rest-frames of the observer and the source.  The standard lensing
formalism lacks such descriptions. In Sec.~\ref{sec:observables},
we introduce the tetrad basis and present the linear-order expression
for the tetrad vectors.
Furthermore, there exists one more critical ingredient missing in the
standard lensing formalism --- {\it check of gauge-invariance}.
In the lensing literature, a gauge condition is adopted (almost exclusively
the conformal Newtonian gauge), and all the calculations are performed
with that gauge condition. However, by fixing the gauge condition, one loses
the ability to check the gauge-invariance of theoretical predictions
computed in that gauge condition. Here we perform all our calculations
with the general metric representation without choosing any gauge conditions,
such that we can explicitly check at each step the gauge-invariance of
our expressions. This procedure greatly helps understand better which part
in our theoretical descriptions can be associated with the physically
meaningful quantities.

The organization of the paper is as follows: In Sec.~\ref{sec:observables},
the basic observables in the observer rest-frame are expressed in relation
to the photon wavevector, and a local tetrad basis is constructed
to provide the connection to the photon wavevector in a FRW coordinate.
In Sec.~\ref{sec:src}, we solve the photon geodesic equation, accounting
for all the relativistic effects and paying particular attention to
the subtleties at the observer position. The source position in a FRW
coordinate is geometrically decomposed to represent the deviation from
the observationally inferred position. In Sec.~\ref{sec:stdlensing},
we generalize the standard weak lensing formalism by using the full
geodesic equation, we demonstrate that the standard formalism is still
incomplete and gauge-dependent. In Sec.~\ref{sec:GIlensing} we present our main
results of the gauge-invariant weak lensing formalism. Using the tetrad
basis at the source position, we construct
the distortion matrix in terms of the physical size and shape
in the source rest-frame, and we compute the lensing observables
in Sec.~\ref{ssec:geo}. In Sec.~\ref{ssec:rotation}, we demonstrate that 
the tetrad basis rotates as it is parallel transported along the photon path.
However, we show that this rotation is gauge-artifact and not observable.
In Sec.~\ref{ssec:EB} we present the lensing E-B decomposition and clarify
its relation to the lensing rotation.
In Sec.~\ref{sec:comparison}, we compare our results to previous work
on the lensing effects due to the gravitational waves in Sec.~\ref{ssec:GW},
on the lensing formalism by a standard ruler in Sec.~\ref{ssec:ruler}, and
on the lensing formalism by a Jacobi mapping approach in 
Sec.~\ref{ssec:jacobi}.
We summarize our findings and discuss the implications of our new
formalism for upcoming surveys in Sec.~\ref{sec:discussion}.

Throughout the paper, we use the Greek indices $\mu,\nu,\rho,\cdots$
for the space-time components in a FRW coordinate with metric~$g_{\mu\nu}$,
in which we use the conformal time~$\eta$ and the Greek indices
$\alpha,\beta,\gamma,\cdots$ to represent the time and the spatial components,
respectively.
The Latin indices $a,b,c,\cdots$ are used to represent the internal 
components in a rest-frame with the Minkowski metric~$\eta_{ab}$,
in which we use the proper-time~$t$ and
the Latin indices $i,j,k,\cdots$ to represent the time and the
spatial components. We summarize our notation convention in 
Table~\ref{table}.

\begin{table*}
\caption{Notation convention used in the paper}
\begin{tabular}{ccc}
\hline\hline
Symbols & Definition of the symbols & Equation\\
\hline
$\omega$, $\nhat$ & photon angular frequency \& propagation direction
& \eqref{eq:rest}\\
$\eta_{ab}$, $e_a^\mu$  & Minkowski metric and local tetrad basis 
($e_0^\mu=u^\mu$, $e_i^\mu$) & \eqref{eq:cond} \\
$\delta e_i^\eta$, $\delta e_i^\alpha$ & perturbations to the
spatial tetrad vectors & \eqref{eq:tetdef}\\
$\VV^\alpha$, $\Omega^i$ & spatial velocity of the observer,
orientation of the local tetrad basis & \eqref{eq:srcv}, \eqref{eq:omei}\\
$n^\mu_\lambda$, $\omega_\lambda$, $u^\mu_\lambda$ &
photon propagation direction and frequency measured by the observer $u^\mu$
&\eqref{eq:direction}\\
$\hat k^\mu$, $\dnu$, $\dea^\alpha$ & conformally transformed photon 
wavevector \& its perturbations & \eqref{eq:affine}, \eqref{eq:confwav}\\
$\dhnu$ & normalization condition for~$\hat k^\mu$ & \eqref{eq:normalization}\\
$\bar\eta_o$, $\bar t_o$ & age of the Universe & \eqref{eq:age}\\
$\delta\eta_o$, $\delta x^\alpha_o$ & coordinate lapse and shift 
of the observer position & \eqref{eq:lapse}, \eqref{eq:shift}\\
$\dz$ & perturbation in the observed redshift~$z$ & \eqref{eq:dz},
\eqref{eq:dzfull}\\
$\bar\eta_z$, $\lambda_z$, $\rbar_z$ & quantities expressed at the
observed redshift~$z$ & \eqref{eq:zzz}\\
$\Delta \lambda_s$ & perturbation in the affine parameter ($\lambda_s=
\lambda_z+\Delta\lambda_s$) & \eqref{eq:del2} \\
$\Delta \eta_s$, $\Delta x^\alpha_s$ & deviation of the source position
from the inferred position &\eqref{eq:deltaeta}, \eqref{eq:del3} \\
$\drr$, $\dtt$, $\dpp$ & geometric decomposition of the spatial deviation
&\eqref{eq:radial}, \eqref{eq:theta} \\
$\Omega^n$, $\Omega^\ttt$, $\Omega^\pp$ & decomposition of the rotation
vector~$\Omega^i$ &\eqref{eq:omedec} \\
$\kappa$, $\kappa_\up{cN}$ & lensing convergence and its expression
in Newtonian gauge &\eqref{eq:kappa}, \eqref{eq:kcN} \\
$\gamma_{1,2}$, ${}_{\pm2}\gamma$, $\gamma_{\alpha\beta}$ & 
lensing shear components & \eqref{eq:gamma1}, \eqref{eq:gammaij} \\
$\Delta s^\alpha$ & perturbation to the extended source size
&\eqref{eq:dsalpha}\\ 
 $\Delta n^\alpha_s$ & deviation of the photon propagation direction~$n^\mu_s$ 
at the source & \eqref{eq:nsrc} \\
$n^i_s=(\ttt_s,\pp_s)$ & photon propagation direction
in the source rest-frame  &\eqref{eq:obssrc} \\
$\Delta\ttt$, $\Delta\pp$ & perturbations in the source angle
$(\ttt_s=\ttt+\Delta\ttt_s$, $\pp_s=\pp+\Delta\pp_s$) & \eqref{eq:sang} \\
$\Delta\ttt^i_s$, $\Delta\pp^i_s$ & perturbations to the source angular
vectors~$\ttt^i_s$ \&~$\pp^i_s$ &\eqref{eq:twob}\\
$\hat\kappa$, $\hat\gamma_{1,2}$, $\hat\omega$ & gauge-invariant
physical lensing observables & \eqref{eq:hkappa}, \eqref{eq:hgamma} \\
$\ee_a^\mu$ & tetrad basis parallel transported along the photon path
& Sec.~(\ref{ssec:rotation}) \\
$n^i_s=\hat n^i_s$ & photon propagation direction in the source rest-frame 
& Sec.~(\ref{ssec:rotation})\\
$\hat\ttt_s^i$, $\hat\pp^i_s$ & basis vectors parallel transported and
Lorentz boosted &\eqref{eq:sang2} \\
$\Delta\hat\ttt_s^i$, $\Delta\hat\pp^i_s$ & perturbations to the basis
vectors parallel transported and Lorentz boosted &\eqref{eq:sang2} \\
$g_{\mu\nu}$, $\AA$, $\BB_\alpha$, $\CC_{\alpha\beta}$
& FLRW metric tensor and its perturbations & \eqref{eq:perturb} \\
$\alpha$, $\beta$, $\varphi$, $\gamma$
& scalar metric perturbations & \eqref{eq:decom}\\
$B_\alpha$, $C_\alpha$, $C_{\alpha\beta}$ & vector and tensor metric
perturbations & \eqref{eq:decom}\\
$\xi^\mu$, $T$, $L$, $L^\alpha$
& coordinate transformation \& its decomposition & \eqref{eq:coord}\\
$\ax$, $\px$, $\pv_\alpha$  & scalar and vector gauge-invariant variables &
\eqref{eq:gigi}\\
\hline\hline
\end{tabular}
\label{table}
\end{table*}

\section{Observables in the Light Propagation}
\label{sec:observables}
We derive the expressions for the basic quantities associated with the
light propagation in terms of the metric perturbations,
clarifying the difference between the FRW frame and the observer frame.
In the former the light propagation is computed, and in the latter the
local observables are defined and measured.

\subsection{Observer rest-frame: Tetrad basis}
The observers perform cosmological observations in the observer rest-frame,
in which the metric is 
Minkowski $\eta_{ab}$ and the time direction is set by the four 
velocity~$u^\mu$ of the observer. This frame is defined only
in the infinitesimal
neighborhood of a given spacetime point of the observer. However, a tangent 
space orthogonal to the time direction~$u^\mu$ of the observer can be
 well defined, by constructing three spacelike vectors $e_i^\mu$, 
where the index~$i$
represents three spatial directions of the observer ($i=1,2,3$).
Together with the time direction $e_0^\mu\equiv u^\mu$, these four vectors
are referred to as a tetrad $e_a^\mu$,
forming an orthonormal basis of the observer
($\eta_{ab}=g_{\mu\nu}e_a^\mu e_b^\nu$).
The Latin indices ($a,b,c,\cdots=0,1,2,3$)
are used to represent the component of a tetrad, 
and they are raised and lowered with the Minkowski metric. 

Cosmological information is measured by observing light from a distant 
source, and the tetrad vectors
serve as a basis for this cosmological observation.
Consider a null vector~$k^\mu$ in a given FRW coordinate, describing the
light propagation. This photon wavevector is measured in the observer
rest-frame as
\beeq
\label{eq:rest}
k^a=e^a_\mu k^\mu=\left(\omega,~\bdi{k}\right)=\omega(1~,-\nhat)~,
\qquad\qquad \omega=|\bdi{k}|~,\qquad |\nhat|=1~,
\eneq
where we expressed the components of the photon wavevector in the observer
rest-frame, in terms of the observable quantities:
the angular frequency $\omega=2\pi\nu$ of the photon and the angular
position~$\nhat$ of the source. In the observer rest-frame, a set of 
angles~$(\ttt,\pp)$ is assigned to the unit directional vector
$n^i=(\sin\ttt\cos\pp,\sin\ttt\sin\pp,\cos\ttt)$. Trivially, these
cosmological observables (e.g., $\omega$, $\nhat$, and so on)
are independent of FRW coordinates, while the
components of the photon wavevector~$k^\mu$ or the tetrad vectors $e_a^\mu$
are coordinate
dependent. This diffeomorphism invariance of any cosmological 
observables was emphasized by \citet{YODU17} in conjunction with their
gauge-transformation properties beyond the linear order in perturbations,
and it can be readily inferred from the above equation 
as the coordinate indices of cosmological observables are
contracted with the coordinate indices of the tetrad vectors and only
the internal indices of the tetrad vectors 
remain. It is emphasized \cite{MIYO18}
that the coordinate indices $\mu,\nu,\cdots$ and the internal tetrad indices
$a,b,\cdots$ transform according to their own (different) symmetry groups.

\subsection{Tetrad basis vectors in a FRW coordinate}
\label{ssec:tetrad}
In general,
the four tetrad vectors in four dimensions have sixteen degrees of freedom,
and ten of which are constrained by the metric tensor~$g_{\mu\nu}$:
\beeq
\label{eq:cond}
\eta_{ab}=g_{\mu\nu}e_a^\mu e_b^\nu~,\qquad\qquad
g^{\mu\nu}=\eta^{ab}e^\mu_ae^\nu_b~,\qquad\qquad e^\mu_a e^a_\nu=\delta^\mu_\nu
~,\qquad\qquad e^\mu_a e^b_\mu=\delta^a_b~.
\eneq
The remaining six degrees of freedom belong to the Lorentz symmetry at a
given spacetime, i.e., three Lorentz boosts and three spatial rotations.
While the tetrad formalism is completely general in choosing the remaining 
symmetry, the Lorentz boosts in our case are already fixed by the observer 
four velocity~$e_0^\mu=u^\mu$. This gauge choice is natural for describing the
cosmological observables, and it applies to all spacetime points, defining
the rest-frame of not only the observer, but also any other ``observers,''
including the sources of our cosmological observables 
(see \cite{MIYO18} for a complete description of the tetrad formalism in
cosmology).

The spatial directions $e_i^\mu$ can be arbitrary due to the remaining symmetry
in rotation in the rest-frame. For convenience, however,
we fix the spatial symmetry by aligning the
spatial tetrad directions with the FRW coordinate directions in a homogeneous
universe as
\beeq
\bar e_0^\mu=\left(\frac1a,~0\right),\qquad \qquad
\bar e_i^\mu=\left(0,~\frac1a\delta^\alpha_i\right),
\eneq
where $\delta^\alpha_i$ is a Kronecker delta and
we assumed a flat FRW universe and chose a rectangular coordinate
(see Appendix~\ref{appendix} for the metric convention). With such choice,
there exists
no further remaining gauge symmetry in the tetrads.
Indeed, this choice is implicitly made in most work in cosmology, which makes
calculations simpler. However, this choice often compounds the internal 
and the coordinate components, which makes the calculations vulnerable 
to errors, when perturbations are considered. We will clarify what
errors were made in previous work, with particular attention to the 
difference in indices --- The
internal components are diffeomorphism invariant and related to what we 
measure.

In our real universe, we have to consider the deviations from the homogeneous
FRW metric.
The time direction~$e_0^\mu$ 
of the tetrad is set by the timelike four velocity 
field~$u^\mu$ as 
\beeq
\label{eq:srcv}
-1=u^\mu u_\mu~,\qquad\qquad
e_0^\mu\equiv u^\mu=\frac1a\left(1-\mathcal{A},~\VV^\alpha\right),
\eneq
at the linear order in perturbations. The spatial directions of the
tetrad can be parametrized as
\beeq
\label{eq:tetdef}
e_i^\mu\equiv\frac1a\left(\delta e^\eta_i,~\delta^\alpha_i
+\delta e^\alpha_i\right), 
\qquad\qquad
\delta e_i^\alpha\equiv -\delta^\alpha_jp^j{}_i~,
\eneq
where $\delta e_i^\eta$ and $\delta e_i^\alpha$ (or $p^j{}_i$) 
are perturbations, capturing the deviation
from the background. The perturbation~$\delta e^\alpha_i$ in the 
spatial component of~$e^\mu_i$ can be 
further split into the symmetric part~$S_{ij}$ and the anti-symmetric
part~$A_{ij}$  as
\beeq
\label{eq:pij}
p_{ij}\equiv S_{ij}+A_{ij}~,\qquad\qquad
S_{ij}\equiv\frac12\left(p_{ij}+p_{ji}\right),\qquad\qquad
A_{ij}\equiv\frac12\left(p_{ij}-p_{ji}\right).
\eneq
The spatial indices~$i,j$ are raised and lowered by the spatial 
part~$\delta_{ij}$ of the Minkowski metric.
In equivalence, the perturbation $\delta e_i^\alpha$ can be split in terms
of FRW
coordinates.\footnote{In complete generality, we can split
the perturbation~$\delta e^\alpha_i$ of the spatial-component of~$e^\mu_i$
into the
symmetric part~$S^\alpha_\beta$ and the anti-symmetric part~$A^\alpha{}_\beta$
as
\beeq
\delta e_i^\alpha\equiv-\delta_i^\beta~p_\beta{}^\alpha{}~,\qquad \qquad
p_\beta{}^\alpha\equiv S_\beta{}^\alpha+A_\beta{}^\alpha~,\qquad\qquad
S_\beta{}^\alpha\equiv\frac12\left(p_\beta{}^\alpha+p^\alpha{}_\beta\right),
\qquad\qquad
A_\beta{}^\alpha\equiv\frac12\left(p_\beta{}^\alpha-p^\alpha{}_\beta\right).
\eneq
This definition is as valid as Eq.~\eqref{eq:pij}.}
The orthonormality condition in Eq.~\eqref{eq:cond}
constrains the time component of the spatial tetrads
\beeq
\delta e_i^\eta=\delta^\alpha_i\left(\VV_\alpha-\BB_\alpha\right),
\eneq
and the symmetric part of the spatial component
\beeq
S_{ij}=\delta^\alpha_i\delta^\beta_j\CC_{\alpha\beta}=\CC_{ij}~.
\eneq
Despite the expression written in terms of the internal indices,
the symmetric part~$S_{ij}$ is {\it not} invariant under diffeomorphism, 
as the coordinate indices are contracted with the metric 
tensor~$\CC_{\alpha\beta}$, which changes under gauge transformations.
The anti-symmetric part~$A_{ij}$ is, however, left unconstrained by
the orthonormality condition, which is why this part was unnoticed
in previous work (e.g., see \cite{JESCHI12,YOO14a,YOZA14}).

To complete the derivation of the tetrad basis vectors, we consider their
gauge-transformation properties. Being a four vector at every spacetime
point, the tetrad vectors transform as vectors under a coordinate
transformation in Eq.~\eqref{eq:coord}, 
and for an infinitesimal coordinate transformation by~$\xi^\mu$, 
this relation dictates the gauge-transformation of the tetrad vectors:
\beeq
\delta_\xi e^\mu_a=-\pounds_\xi e^\mu_a   \quad\mapsto\quad
\frac1a\delta_\xi\left(\delta e^\mu_a\right)=-\pounds_\xi\bar e^\mu_a+\OO(2)~,
\eneq
where $\pounds_\xi$ is the Lie derivative.
With the gauge-transformation of the metric perturbations in 
Appendix~\ref{appendix},
the anti-symmetric part has to transform as
\beeq
\label{eq:aij}
\delta_{\alpha j}\tilde A^j{}_i
=\delta_{\alpha j}A^j{}_i-L_{[\alpha,\beta]}\delta_i^\beta~,
\eneq
where the scalar part of the spatial transformation~$\LL^\alpha$ cancels
and only the vector part~$L^\alpha$ remains.
Given the constraint from the gauge transformation, the anti-symmetric part
of the spatial tetrad vectors can be expressed in terms of the metric
perturbations and the rotation of the spatial tetrad vectors as
\beeq
\label{eq:omei}
\delta_{\alpha j}A^j{}_i\equiv C_{[\alpha,\beta]}\delta_i^\beta
+\epsilon_{\alpha ij}~\Omega^j~,
\eneq
where $\epsilon_{ijk}$ is the Levi-Civita symbol,
the rotation vector~$\Omega^i$ captures the residual symmetry in 
spatial rotation and is invariant under diffeomorphism.
The anti-symmetric part, often ignored in previous work, is composed
of the vector perturbation~$C_\alpha$ and the spatial rotation~$\Omega^i$ 
of the tetrad vectors. Fortunately, the errors in 
missing the anti-symmetric part in the tetrad
expressions are relatively innocuous --- As a choice of internal gauge, 
the spatial rotation can be set zero $\Omega^i=0$,
and the perturbation calculations
are performed in most cases without vector perturbations
or by choosing a spatial gauge $C_\alpha\equiv0$. In this work, we
will keep the rotation in general, and the implication of the spatial
rotation~$\Omega^i$ will be discussed at length in Sec.~\ref{ssec:rotation}.

Accounting for the symmetric and the anti-symmetric parts,
we present the complete expression for the spatial tetrad vectors 
at the linear order \cite{MIYO18}:
\beeq
\label{eq:stetrad}
e_i^\mu=\frac1a\left[\delta^\beta_i\left(\VV_\beta-\BB_\beta\right),~
\delta^\alpha_i-\delta^\beta_i\left(\varphi~\delta^\alpha_\beta
+\CCG^\alpha{}_{,\beta}+C^\alpha_\beta
\right)-\epsilon^\alpha{}_{ij}\Omega^j\right]~.
\eneq
Given the six degrees of freedom, we fixed the three Lorentz boosts by setting
the timelike direction $e_0^\mu\equiv u^\mu$ as the four velocity field,
but we left unspecified the remaining three spatial rotations~$\Omega^k$
in the spatial tetrad vectors at the perturbation level. All degrees
of freedom are already fixed at the background.

\subsection{Photon wavevector in a FRW coordinate}
\label{ssec:photon}
Having derived the local tetrad vectors, we are now in a position to relate
the local observables to the photon wavevector~$k^\mu$
in the FRW frame. In the observer rest-frame,
the observer measures the photon frequency~$\nu$ and the angular 
position~$\nhat$ of the source. These basic observables can be used 
to construct a photon wavevector $k^a=e^a_\mu k^\mu$ in the rest-frame
as in Eq.~\eqref{eq:rest},
and they are invariant under diffeomorphism. Using the tetrad expression, 
we can derive the photon wavevector in a FRW coordinate 
\beeq
\label{eq:photon}
k^\mu=e_a^\mu k^a={\omega\over a}\left[
1-\mathcal{A}-n^i\delta^\beta_i\left(\VV_\beta-\BB_\beta\right),
-n^i\delta_i^\alpha+\VV^\alpha+n^i\delta_i^\beta
\left(\varphi~\delta^\alpha_\beta+\CCG^\alpha{}_{,\beta}
+C^\alpha_\beta\right)+\epsilon^\alpha{}_{ij}n^i\Omega^j\right]~,
\eneq
and the observed direction of the photon propagation in a FRW coordinate
can be derived in a similar way as
\beeq
\label{eq:direction}
n^\mu= n^ie_i^\mu=-{k^\mu\over\omega}+u^\mu~,\qquad\qquad 0=n^\mu u_\mu~,
\qquad\qquad 1=n^\mu n_\mu~,
\eneq
where the explicit expression can be readily inferred from 
Eq.~\eqref{eq:stetrad}. In the absence of
perturbations, the spatial components of the photon wavevector~$k^\mu$
or the photon propagation direction~$n^\mu$
are proportional to the observed direction~$n^i$ in the observer rest-frame,
because the spatial tetrad vectors
are by construction aligned with a FRW coordinate. 

However, the presence
of perturbations changes their expressions in a FRW coordinate. This is a
general relativistic generalization of a Lorentz boost in special relativity,
with which the observer moving with a relative velocity measures different
frequency and different propagation direction. 
Compared to previous work \cite{YOO14a},
the difference in Eq.~\eqref{eq:photon} arises solely due to the missing 
anti-symmetric part~$A_{ij}$ in the spatial tetrad vectors, and that is, the
vector perturbation and the spatial rotation.
We emphasize that all the quantities above are evaluated at the observer 
position and the expression is valid only at the observer position, because
the angular frequency~$\omega$ and the propagation direction~$n^i$ are the
quantities measured by the observer, not a field defined everywhere.

Nevertheless, it is useful to have such expressions along the photon path
by generalizing the above equations.
Given the observables $(\omega,n^i)$ in the observer rest-frame
described by~$u^\mu$, we completely fixed the photon wavevector~$k^\mu$
in Eq.~\eqref{eq:photon} at the observer position
by setting the orientation~$\Omega^i$ 
of the spatial tetrad~$e_i^\mu$. The photon wavevector is subject to the
geodesic equation and the null condition, such that it is completely
determined in a FRW coordinate along the null path described by the observables
$(\omega,n^i)$. At any point~$x^\mu_\lambda$
along the null path (parametrized by~$\lambda$), we will need to specify
an ``observer'' with four velocity~$u^\mu_\lambda$,
defining the timelike direction, in which another
``observation'' will be performed. This observer at~$x^\mu_\lambda$ will
measure the frequency $\omega_\lambda=-\left(u^\mu k_\mu\right)_\lambda$
in the rest-frame (different from~$\omega$ at origin),
and hence the ``observed direction~$n^\mu_\lambda$ ''
of this observer in a FRW coordinate
is determined as in Eq.~\eqref{eq:direction}. However, as evident in
Eq.~\eqref{eq:direction}, the observed direction~$n^i_\lambda$ in the observer
rest-frame depends on the choice of the spatial tetrad~$e^\mu_i$ 
at~$x^\mu_\lambda$, i.e., the rotation~$\Omega^i_\lambda$.
This reflects the freedom to choose local coordinate directions, on which
the observed angle $(\ttt,\pp)_\lambda$ depends.

\section{Photon Geodesic Path and Source Position in a FRW Coordinate}
\label{sec:src}
Here we solve the geodesic equation to obtain
the source position in a FRW coordinate and derive the geometric
distortions of the source position from the observed position.
Compared to the previous work, we clarify the change
in the calculations due to the missing anti-symmetric part of the spatial 
tetrad vectors and the spatial coordinate shift at the observer position.

\subsection{Conformal transformation of the FRW metric}
\label{ssec:conformal}
To facilitate the computation of the null geodesic path in a FRW coordinate,
we perform a conformal transformation
$g_{\mu\nu}\mapsto\hat g_{\mu\nu}$:
\beeq
\hat g_{\mu\nu}\equiv\frac1{a^2}g_{\mu\nu}
=-\left(1+2\mathcal{A}\right)
d\eta^2-2\BB_\alpha dx^\alpha d\eta+\left[(1+2\varphi)
\delta_{\alpha\beta}+2\gamma_{,\alpha\beta}+2C_{(\alpha,\beta)}+
2C_{\alpha\beta}\right]dx^\alpha dx^\beta~.
\eneq
Since the null geodesic path ($ds^2=0$)
remains unaffected by the conformal transformation, we can utilize the
geodesic equation in the conformally transformed metric to derive the
null path~$x^\mu$. While the null path~$x^\mu(\Lambda)$
can be parametrized by any affine parameter~$\Lambda$, 
we can physically fix the affine parameter~$\Lambda$ 
by demanding that the tangent vector along the path is the photon 
wavevector,\footnote{Note that {\it not} all tangent vectors of a given path
correspond to the photon wavevector. So, the condition that the tangent
vector in Eq.~\eqref{eq:tangent} is the photon wavevector completely fixes
the parametrization of the path.}
\beeq
\label{eq:tangent}
k^\mu(\Lambda)={dx^\mu\over d\Lambda}~,
\eneq
and Eq.~\eqref{eq:photon} is satisfied at the observer position 
as the initial condition for the photon wavevector. In addition,
the photon wavevector should meet the null condition
$0=k^\mu k_\mu$ and  the geodesic equation $0=k^\nu k^\mu{}_{;\nu}$
at any point along the path.

With the conformal transformation, the geometry of the spacetime
manifold changes,
and the covariant derivatives in two different manifolds are not 
identical in order to satisfy their own metric compatibility:
\beeq
0=\nabla_\rho g_{\mu\nu}~,\qquad\qquad 0=\hat \nabla_\rho \hat g_{\mu\nu}~,
\eneq
where quantities in the conformally transformed metric are represented 
with hat. The metric compatibility condition in two manifolds
implies \cite{WALD84} that the covariant derivatives are
related to each other with the connecting tensor as
\beeq
\hat\nabla_\nu k^\mu=\nabla_\nu k^\mu+C^\mu_{\nu\rho}k^\rho~,\qquad\qquad
C^\mu_{\nu\rho}\equiv\HH\left(g_{\nu\rho}g^{\mu\eta}-\delta^\mu_\nu
\delta^\eta_\rho-\delta^\mu_\rho\delta^\eta_\nu\right)~.
\eneq
Hence, the geodesic equation is not satisfied for the photon 
wavevector~$k^\mu$ with $\hat\nabla_\mu$ in the conformally transformed
metric. However, by re-parameterizing the photon path~$x^\mu(\lambda)$ 
with different affine 
parameter~$\lambda$, we can derive the conformally transformed 
wavevector~$\hat k^\mu$ for the same null path that satisfies the geodesic
equation $0=\hat k^\nu\hat\nabla_\nu \hat k^\mu$ in the conformally transformed
metric:
\beeq
\label{eq:affine}
\hat k^\mu={dx^\mu\over d\lambda}=\NC a^2k^\mu~,\qquad\qquad
{d\Lambda\over d\lambda}=\NC a^2~,
\eneq
where the proportionality constant~$\NC$ is left unconstrained in the
conformal transformation, because the metric compatibility constrains only
the derivative of $d\Lambda/d\lambda$ \cite{WALD84}.

Given the conformal transformation, the choice of the normalization~$\NC$
is completely free. With Eqs.~\eqref{eq:affine} and~\eqref{eq:photon},
it appears natural to choose the normalization to fix the 
combination~$\NC a\omega$ that is constant everywhere in the 
background.\footnote{Furthermore,
the ``observed frequency~$\omega$'' 
at any points other than the observer position
requires a specification of the observer four velocity~$u^\mu$.
However, this is completely fixed at the background as $u^\mu=(1,0)/a$.}
Therefore, we fix the normalization factor~$\NC$ by setting the product
at the observer position \cite{YOO14a,YOZA14,YOSC16}
\beeq
\label{eq:normalization}
1\equiv \NC a\omega~~\up{at}~~x^\mu(\lambda_o)=x^\mu_o~,
\eneq
where the subscript~$o$ represents the observer position. The presence of
perturbations makes the combination
$\NC a\omega$ vary as a function of position, while
the normalization constant~$\NC$ is still a constant. With such condition,
the conformally transformed photon wavevector can be parametrized as
\beeq
\label{eq:confwav}
\hat k^\mu\equiv(1+\dnu~,-n^i\delta_i^\alpha-\dea^\alpha)~,
\eneq
and the perturbations to the photon wavevector at the observer position are
then
\bear
\label{eq:init}
\dnu_o&=&\dhnu_o
-\mathcal{A}_o-n^i\delta^\beta_i\left(\VV_\beta-\BB_\beta\right)_o
=\dhnu_o
-\left[\ax+\VP+{d\over d\lambda}\left({\chi\over a}\right)+H\chi\right]_o~,\\
\label{eq:inita}
\dea^\alpha_o&=&n^i\delta_i^\alpha\dhnu_o-\VV_o^\alpha-n^i\delta_i^\beta
\left(\varphi~\delta^\alpha_\beta+\CCG^\alpha{}_{,\beta}
+C^\alpha_\beta\right)_o-\epsilon^\alpha{}_{ij}n^i
\Omega^j_o\\
&=&n^i\delta_i^\alpha\left(\dhnu-\px-H\chi\right)_o-V^\alpha_o-\pv^\alpha_o
-C^\alpha_{\beta o}\delta^\beta_in^i-\epsilon^\alpha{}_{ij}n^i\Omega^j_o
+\left({d\over d\lambda}\CCG^\alpha\right)_o
~,\nonumber
\enar
where we define the perturbation~$\dhnu$
in the observed frequency in the conformally transformed metric,
in terms of the product
\beeq
\label{eq:dhnu}
\NC a\omega=-\NC a \left(u_\mu k^\mu\right)=-\hat u_\mu\hat k^\mu=
1+\dnu+\mathcal{A}+\left(\VV_\alpha-\BB_\alpha\right)n^\alpha\equiv 1+\dhnu~,
\eneq
because the four velocity in the conformally transformed metric is
$\hat u^\mu=au^\mu$ and $\hat u_\mu=\hat g_{\mu\nu}\hat u^\nu=u_\mu/a$,
justifying the notation. Our choice of the normalization condition 
in Eq.~\eqref{eq:normalization}
becomes $\dhnu_o=0$ at the observer position, but we keep the term~$\dhnu_o$
in general. However, we stress that the choice of the normalization~$\dhnu$
in~$\dea^\alpha_o$ only affects the component in proportion to~$n^i$,
while the rotation component is perpendicular to~$n^i$.
With $\dhnu_o=0$, the perturbations to the photon wavevector at the observer
position transform as
\beeq
\widetilde{\dnu}_o=\dnu_o+\left({d\over d\lambda}T+\HH T\right)_o~,\qquad\qquad
\widetilde{\dea}{}^\alpha_o=\dea^\alpha_o+
\left(\HH T n^i\delta_i^\alpha-{d\over d \lambda}\LL^\alpha\right)_o~,
\eneq
where $d/d\lambda=\partial_\eta-n^i\delta_i^\alpha\partial_\alpha$
is the derivative along the photon path in Eq.~\eqref{eq:dlambda}.
Their gauge-transformation properties in general can be derived from
the transformation of the photon wavevector~$k^\mu$ with additional constraint
that the normalization condition~$\dhnu$ is imposed at one physical point~$p$
in any coordinate systems, i.e., Eq.~\eqref{eq:dhnu} has the same value
at the same physical point~$p$:
\beeq
\label{eq:trwav}
\widetilde{\dnu}=\dnu+2\HH T-\HH_pT_p+{d\over d\lambda}~T~,  \qquad\qquad
\widetilde{\dea}{}^\alpha=\dea^\alpha+\left(2\HH T-\HH_pT_p\right)n^i
\delta_i^\alpha-{d\over d\lambda}\LL^\alpha~.
\eneq
For our choice of the normalization condition, the fixed physical point~$p$
is the observer position. The corrections at~$p$ in Eq.~\eqref{eq:trwav}
were neglected in Eq.~(2.21) in \cite{YOO14a}.

In addition, we define the observed angular vector~$n^\alpha$ in a FRW
coordinate as
\beeq
\label{eq:nalpha}
n^\alpha\equiv n^i\delta_i^\alpha~.
\eneq
However, it should be noted that the internal index~$i$ is invariant under
diffeomorphism. We also define two additional vectors
$\ttt^\alpha$ and~$\pp^\alpha$ in terms of two unit directional
vectors~$\ttt^i$ and~$\pp^i$ perpendicular to~$n^i$
in the observer rest-frame:
\beeq
\label{eq:ttpp}
\ttt^i=(\cos\ttt\cos\pp,\cos\ttt\sin\pp,-\sin\ttt)~,\qquad\qquad
\pp^i=(-\sin\pp,\cos\pp,0)~.
\eneq
Furthermore, we will use the following notation in connection
to the directional vector~$n^\alpha$:
\beeq
\BB_\parallel\equiv\BB_\alpha n^\alpha~,\qquad\qquad
\CC_\parallel\equiv\CC_{\alpha\beta}n^\alpha n^\beta~,\qquad\qquad
\CC_\parallel^\alpha\equiv\CC^\alpha_\beta n^\beta~,
\eneq
generally applicable to any vectors or tensors.

\subsection{Observer position in a FRW coordinate}
\label{ssec:shift}
In a homogeneous universe, the spatial coordinate of the observer or any other
observers can be set  $\bar x^\alpha_o=0$,
and the (conformal) time coordinate is uniquely set to be~$\bar\eta_o$
in relation to the age~$\bar t_o$ of the Universe:
\beeq
\label{eq:age}
\bar \eta_o=\int_0^\infty{dz\over H(z)}~,\qquad\qquad
\bar t_o=\int_0^\infty{dz\over H(z)(1+z)}~,
\eneq
where we used bar to indicate that the coordinate position of the
observer is obtained in a homogeneous universe.
Indeed, this can be readily derived by considering the motion of
a free-falling observer with four velocity~$u^\mu$. The timelike four
velocity can be parametrized in terms of the proper time~$\tau$ measured by the
observer in the rest-frame, and the FRW coordinates of the observer can be
obtained by integrating the four velocity over the proper time:
\beeq
x^\mu_{\tau_f}-x^\mu_{\tau_i}=\int_{\tau_i}^{\tau_f}d\tau~u^\mu~.
\eneq
In a homogeneous universe, in which the observer four velocity 
is $u^\mu=(1/a,0)$, the time coordinate~$\bar\eta_o$ of the observer today 
can be obtained by setting $\tau_i=0$ in the above equation.

In the presence of perturbations, the observer motion deviates from the
static motion in a homogeneous universe. Consequently, the coordinates
of the observer today also deviate from $\bar x^\mu_o=(\bar\eta_o,0)$
in a homogeneous universe --- the (time) coordinate lapse is \cite{YOO14b}
\beeq
\label{eq:lapse}
\eta_o\equiv\bar\eta_o+\delta\eta_o~,\qquad\qquad
\delta\eta_o=\frac1{a_o}\delta t_o=-\frac1{a_o}\int_0^{\bar t_o}d\tau~
\mathcal{A}=-\int_0^{\bar t_o}dt~\mathcal{A}~,
\eneq
and the (spatial) coordinate shift is \cite{YODU17}
\beeq
\label{eq:shift}
x^\alpha_o\equiv\bar x^\alpha_o+\delta x^\alpha_o~,\qquad\qquad
\bar x^\alpha_o=0~,\qquad\qquad
\delta x^\alpha_o=\int_0^{\bar t_o}dt~\frac1a\VV^\alpha~,
\eneq
where we changed the integration over the motion of the observer
in terms of the proper time~$d\tau$ to the integration over the 
time coordinate at a fixed spatial coordinate,
valid at the linear order in perturbations.
Using the geodesic condition of the observer motion
\beeq
0=a^\alpha=u^\nu\nabla_\nu u^\alpha=\left[\mathcal{A}
-\left(av\right)^\cdot\right]
^{,\alpha}+\left(av^\alpha\right)^\cdot~,
\eneq
the coordinate lapse can be further simplified as
\beeq
\delta t=-av~,\qquad\qquad \delta\eta=-v~.
\eneq

The coordinate lapse~$\delta \eta_o$ and coordinate
 shift~$\delta x^\alpha_o$
represent the deviation of the observer position~$x^\mu_o$ from the position
$\bar x^\mu_o=(\bar\eta_o,0)$ in a homogeneous universe.
For a coordinate transformation in Eq.~\eqref{eq:coord}, 
the observer position and
its deviations transform as
\beeq
\tilde x^\mu_o=x^\mu_o+\left(T,~\LL^\alpha\right)_o~,\qquad\qquad
\widetilde{\delta x}^\mu_o=\delta x^\mu_o+\left(T,~\LL^\alpha\right)_o~.
\eneq
From the gauge transformation properties in Appendix~\ref{appendix}, we can
readily show that these deviations in Eqs.~\eqref{eq:lapse} 
and~\eqref{eq:shift} satisfy the transformation properties above.
In the comoving-synchronous gauge $(v=\mathcal{A}=\VV^\alpha=0$), the observer
position~$x^\mu_o$ is identical to that~$\bar x^\mu_o$ in a homogeneous
universe. However, this is not valid in general.
At the linear order in perturbations, the coordinate
shift~$\delta x^\alpha_o$  drops out in 
the expressions of the cosmological observables such as the luminosity
distance~$\dL$ in Eq.~\eqref{eq:ddD}, the lensing 
shear~$\hat\gamma_{\alpha\beta}$ in Eq.~\eqref{eq:hgamma},
and the lensing rotation~$\hat\omega$ in Eq.~\eqref{eq:homega},
such that there is no systematic error in missing the
 coordinate shift~$\delta x^\alpha_o$. However, the coordinate
lapse~$\delta\eta_o$ has significant impact \cite{BIYO16,BIYO17}, even
at the linear order. The change in the observer time coordinate
affects the distortion~$\dz$ in the observed redshift in Eq.~\eqref{eq:dz},
because it changes the ratio of the cosmic expansion.
The radial distortion~$\drr$ in Eq.~\eqref{eq:radial} is affected due to the
change in the length of the null path, while the angular 
distortions~$\dtt$ and~$\dpp$ in Eq.~\eqref{eq:theta} are not affected
at the linear order. Therefore, the coordinate lapse~$\delta\eta_o$ appears
in the observable quantities such as the luminosity distance~$\dL$, and
it was shown \cite{BIYO16} that the absence of~$\delta\eta_o$ in the
luminosity distance calculations breaks the gauge invariance and the
equivalence principle, causing the infrared divergences in the variance
of the luminosity distance. We must emphasize that though one can set
zero the coordinate lapse and shift, this choice corresponds to a gauge
choice, i.e., the comoving-synchronous gauge. This means that one cannot
set the lapse and shift zero and choose other gauge conditions such as
the conformal Newtonian gauge than the comoving-synchronous gauge.

\subsection{Photon geodesic equation and observed redshift}
The photon wavevector in the conformally transformed metric trivially
satisfies the geodesic equation in a homogeneous universe.
In the presence of perturbations, the perturbations $(\dnu,\dea^\alpha)$ 
to the photon wavevector~$\hat k^\mu$ are constrained by 
the temporal and the spatial geodesic equations
\beeq
0=\hat k^\nu\hat\nabla_\nu\hat k^\eta={d\over d\lambda}
\dnu+\delta\hat\Gamma^\eta~,
\qquad\qquad
0=\hat k^\nu\hat\nabla_\nu\hat k^\alpha=-{d\over d\lambda}\dea^\alpha
+\delta\hat\Gamma^\alpha~,
\eneq
where we defined the derivative along the photon path with respect to the
affine parameter
\beeq
\label{eq:dlambda}
{d\over d\lambda}=\hat k^\mu{\partial\over\partial x^\mu}=
\left({\partial \over \partial\eta}-n^\alpha{\partial\over\partial x^\alpha}
\right)+\left(\dnu{\partial \over \partial\eta}
-\dea^\alpha{\partial\over\partial x^\alpha}\right)~,
\eneq
and the perturbations in the geodesic equations
\bear
\delta\hat\Gamma^\eta&\equiv&\hat\Gamma^\eta_{\mu\nu}\hat k^\mu\hat k^\nu=
\AA'-2\AA_{,\alpha}n^\alpha+\left(\BB_{\alpha,\beta}
+\CC'_{\alpha\beta}\right)n^\alpha n^\beta \\
&=&
{d\over d\lambda}\left[2\ax
+2H\chi+{d\over d\lambda}\left({\chi\over a}\right)\right]
-\left(\ax-\px\right)'+\left(\pv_{\alpha,\beta}
+C'_{\alpha\beta}\right)n^\alpha n^\beta
~,\nnn
\delta\hat\Gamma^\alpha&\equiv&\hat\Gamma^\alpha_{\mu\nu}\hat k^\mu\hat k^\nu
=\AA^{,\alpha}-\BB^{\alpha\prime}-\left(\BB_\beta{}^{,\alpha}
-\BB^\alpha{}_{,\beta}+2\CC^{\alpha\prime}_\beta\right)n^\beta
+\left(2\CC^\alpha_{\beta,\gamma}-\CC_{\beta\gamma}{}^{,\alpha}\right)n^\beta
n^\gamma\\
&=&\left(\ax-\px\right)^{,\alpha}-\pv_\beta{}^{,\alpha}n^\beta
-C_{\beta\gamma}{}^{,\alpha}n^\beta n^\gamma
-{d\over d\lambda}\left(2\px n^\alpha+\pv^\alpha+2C^\alpha_\beta n^\beta
+2H\chi n^\alpha\right)
+{d^2\over d\lambda^2}\CCG^\alpha~.\nonumber
\enar
These perturbations were also introduced in 
\cite{YOZA14,YOO14a,YOSC16}, but there was a typo
in Eq.~(2.26) for the expression of~$\delta\hat\Gamma^\eta$ in \cite{YOO14a}.
Note that we already chose a rectangular coordinate for our flat FRW 
coordinate, in which the Christoffel symbols vanish in the background.
In addition to the geodesic equation,
the perturbations to the photon wavevector are subject to
the null condition $0=\hat k^\mu\hat k_\mu$:
\beeq
\label{eq:null}
n^\alpha\dea_\alpha=\dnu+\AA-\BBP-\CCP~.
\eneq
With the explicit expressions of the geodesic equations, we integrate them 
over the affine parameter to obtain the perturbations 
$(\dnu,\dea^\alpha)_\lambda$ along the photon path~$x^\mu_\lambda$:
\bear
\label{eq:dnu}
\dnu_\lambda-\dnu_o&=&-\int_0^\lambda d\lambda'~\delta\hat\Gamma^\eta\\
&=&
-\bigg[2\ax+2H\chi+{d\over d\lambda}\left({\chi\over a}\right)\bigg]^\lambda_
{\lambda_o}+\int_{\lambda_o}^{\lambda}d\lambda'
\bigg[(\ax-\px)'-\left(\pv_{\alpha,\beta}+C'_{\alpha\beta}\right)n^\alpha 
n^\beta\bigg]~\nnn
&=&
-\left[2\ax-\pvP
+2H\chi+{d\over d\lambda}\left({\chi\over a}\right)\right]^\lambda_{\lambda_o}
-\int_0^{\rbar_\lambda} d\rbar\left(\ax-\px-\pvP-\TCP\right)'~, \nonumber
\enar
and 
\bear
\label{eq:dea}
\dea^\alpha_\lambda-\dea^\alpha_o&=&\int_0^\lambda d\lambda'
~\delta\hat\Gamma^\alpha 
=-\left[2\px n^\alpha+\pv^\alpha+2C^\alpha_\beta n^\beta+2H\chi n^\alpha
-{d\over d\lambda}\CCG^\alpha\right]^\lambda_{\lambda_o} \nnn
&&-\int_0^{\rbar_\lambda} d\rbar\left[\left(\ax-\px\right)^{,\alpha}
-\pv_\beta{}^{,\alpha}n^\beta-C_{\beta\gamma}{}^{,\alpha}n^\beta
n^\gamma\right]~,
\enar
where the quantities in the square bracket are evaluated at the source
and the observer positions parametrized by~$\lambda$ and~$\lambda_o$.
Note that 
the derivative~$d\lambda$ along the photon path was considered
only at the background level and we replaced it with the integration 
over the comoving distance~$d\rbar$, all of which are valid only 
when the integrands are at the linear order in perturbations.
The perturbations $(\dnu,\dea^\alpha)_o$ at the observer position
are fixed in Eq.~\eqref{eq:init} as the initial condition.

Before we proceed to obtain the source position~$x^\mu_s$
by integrating the geodesic
equations once more over the affine parameter, we derive the expression
for the observed redshift. The light emitted in the rest-frame of the source
travels across the Universe, and its wavelength is stretched due to
the expansion. With reference to the rest-frame wavelength 
or the emission frequency $\omega_s$ in the source rest-frame,
the observed redshift~$z$ is constructed by
using the observed frequency~$\omega_o$ at the observer as
\beeq
\label{eq:dz}
1+z\equiv{\omega_s\over\omega_o}={k^0_s\over k^0_o}
={\left(u_\mu k^\mu\right)_s\over \left(u_\mu k^\mu\right)_o}
={a_o\over a_s}\left(1+\dhnu_s-\dhnu_o\right)\equiv{1+\dz\over a_s}~,
\eneq
where we used Eq.~\eqref{eq:dhnu} and we defined the perturbation~$\dz$
in the observed redshift. In addition to the cosmic expansion, the
photon wavelength (hence the observed redshift)
is affected by the peculiar velocity, the gravitational
redshift, and so on, and the perturbation~$\dz$ in the observed redshift
captures such effects of
inhomogeneities. Noting that the observer time-coordinate is
$\eta_o=\bar\eta_o+\delta\eta_o$ in Eq.~\eqref{eq:lapse} and
the expression for the perturbation~$\dnu$ is Eq.~\eqref{eq:dnu},
we can derive the expression for~$\dz$ as
\beeq
\label{eq:dzfull}
\dz\equiv\HH_o\delta\eta_o+\dhnu_s-\dhnu_o
=-H\chi+\left(\HH\delta\eta+H\chi\right)_o
+\left[\VP-\ax+\pvP\right]^{\lambda_s}_{\lambda_o}
-\int_0^{\rbar_s} d\rbar\left(\ax-\px-\pvP-\TCP\right)'~.
\eneq
By converting to gauge-invariant variables,
we isolated the gauge-dependent term in~$\dz$, and it transforms as
$\widetilde{\dz}=\dz+\HH T$, with which we can define a gauge-invariant
variable $\dz_\chi\equiv\dz+H\chi$.
Note that while at the observer position~$\lambda_o$ the perturbation~$\dz$
in the limit $z=0$ is non-vanishing 
\beeq
\lim_{\lambda_s\RA\lambda_o}\dz=\HH_o\delta\eta_o\neq0~,
\eneq
by definition in 
Eq.~\eqref{eq:dz}, the observed redshift is indeed zero $(z=0$), because
$a_s=a_o=1+\HH_o\delta\eta_o$ in this case, canceling the non-vanishing
part in the perturbation~$\dz$.

\subsection{Source position along the photon geodesic path}
\label{ssec:path}
With the photon wavevector in the conformally transformed metric and its
initial condition at the observer position, we will integrate the photon
wavevector over the affine parameter to derive the source position along the
photon geodesic path. The photon path is a straight line in a homogeneous
universe, and the inhomogeneities in the real universe deflect the photon
path from a straight line. We will begin the calculations by considering a
homogeneous universe first and thereby obtaining the relation to the
affine parameter set by our normalization condition in 
Eq.~\eqref{eq:normalization}.
Any position $x^\mu_\lambda$ along the photon path will be marked by
the affine parameter~$\lambda$, and we will use bar to indicate that
the position is derived in a homogeneous universe:
\beeq
\bar x^\mu_\lambda-\bar x^\mu_o
=\int_0^\lambda d\lambda'~\hat{\bar k}^\mu_{\lambda'}
=\left(\lambda~,-\lambda~n^\alpha\right)~, 
\eneq
where we set to zero the affine parameter at the observer $\lambda_o=0$ and 
the position of the observer in a homogeneous universe is uniquely set
$\bar x^\mu_o\equiv\left(\bar\eta_o,0\right)$.
As a coordinate in the world-line manifold, the affine parameter is defined by 
the above equation as
\beeq
\lambda\equiv\bar\eta_\lambda-\bar\eta_o=-\rbar_\lambda~,
\eneq
hence the spatial position becomes 
\beeq
\bar x^\alpha_\lambda=-\lambda~ n^\alpha=\rbar_\lambda n^\alpha~.
\eneq
The observed redshift is the only way we can assign a physically meaningful
distance to cosmological objects. Since the comoving distance~$\rbar$ is 
often defined in terms of a redshift parameter~$z$,  we define the
affine parameter~$\lambda_z$ and the time coordinate~$\bar\eta_z$
of the source in the background in terms of the observed redshift~$z$ as
\beeq
\label{eq:zzz}
\lambda_z\equiv \bar \eta_z-\bar\eta_o=-\rbar_z~,
\qquad\qquad 1+z={a(\bar\eta_o)
\over a(\bar \eta_z)}~,\qquad\qquad \rbar_z=\int_0^z{dz'\over H(z')}~.
\eneq
We have used bar for the position 
$\bar x^\mu_\lambda=(\bar\eta,\bar x^\alpha)_\lambda$
along the photon path to indicate that this position is evaluated in a
homogeneous universe, given the observed angle~$n^i$ in the observer
rest-frame. We prefer to use this notation, because
the position along the photon path changes in the presence of perturbations,
or the mapping from a world-line manifold changes from
$\bar x^\mu_\lambda$ to~$x^\mu_\lambda$, 
not because we want to introduce perturbations to a coordinate system.

With ($\dnu$, $\dea^\alpha$) in Eqs.~\eqref{eq:dnu} and~\eqref{eq:dea}, 
we can derive the 
expression for any position~$x^\mu_\lambda$ along the path by integrating 
the perturbations over the affine parameter. First, we integrate
the temporal part of the photon wavevector:
\bear
\label{eq:del}
\delta\eta_\lambda-\delta\eta_o&=&\lambda\left(\dhnu
+\ax-\VP-\pvP+H\chi\right)_o
-\left({\chi\over a}\right)+\left({\chi\over a}\right)_o
+\int_0^{\rbar_\lambda}
d\rbar\left(2\ax+2H\chi-\pvP\right) \nnn
&&+\int_0^{\rbar_\lambda} d\rbar~(\rbar_\lambda-\rbar)
\left(\ax-\px-\pvP-\TCP\right)'~,
\enar
such that the time coordinate of the source position is
$\eta_s=\bar\eta_s+\delta\eta_s$.  However, since the distance is more
physically related to the observed redshift, we first
relate the affine parameter~$\lambda_s$
at the source position to the observed redshift as
\beeq
\label{eq:del2}
\lambda_s=\bar\eta_s-\bar\eta_o\equiv \lambda_z+\Delta\lambda_s~,
\eneq
where we defined the residual deviation~$\Delta\lambda_s$ of the
affine parameter~$\lambda_s$ from~$\lambda_z$. Then the time-coordinate
of the source position becomes
\beeq
\label{eq:etas}
\eta_s=\bar\eta_z+\delta\eta_s+\Delta\lambda_s\equiv\bar\eta_z+\Delta\eta_s~,
\eneq
and the definition of the perturbation~$\dz$ in Eq.~\eqref{eq:dz}
yields
\beeq
\label{eq:deltaeta}
\Delta\eta_s={\dz\over\HH}~,\qquad\qquad \widetilde{\Delta\eta}_s=\Delta
\eta_s+T_s~.
\eneq
Naturally, when expressed at the observed redshift, only~$\Delta\eta_s$
will appear in the equations.\footnote{Note that the presence of~$\dhnu$ 
in Eq.~\eqref{eq:del}
in~$\delta\eta_\lambda$ implies an undetermined split of~$\bar\eta_\lambda$ 
and~$\delta\eta_\lambda$, rather than
the arbitrariness (or the normalization factor~$\NC$) in the conformal 
transformation, and indeed $\dhnu$ disappears in~$\Delta\eta_s$, when we used
the coordinate~$\bar\eta_z$ in terms of the observed redshift.}
In the limit~$z=0$, $\Delta\eta_s=\delta\eta_o$,
and $\Delta\lambda_s=0$.
We prefer to work with $(\bar\eta_z,\Delta\eta_s)$,
rather than $(\bar\eta_s,\delta\eta_s)$, because only $\Delta\eta_s$ 
gauge-transforms in the former combination, 
while $(\bar\eta_s,\delta\eta_s)$ are both affected by a
coordinate transformation. 
Note that this choice is made by convenience and we can use $\delta\eta_s$
derived in Eq.~\eqref{eq:del} to reproduce the relations in 
Eqs.~\eqref{eq:etas} and~\eqref{eq:deltaeta}.
Compared to the time coordinate~$\bar\eta_z$ we infer from the observed
redshift, the difference~$\Delta\eta_s$ in the source coordinate~$\eta_s$
naturally depends on the time coordinate~$\eta_o$ of the observer
and hence the coordinate lapse~$\delta\eta_o$. However, it is independent
of the rotation $\Omega^i_o$ of the local tetrad bases.

Next, we integrate the spatial part of the photon wavevector to obtain the
source position in a FRW coordinate. As mentioned, it proves convenient 
to express the source position~$x^\mu_s$ around the
position~$\bar x^\mu_z$ inferred from the observed redshift and angle.
Having computed 
the time distortion~$\Delta\eta_s$ in Eq.~\eqref{eq:deltaeta}, we will
compute the spatial distortion~$\Delta x^\alpha_s$ of the source
position as 
\beeq
\label{eq:del3}
x_s^\alpha\equiv\bar x^\alpha_z+\Delta x^\alpha_s=
\delta x^\alpha_o-\lambda_s n^\alpha-\int_0^{\lambda_s} d\lambda~
\dea^\alpha=\bar x^\alpha_z+\delta x^\alpha_o
-\Delta\lambda_s n^\alpha+\int_0^{\rbar_z} d\rbar~\dea^\alpha~.
\eneq
Given the privileged direction~$n^\alpha$, we decompose the spatial distortion
into the radial~$\drr$ distortion and the transverse distortion
$\delta x_\perp^\alpha$ as
\beeq
\Delta x^\alpha_s\equiv \drr~n^\alpha+\Delta x_\perp^\alpha~,\qquad
\qquad \drr=n_\alpha \Delta x^\alpha_s~,\qquad\qquad 0=n_\alpha \Delta x_\perp
^\alpha~,
\eneq
where the radial distortion is \cite{YOO10,YOO14a,YOZA14}
\bear
\label{eq:radial}
\delta r&=&
n_\alpha\delta x_o^\alpha-\Delta\lambda_s
-\left[{\chi\over a}+\CCG_\parallel\right]^z_o 
+\int_0^{\rbar_z}d\rbar~\left(\dnu+\ax-\px-\pvP-C_\parallel\right) \nnn
&=&\left(\chi_o+\delta\eta_o\right)-{\dzg\over\HH_z}+\int_0^{\rbar_z}d\rbar
\left(\ax-\px-\pvP-C_\parallel\right)+n_\alpha\left(\delta x^\alpha+\CCG^\alpha
\right)_o-n_\alpha\CCG^\alpha_s~,~~~~~~~~~~~~
\enar
where we used the null condition in Eq.~\eqref{eq:null} and the 
distortion in the time coordinate in Eq.~\eqref{eq:deltaeta}.
In \cite{YOO14a}, there was a typo in the final equation~(3.5), 
in addition to the missing term of the coordinate shift~$\delta x^\alpha_o$
at the observer position. The expression for~$\drr$ is arranged in terms
of gauge-invariant variables, isolating the gauge-dependent term
$n_\alpha\CCG^\alpha_s$, such that 
$\widetilde{\drr}=\drr+n_\alpha\LL^\alpha_s$.
Being the spatial distortion, the radial distortion~$\drr$ depends on
the spatial position~$x^\alpha_o$ of the observer and hence the coordinate
shift~$\delta x^\alpha_o$, but again it is independent of the rotation
$\Omega^i_o$ of the spatial tetrad vectors (same
for the distortion~$\Delta\eta_s$ in the time coordinate).

For the transverse components $\Delta x_\perp^\alpha$ 
of the spatial distortions, we have to
integrate $\dea^\alpha_\lambda$ in Eq.~\eqref{eq:dea} 
over the affine parameter as 
\bear
&&
\int_0^{\rbar_z}d\rbar~\dea^\alpha_\lambda=
\rbar_z\left[\dea^\alpha+2H\chi n^\alpha-{d\over d\lambda}\CCG^\alpha
+2\px n^\alpha+\pv^\alpha+2C^\alpha_\beta n^\beta\right]_o 
-\CCG^\alpha_s+\CCG^\alpha_o~\\
&&\hspace{-20pt}
-\int_0^{\rbar_z}d\rbar\left[2H\chi n^\alpha+2\px n^\alpha+\pv^\alpha+
2C^\alpha_\beta n^\beta\right] 
-\int_0^{\rbar_z}d\rbar(\rbar_z-\rbar)\left[
(\ax-\px)^{,\alpha}-\pv_\beta{}^{,\alpha}n^\beta-C_{\beta\gamma}{}^{,\alpha}
n^\beta n^\gamma\right]~.\nonumber
\enar
Further splitting the transverse distortion by using $\ttt^\alpha$ 
and~$\pp^\alpha$ in Eq.~\eqref{eq:nalpha}
\beeq
\Delta x_\perp^\alpha\equiv \rbar_z\left(\dtt,~\sin\ttt~\dpp\right)~,
\qquad \rbar_z\dtt=\ttt_\alpha\Delta x^\alpha_s~,\qquad
\rbar_z\sin\ttt~\dpp=\pp_\alpha\Delta x^\alpha_s~,
\eneq
we derive the angular distortions as \cite{YOO10,YOO14a,YOZA14}
\bear
\label{eq:theta}
\rbar_z\dtt&=&
\rbar_z\ttt_\alpha\left[-V^\alpha+C^\alpha_\beta n^\beta
-\epsilon^\alpha{}_{ij}n^i\Omega^j\right]_o
-\int_0^{\rbar_z}d\rbar~\ttt_\alpha\left[\pv^\alpha+2C^\alpha_\beta 
n^\beta\right] \\
&&
-\int_0^{\rbar_z}d\rbar(\rbar_z-\rbar)\ttt_\alpha\left[
(\ax-\px)^{,\alpha}-\pv_\beta{}^{,\alpha}n^\beta-C_{\beta\gamma}{}^{,\alpha}
n^\beta n^\gamma\right]
+\ttt_\alpha\left(\delta x^\alpha+\CCG^\alpha\right)_o
-\ttt_\alpha\CCG^\alpha_s~, \nonumber
\enar
where we used $\dea^\alpha_o$ in Eq.~\eqref{eq:init}.
For the azimuthal distortion $\sin\ttt~\dpp$, the above equation
can be used with~$\ttt^\alpha$ replaced with~$\pp^\alpha$.
The derivative in the second integration 
in Eq.~\eqref{eq:theta} cannot be pulled out as in equation~(3.6) in 
\cite{YOO14a} for the vector and the tensor perturbations, because it is 
the derivative with respect to the observed angle due to~$\ttt_\alpha$:
\beeq
\pv_{\beta,\alpha}\ttt^\alpha n^\beta={1\over\rbar}\left({\partial\over\partial
\ttt}\pvP-\pv_\alpha\ttt^\alpha\right)~,
\qquad\qquad
C_{\beta\gamma,\alpha}\ttt^\alpha n^\beta n^\gamma={1\over\rbar}\left(
{\partial\over\partial\ttt}\TCP-2C_{\alpha\beta}n^\alpha \ttt^\beta\right)~.
\eneq
With such expressions, we can further simplify the angular distortions as
\bear
\label{eq:theta2}
\dtt&=&\ttt_\alpha\left[-V^\alpha+C^\alpha_\beta n^\beta
-\epsilon^\alpha{}_{ij}n^i\Omega^j\right]_o
-\int_0^{\rbar_z}{d\rbar\over\rbar}~\ttt_\alpha
\left[\pv^\alpha+2C^\alpha_\beta n^\beta\right] \\
&&
-\int_0^{\rbar_z}d\rbar\left({\rbar_z-\rbar\over\rbar_z\rbar}\right)
{\partial\over\partial\ttt}\left(\ax-\px-\pvP-\TCP\right)
+{\ttt_\alpha\over\rbar_z}\left[\left(\delta x^\alpha+\CCG^\alpha\right)_o
-\CCG^\alpha_s\right]~. \nonumber
\enar
The expression for~$\dtt$ is again arranged in a gauge-invariant way,
that $\rbar_z\widetilde{\dtt}=\rbar_z\dtt+\ttt_\alpha\LL^\alpha_s$.
Similarly to the radial distortion, the transverse distortions depend 
on the observer 
position~$x^\alpha_o$ (and the coordinate shift $\delta x^\alpha_o$).
Furthermore, since they are expressed in terms of the observed angle 
$(\ttt,\pp)$ in the observer rest-frame, it is also affected by the orientation
of the local tetrad basis as
\beeq
\rbar_z\dtt \ni \rbar_z\Omega^\pp_o~,\qquad\qquad
\rbar_z\sin\ttt~\dpp \ni -\rbar_z\Omega^\ttt_o~,
\eneq
where for convenience we decompose the gauge-invariant vector for the
rotation of the local tetrad basis as
\beeq
\label{eq:omedec}
\Omega^i\equiv n^i\Omega^n+\ttt^i\Omega^\ttt+\pp^i\Omega^\pp~,\qquad\qquad
\epsilon^\alpha{}_{ij}n^i\Omega^j
=-\ttt^\alpha\Omega^\pp+\pp^\alpha\Omega^\ttt~.
\eneq
The implication is clear, such that if the local tetrad basis is rotated
($\Omega^i_o\neq0$) against a FRW coordinate, the angular source position
would be further rotated, given the observed angle $(\ttt,\pp)$.

In summary, the source position~$x^\mu_s$, given the observed redshift~$z$
and the observed angle~$n^i$, is expressed as the sum of the 
position~$\bar x^\mu_s$ inferred from these observables and the 
deviation~$\Delta x^\mu_s$ around it:
\beeq
\label{eq:dx}
x^\mu_s(z,\ttt,\pp)
=\left(\bar\eta_z+\Delta\eta_s,\rbar_z+\drr,\ttt+\dtt,\pp+\dpp\right)
=\bar x^\mu_z+\Delta x^\mu_s~,\qquad\qquad 
\bar x^\mu_z=\left(\bar\eta_z,\rbar_z n^i\right)~,
\eneq
where the components of the source position is written in 
a spherical coordinate. For a coordinate transformation
at the source position~$x^\mu_s$ in Eq.~\eqref{eq:coord},
the deviation~$\Delta x^\mu_s$ must gauge transform as
\beeq
\widetilde{\Delta x}{}^\mu_s=\Delta x^\mu_s+\xi^\mu_s~,\qquad
\widetilde{\Delta\eta}_s=\Delta\eta_s+T_s~,\qquad
\widetilde{\Delta x}{}^\alpha_s=\Delta x^\alpha_s+\LL^\alpha_s~,
\eneq
because the position~$\bar x^\mu_z$ inferred from the observables
remains unchanged. These transformation properties are indeed satisfied
in Eqs.~\eqref{eq:radial}, \eqref{eq:theta}, and~\eqref{eq:deltaeta}.
In the limit $z=0$, the source position becomes $x^\alpha_s=x^\alpha_o=\delta
x^\alpha_o$ in Eq.~\eqref{eq:shift}, and hence we have 
$\drr=n_\alpha\delta x^\alpha_o$ and $\Delta x^\alpha_\perp=\delta x^\alpha
_{\perp o}\neq0$. With $\rbar_z=0$ in the limit~$z=0$, the angular distortion
$(\dtt,\dpp)$ is not defined at the observer position.

\section{Standard Weak Lensing Formalism and Its Gauge Issues}
\label{sec:stdlensing}
We have derived the source position in a given FRW coordinate by solving
the full geodesic equation at the linear order in perturbations, including
the vector and the tensor type perturbations.
Here we briefly review the standard weak lensing formalism and follow
the lensing formalism with the fully relativistic solution $\Delta x_s^\mu$ in
Sec.~\ref{sec:src} to derive the expression for the gravitational lensing
convergence~$\kappa$, the shear~$\gamma$, and the 
rotation~$\omega$.\footnote{While the notation for the photon angular 
frequency is the same, we keep the standard notation~$\omega$ for the
rotation, because they are unlikely to be confused in our discussion.}
We demonstrate that the standard weak lensing formalism is built upon
the unobservable (coordinate-dependent) source position~$x^\mu_s$ and hence
it has a limitation in its relativistic generalization, which will be resolved
in Sec.~\ref{sec:GIlensing}.

\subsection{Short review of the standard weak lensing formalism}
\label{ssec:std}
In classical mechanics, the gravitational interaction due to a point
mass~$M$ provides a perturbation along the transverse direction to a
test particle moving with the relative speed~$v_\up{rel}$:
\beeq
\Delta v_\perp={2GM\over b~v_\up{rel}}~,
\eneq
where $G$ is the Newton's constant and~$b$ is the transverse separation
(or the impact parameter). The prediction for the light
deflection angle~$\hat\alpha$
in Einstein's general relativity is well-known to follow the same result
in classical mechanics, but with additional factor two:
\beeq
\hat\alpha
={4GM\over b~c^2}=8.155\times10^{-3}~\up{arcsec}\left({M\over
M_\odot}\right)\left({b\over\up{AU}}\right)^{-1}~.
\eneq
This light deflection due to a point mass can be generalized to derive 
the standard weak lensing formalism by considering the gravitational
potential fluctuation $\psi=-GM/r$ due to a point mass (this
indeed corresponds to the metric fluctuation~$\alpha$ in 
Eq.~\eqref{eq:decom}).
The lensing potential~$\Phi$ is the line-of-sight integration of the
metric fluctuation, and its angular derivative gives the relation between
the source angular position~$\hat{\bm{s}}$ and the observed 
position~$\hat{\bm{n}}$, so called, the lens equation
(see, e.g., \cite{BLNA86,BLNA92,SCEHFA92,KOSCWA05} for general reviews)
\beeq
\label{eq:basic}
\hat{\bm{s}}=\hat{\bm{n}}-\hat\nabla\Phi~,\qquad\qquad\qquad
\Phi=\int_0^{\rbar_s}d\rbar\left({\rbar_s-\rbar\over\rbar_s\rbar}\right)
~2\psi~,
\eneq
where $\hat\nabla$ is the angular gradient.
Since the observed source position is $\hat{\bm{n}}=(\ttt,\pp)$ in a
spherical coordinate, the deflection angle in the standard lensing formalism
would naturally correspond to the angular distortion of the source position:
\beeq
\hat\nabla\Phi~~\mapsto~-\left(\dtt,\dpp\right)~.
\eneq

Using the lens equation, the distortion matrix~$\DD$ (or sometimes
called the amplification matrix) is defined as
\beeq
\label{eq:lensD}
\DD_{ij}\equiv{\partial s_i\over\partial n_j}=\mathbb{I}_{ij}-
\left(\begin{array}{cc}\Phi_{11}&\Phi_{12}\\ \Phi_{21}&\Phi_{22}\end{array}
\right),\qquad\qquad\qquad
\Phi_{ij}\equiv\hat\nabla_j\hat\nabla_i\Phi~,
\eneq
where $\mathbb{I}$ is the two-dimensional identity matrix and
we defined a short hand notation for the angular derivatives of the
lensing potential. The distortion matrix is conventionally decomposed 
into the trace, the traceless symmetric and the anti-symmetric matrices:
\beeq
\label{eq:DD}
\DD\equiv
\mathbb{I}-\left(\begin{array}{cc}\kappa&0\\0&\kappa\end{array}\right)
-\left(\begin{array}{cc}\gamma_1&\gamma_2\\\gamma_2&-\gamma_1\end{array}\right)
-\left(\begin{array}{cc}0&\omega\\-\omega&0\end{array}\right)
~,\qquad\qquad \up{det}~\DD=\left(1-\kappa\right)^2-\gamma^2+\omega^2~,
\eneq
where the trace is the gravitational lensing convergence~$\kappa$ and 
the symmetric traceless part is the lensing shear~$\gamma=\sqrt{\gamma_1^2
+\gamma_2^2}$~:
\bear
\kappa&=&1-\frac12\up{Tr}~\DD=\frac12\left(\Phi_{11}+\Phi_{22}\right)~,\qquad
\qquad\qquad \omega={\DD_{21}-\DD_{12}\over2}=0~,\\
\gamma_1&=&{\DD_{22}-\DD_{11}\over2}=\frac12\left(\Phi_{11}-\Phi_{22}\right)~,
\qquad\qquad\qquad \gamma_2=-{\DD_{12}+\DD_{21}\over2}=\Phi_{12}=\Phi_{21}~.
\enar
Since the distortion matrix in Eq.~\eqref{eq:lensD}
is symmetric, the rotation~$\omega$ vanishes
in the standard formalism at all orders.

While the distortion matrix is defined in terms of angles, it is often 
assumed in literature that the line-of-sight direction is along $z$-axis
($\hat{\bm{n}}/\!\!/\hat{\bm{z}}$, i.e., $\ttt=0$),
and two angles are aligned with $x$-$y$ plane.\footnote{Note that in the
standard formalism there is no distinction between the FRW coordinate and
the (internal) observer coordinate.}
In such a setting, consider
two small angular vectors at the source position subtended
respectively by~$d\ttt$ and~$d\pp$ at the observer position 
\beeq
\Delta s_i^{d\ttt}=\DD_{i1}d\ttt~,\qquad\qquad \Delta s_i^{d\pp}=\DD_{i2}d\pp~.
\eneq
The solid angle at the source subtended by these two angular vectors
is then related to the solid angle at the observer as
\beeq
d\Omega_s=\left|\Delta\bm{s}^{d\ttt}\times\Delta\bm{s}^{d\pp}\right|
=\up{det}~\DD~d\ttt d\pp=\up{det}~\DD~d\Omega_o~,
\eneq
and hence the gravitational lensing magnification~$\mu$ is then
\beeq
\mu^{-1}\equiv {d\Omega_s\over d\Omega_o}=\up{det}~\DD~.
\eneq
For this reason,
the distortion matrix is often called the inverse magnification matrix.
Using the Poisson equation in cosmology,
\beeq
\label{eq:poisson}
\nabla^2\psi=4\pi G\bar\rho a^2\delta_m={3H_o^2\over2}\Omega_m{\delta_m\over
a}~,
\eneq
the gravitational lensing convergence can be computed in terms of
the matter density fluctuation~$\delta_m$ in the comoving gauge as
\beeq
\kappa=\int_0^{\rbar_s}d\rbar~{\left(\rbar_s-\rbar\right)\rbar\over\rbar_s}
\nabla_\perp^2\psi={3H_o^2\over2}\Omega_m\int_0^{\rbar_s}d\rbar~
{\left(\rbar_s-\rbar\right)\rbar\over\rbar_s}{\delta_m\over a}~,
\eneq
where we used  
\beeq
\nabla^2=\nabla_\perp^2+{\partial^2\over\partial \rbar^2}+{2\over\rbar}
{\partial\over\partial\rbar}~,
\eneq
and ignored the boundary terms.

The standard lensing formalism is based on the lens equation and the
lensing potential in Eq.~\eqref{eq:basic}. However, the source angular
position $\hat{\bm{s}}=(\ttt+\dtt,\pp+\dpp)$ is gauge-dependent, and the 
lensing potential that is responsible for the angular distortion $(\dtt,\dpp)$
is ill-defined. Indeed, we already know that $2\psi$ in Eq.~\eqref{eq:basic}
should be $(\ax-\px)$ to match the leading terms for~$\dtt$ in 
Eq.~\eqref{eq:theta} and the Poisson equation in Eq.~\eqref{eq:poisson}
is indeed an Einstein equation with~$\psi$ there replaced 
by~$-\px$.\footnote{Additional condition of a vanishing anisotropic pressure
is needed to guarantee $\ax=-\px$ and hence the consistency in the lensing
equation.} Furthermore, there exist no contributions from the vector
and the tensor perturbations in the standard lensing formalism. Finally,
while the derivations in this subsection assume no linearity, all formulas 
of the standard lensing formalism turn out to be valid only at the linear order
in perturbations.

\subsection{Relativistic generalization of the standard weak lensing formalism}
\label{ssec:relens}
Despite the issues listed in Sec.~\ref{ssec:std}, the standard lensing
formalism can be readily generalized and be put in the general relativistic
framework.
Gravitational lensing exclusively deals with the distortion in 
the source angular position $\bdi{\hat s}=(\ttt+\dtt,\pp+\dpp)$, compared to
the observed angular position 
$\bdi{\hat n}=(\ttt,\pp)$. So we can generalize the 
standard lensing formalism by replacing the deflection angle $-\hat\nabla\Phi$
due to the lensing potential by the angular distortion $(\dtt,\dpp)$
with all the relativistic effects taken into account. However,
already apparent in the calculations in Sec.~\ref{sec:src}, 
the source position~$x^\mu_s$ is unobservable
and gauge-dependent. Derived based on the unobservable quantities,
the distortion matrix and the lensing observables
in the generalized lensing formalism will be gauge-dependent.
Here we prove this by explicitly computing the distortion matrix and
the lensing observables with the angular distortion~$(\dtt,\dpp)$
in Eq.~\eqref{eq:theta}.

First, let us consider the ``generalized'' lens equation
\beeq
\label{eq:dsang}
s^i=n^i+\dtt~\ttt^i+\sin\ttt~\dpp~\pp^i~.
\eneq
Since all the relativistic effects are accounted for in the distortion
of the source position, this lens equation is indeed a generalization
of the lens equation~\eqref{eq:basic}. However, 
as shown in Eq.~\eqref{eq:theta2}, the source
angular position gauge-transforms as
\beeq
\tilde s^i=s^i+{1\over\rbar_z}\left(\ttt_\alpha\LL^\alpha_s~\ttt^i
+\pp_\alpha\LL^\alpha_s~\pp^i\right)~.
\eneq
Despite this incompleteness, we proceed with this simple relativistic
generalization for now.
Consider an extended source in the sky that appears subtended
by an infinitesimal size $(d\ttt,d\pp)$ in angle centered 
at the observed angle $(\ttt,\pp)$. Since the angular position of the source
is $\hat{\bm{s}}=(\ttt+\dtt,\pp+\dpp)$ in a spherical coordinate,
the infinitesimal size of the source is then
\bear
\label{eq:infds}
ds^i&=&\ttt^i
\left[d\ttt+\left(d\ttt{\partial\over\partial\ttt}+d\pp{\partial\over\partial
\pp}\right)\dtt-d\pp~\cos\ttt~\sin\ttt~\dpp\right]~\nnn
&&
+\pp^i\left[\sin\ttt~d\pp\left(1+\cot\ttt~\dtt+{\partial\over\partial\pp}
\dpp\right)+d\ttt{\partial\over\partial\ttt}\left(\sin\ttt~\dpp\right)\right]~,
\enar
where we computed the variation of the source position and ignored the
variation along the radial direction~$n^i$. Note that these angular
sizes~$(d\ttt,d\pp)$ should
not be confused with the angular distortion~$(\dtt,\dpp)$.
Expressing the angular extent of the source along the {\it observed}
angular directions~$\ttt^i$ and~$\pp^i$, the distortion matrix 
can be derived as
\bear
\label{eq:DD2}
\left(\begin{array}{c}ds_\ttt \\ \sin \ttt ~ds_\pp\end{array}\right)
&=&\left(\begin{array}{ccc}1+{\partial\over\partial\ttt}\dtt &
&{1\over\sin\ttt}{\partial\over\partial\pp}\dtt-\cos\ttt~\dpp \\ 
\cos\ttt~\dpp+\sin\ttt{\partial\over\partial\ttt}\dpp&&
1+\cot\ttt~\dtt+{\partial\over\partial\pp}\dpp\end{array}\right)
\left(\begin{array}{c}d\ttt \\ \sin\ttt ~d\pp\end{array}\right) \nnn
&\equiv&\left(\begin{array}{cc}\DD_{11}&\DD_{12}\\ \DD_{21}&\DD_{22}
\end{array}\right)
\left(\begin{array}{c}d\ttt \\ \sin\ttt d\pp\end{array}\right)~.
\enar
In this way, we generalized the standard weak lensing formalism by using
the angular distortion $(\dtt,\dpp)$ computed from the geodesic equation.
The distortion matrix computed in Eq.~\eqref{eq:DD2}
is {\it exact} for an infinitesimal size
of the source, provided that the angular distortion~$(\dtt,\dpp)$ is also
{\it exact} (computed at all orders in perturbations). However,
we defined the angular extent $(ds_\ttt,ds_\pp)$
of the source along the observed angular directions, 
rather than the angular directions set by the
source position~$s^i=(\ttt+\dtt,\pp+\dpp)$, i.e., the left-hand side
of Eq.~\eqref{eq:DD2} should be $(ds_\ttt,~\sin s_\ttt~ds_\pp)$.
This is indeed the ambiguity in the standard weak lensing formalism, and
we show that the distortion matrix defined this way reproduces
the standard results for the convergence and the shear. However, it is clear
that the distortion matrix is {\it not} physically well-defined in
the standard weak lensing formalism.

Given the distortion matrix in Eq.~\eqref{eq:DD2} and the angular distortion
$(\dtt,\dpp)$ in Eq.~\eqref{eq:theta}, 
it is straightforward to compute the lensing
observables. First, the gravitational lensing convergence is
\bear
\label{eq:kappa}
-2\kappa&\equiv&-2\left(1-\frac12\up{Tr}~\DD\right)=
\left(\cot\ttt+{\partial\over\partial\ttt}\right)\dtt+
{\partial\over\partial\pp}\dpp \\
&=&(2\VP-3\TCP)_o
+\int_0^{\rbar_z}{d\rbar\over\rbar}~\left(2n_\alpha
-\hat\nabla_\alpha\right)\left(\pv^\alpha+2C^\alpha_\beta n^\beta\right)
-{2n_\alpha\over\rbar_z}\left(\CCG^\alpha+\delta x^\alpha\right)_o
+{2n_\alpha\CCG^\alpha\over\rbar_z}\nnn
&-&\int_0^{\rbar_z}d\rbar\left({\rbar_z-\rbar\over \rbar_z \rbar}\right)
\hat\nabla^2\left(\ax-\px-\pvP-\TCP\right)
-{1\over\rbar_z}\hat\nabla_\alpha\CCG^\alpha~,\nonumber
\enar
where all the terms are arranged in terms of gauge-invariant variables
except two terms multiplied with~$\CCG^\alpha$,
such that the gravitational lensing
convergence gauge-transforms as
\beeq
\tilde \kappa=\kappa+{n_\alpha\LL^\alpha\over\rbar_z}
-{1\over 2\rbar_z}\hat\nabla_\alpha\LL^\alpha~.
\eneq
There exist typos for the expression~$\kappa$ in \cite{YOO14a},
in addition to the spatial shift~$\delta x^\alpha_o$
of the observer position. The absence of the rotation~$\Omega^i_o$ in
Eq.~\eqref{eq:kappa} represents that the gravitational lensing convergence is 
independent of rotation of the local tetrad basis. This is indeed true
for the rotation along the line-of-sight direction, but not for the other
directions beyond the linear order in perturbations.

Most works on weak lensing in literature consider {\it only} 
the scalar contribution
and adopts the conformal Newtonian gauge, in which $\CCG^\alpha\equiv0$.
While there exist the vector and the tensor contributions to~$\kappa$,
one can safely assume that the initial conditions are devoid of any
vector or tensor contributions at the linear order in perturbations.
Under such assumptions, the gravitational lensing convergence is
\beeq
\label{eq:kcN}
-2\kappa_\up{cN}=2V_{\parallel o}-{2n_\alpha \delta x^\alpha_o\over\rbar_z}
-\int_0^{\rbar_z}d\rbar\left({\rbar_z-\rbar\over \rbar_z \rbar}\right)
\hat\nabla^2\left(\ax-\px\right)~.
\eneq
The first term is the velocity contribution at the observer, and the second
term is the contribution due to the coordinate shift of the observer
position. These two contributions are often missing in literature.
While the first term is sometimes considered in the calculations
(e.g., \cite{SCJE12a}),
the second contribution was never considered in literature. 
Since the conformal Newtonian gauge leaves no gauge 
freedom, the expression of~$\kappa_\up{cN}$ is gauge-invariant.\footnote{Note
that the gauge-invariance is not a sufficient condition for an observable
quantity. One can choose the comoving gauge condition and compute the lensing
convergence~$\kappa_\up{com}$. The expression of~$\kappa_\up{com}$ is again
gauge-invariant, but the numerical values are different, i.e.,
$\kappa_\up{cN}\neq\kappa_\up{com}$.} However,
as expected in Eq.~\eqref{eq:kappa}, the lensing
convergence~$\kappa$ (or~$\kappa_\up{cN}$) is
{\it not} the correct lensing observable we measure in surveys.

The next lensing observable is the gravitational
lensing shear, or the traceless symmetric
part of the distortion matrix. According to the decomposition of the
distortion matrix in Eq.~\eqref{eq:DD}, two independent shear components 
can be computed as
\bear
\label{eq:gamma1}
2\gamma_1&\equiv&\DD_{22}-\DD_{11}=
\left(\cot\ttt-{\partial\over\partial\ttt}\right)\dtt
+{\partial\over\partial\pp}\dpp \\
&=&
\left(\pp^\alpha\pp^\beta-\ttt^\alpha\ttt^\beta\right)
\left[\left(C_{\alpha\beta}\right)_o-\CCG_{\alpha,\beta}
-\int_0^{\rbar_z}d\rbar
{\partial\over\partial x^\beta}\left(\pv_\alpha
+2C_{\alpha\gamma}n^\gamma\right)\right]\nnn
&&
+\int_0^{\rbar_z}d\rbar\left({\rbar_z-\rbar\over\rbar_z\rbar}\right)
\bigg[
\left({\partial^2\over\partial\ttt^2}-\cot\ttt{\partial\over\partial\ttt}
-{1\over\sin^2\ttt}{\partial^2\over\partial\pp^2}\right)\left(\ax-\px
-\pvP-\TCP\right)~, \nonumber
\enar
and
\bear
\label{eq:gamma2}
-2\gamma_2&\equiv&\DD_{12}+\DD_{21}=\sin\ttt{\partial\over\partial\ttt}\dpp
+{1\over\sin\ttt}{\partial\over\partial\pp}\dtt\\
&=&
\left(\ttt^\alpha\pp^\beta+\ttt^\beta\pp^\alpha\right)\left[
\left(C_{\alpha\beta}\right)_o-\CCG_{\alpha,\beta}
-\int_0^{\rbar_z}d\rbar{\partial\over\partial x^\beta}\left(\pv_\alpha
+2C_{\alpha\gamma}n^\gamma\right)\right]\nnn
&&
+\int_0^{\rbar_z}d\rbar\left({\rbar_z-\rbar\over\rbar_z\rbar}\right)
\bigg[-
{2\over\sin\ttt}\left({\partial^2\over\partial\ttt\partial\pp}-\cot\ttt
{\partial\over\partial\pp}\right)\left(\ax-\px-\pvP-\TCP\right)~.\nonumber
\enar
The expressions for the gravitational lensing shear are more complicated
due to their rotational properties.
Since the shear components~$\gamma_1$ and~$\gamma_2$ 
are confined to a two-dimensional
plane perpendicular to the line-of-sight direction, they
transform under rotation around the line-of-sight direction,
but the shear components are
invariant under 180 degree rotation. This implies that they
can be expressed in terms of the spin~$\pm2$ objects~${}_{\pm2}\gamma$ as
\beeq
{}_{\pm2}\gamma\equiv\gamma_1\pm i\gamma_2~,\qquad\qquad
2\gamma_1={}_{+2}\gamma+{}_{-2}\gamma~,\qquad\qquad
2i\gamma_2={}_{+2}\gamma-{}_{-2}\gamma~,
\eneq
where a function~${}_sf(\ttt,\pp)$
with spin~$s$ transforms under a clock-wise rotation of axes by~$\Xi$ as 
(see, e.g., \cite{HUWH97,ZASE97,HU00,CAHEKI05})
\beeq
\widetilde{{}_sf}(\ttt,\pp)=e^{-is\Xi}{}_sf(\ttt,\pp)~.
\eneq
Using the basis vectors~$m_\pm^i$ with spin~$\mp1$ 
under rotation that describe objects with spin~$\pm1$
\beeq
m_\pm^i={1\over\sqrt2}\left(\ttt^i\mp i\pp^i\right)~,\qquad\qquad
0=m_+^im_+^i=m_-^im_-^i~,\qquad\qquad 1=m_+^im_-^i~,
\eneq
we can construct the basis matrices with spin~$\mp2$ 
\beeq
m_\pm^im_\pm^j=\frac12\left[\ttt^i\ttt^j-\pp^i\pp^j\mp i\left(\ttt^i\pp^j
+\ttt^j\pp^i\right)\right]~,
\eneq
and their variants
\bear
&&
m^i_\pm m^j_\pm\partial_j
={1\over 2\rbar}\left(\ttt^i{\partial\over\partial\ttt}-{\pp^i\over\sin\ttt}
{\partial\over\partial\pp}\right)\mp{i\over2\rbar}\left({\ttt^i\over\sin\ttt}
{\partial\over\partial\pp}+\pp^i{\partial\over\partial\ttt}\right)~,\\
&&
m_\pm^im_\pm^j\partial_i\partial_j={1\over2\rbar^2}
\left({\partial^2\over\partial\ttt^2}-\cot\ttt{\partial\over\partial\ttt}
-{1\over\sin^2\ttt}{\partial^2
\over\partial\pp^2}\right)\mp{i\over \rbar^2\sin\ttt}\left({\partial^2
\over\partial\ttt\partial\pp}-\cot\ttt{\partial\over\partial\pp}\right)~,~~~
\enar
where we used the internal indices for their expressions, 
as they are defined in the observer rest-frame. However, as in 
Eq.~\eqref{eq:nalpha}, we will project the internal indices into a FRW
coordinate with the background relation~$\delta^\alpha_i$.
The spin~$\pm2$ shear components ${}_{\pm2}\gamma$ can then be expressed as
\beeq
{}_{\pm2}\gamma\equiv m_\mp^\alpha m_\mp^\beta\gamma_{\alpha\beta}~,
 \qquad\qquad
\gamma_{\alpha\beta}={}_{+2}\gamma ~m^\alpha_+m^\beta_+
+{}_{-2}\gamma ~m^\alpha_-m^\beta_-={}_{\pm2}\gamma~m^\alpha_\pm m^\beta_\pm~,
\eneq
where the shear matrix is \cite{SCJE12a}
\bear
\label{eq:gammaij}
\gamma_{\alpha\beta}&=&-\left(C_{\alpha\beta}\right)_o
+\CCG_{\alpha,\beta}
-\int_0^{\rbar_z}d\rbar\left({\partial\over\partial 
x^\beta}\right)\left(\pv_\alpha+2C_{\alpha\gamma}n^\gamma\right)~\\
&&
+\int_0^{\rbar_z}d\rbar\left({\rbar_z-\rbar\over\rbar_z\rbar}\right)
\left[\rbar^2\left({\partial^2\over\partial x^\alpha\partial x^\beta}\right)
\left(\ax-\px-\pvP-\TCP\right)\right]~.\nonumber
\enar
It is apparent that the shear matrix~$\gamma_{\alpha\beta}$ (or the
shear components $\gamma_1$, $\gamma_2$) is gauge-dependent, and they transform
as
\beeq
\tilde\gamma_{\alpha\beta}=\gamma_{\alpha\beta}+\LL_{\alpha,\beta}~,
\qquad
2\tilde\gamma_1=2\gamma_1+\LL_{\pp,\pp}-\LL_{\ttt,\ttt}~,
\qquad
-2\tilde \gamma_2=-2\gamma_2+\LL_{\ttt,\pp}+\LL_{\pp,\ttt}~.
\eneq
The shear matrix is independent of the rotation~$\Omega^i_o$, because
it only contributes to the second-order shear matrix. The expression for the
shear matrix in Eq.~\eqref{eq:gammaij} is again {\it incomplete} and 
gauge-dependent. However, when the conformal Newtonian gauge is adopted
and {\it only}
the scalar perturbations are considered, this expression corresponds
to the correct expression~$\hat\gamma_{\alpha\beta}$ in Eq.~\eqref{eq:hgamma}.
However, 
when the tensor modes are considered, the shear matrix~$\gamma_{\alpha\beta}$
contains the tensor 
contribution at the observer, but not at the source position.
This missing contribution is called the FNC term \cite{SCJE12b}
(in relation to the metric shear \cite{DOROST03}), and the
absence of the FNC term at the source position also indicates that
Eq.~\eqref{eq:gammaij} is incomplete.

One remaining component of the lensing observables is the rotation~$\omega$.
The anti-symmetric part of the distortion matrix in Eq.~\eqref{eq:DD2}
can be obtained in a similar way as
\bear
\label{eq:rot}
2\omega&\equiv&\DD_{21}-\DD_{12}
=\sin\ttt{\partial\over\partial\ttt}\dpp-{1\over
\sin\ttt}{\partial\over\partial\pp}\dtt+2\cos\ttt~\dpp \\
&=&2\Omega_o^n+\int_0^{\rbar_z}d\rbar~{1\over\rbar}\left({\ttt_\alpha
\over\sin\ttt}{\partial\over\partial\pp}-\pp_\alpha{\partial\over\partial\ttt}
\right)\left(\pv^\alpha+2C^\alpha_\beta n^\beta\right)
+\CCG_{\ttt,\pp}-\CCG_{\pp,\ttt}~\nnn
&=&2\Omega^n_o-\int_0^{\rbar_z}d\rbar~
\bdi{n}\cdot\nabla\times
\left(\pv^\alpha+2C^\alpha_\parallel\right)
-\bdi{n}\cdot\nabla\times\bm{\CCG}~. \nonumber
\enar
The rotation naturally changes in proportion to~$\Omega^n_o$, or the rotation 
of the local tetrad basis along the line-of-sight direction. In fact, the
physical rotation should be against the local tetrad basis, absorbing
the rotation~$\Omega^n_o$ of the local basis. The expression for the rotation
is gauge-dependent due to the presence of~$\CCG^\alpha$, and it transforms
as
\beeq
2\tilde\omega=2\omega+\bdi{n}\cdot\nabla\times\bm{\LL}~.
\eneq
Furthermore, the rotation is non-vanishing in the presence of the vector
and the tensor perturbations, while it vanishes in their absence.
This is also known as the Skrotsky effect \cite{SKROT57} or the gravitational
Faraday effect. However, this contribution to the rotation along the 
line-of-sight is artificial,
arising from the rotation of the tetrad basis against the global coordinate.
Indeed the tetrad basis vectors
are parallel transported along the geodesic, and
hence its contribution to the rotation vanishes, as the lensing
images are always
compared against the local tetrad basis. We come back to this issue
in Sec.~\ref{ssec:rotation}.

\section{Gauge-Invariant Formalism of Weak Lensing}
\label{sec:GIlensing}
Here we present the full gauge-invariant formalism of cosmological weak
lensing and resolve the issues associated with the lensing rotation.

\subsection{Geometric approach to weak lensing}
\label{ssec:geo}
In Sec.~\ref{sec:src} we have derived the source position~$x^\mu_s$ 
in a FRW coordinate, given the observed redshift~$z$ and the angular 
position~$n^i$ in the rest-frame of the observer. The source position
is different from the inferred position~$\bar x^\mu_z$, and this difference
is geometrically decomposed as the radial distortion~$\drr$, 
the angular distortion~$(\dtt,\dpp)$, and the time distortion~$\Delta\eta_s$
of the source position. Here we extend
the standard weak lensing formalism by applying the geometric approach
and checking the gauge-invariance of the lensing observables.

In order to develop a gauge-invariant formalism of weak lensing, 
we need to separate
observable quantities and unobservable quantities in weak lensing. The
first and the foremost is the source angular position~$s^i$ that is {\it not}
observable. While we measure the source position~$n^i$ in the rest-frame,
the source position~$x^\mu_s$ is constructed in a FRW coordinate by 
tracing the photon path backward in time, and its geometric deviations
from the inferred position are gauge-dependent as derived in 
Sec.~\ref{sec:src}. So the source angular position in a {\it FRW coordinate}
can be legitimately expressed as 
\beeq
s^\alpha=n^\alpha+\dtt~\ttt^\alpha+\sin\ttt~\dpp~\pp^\alpha~,
\eneq
but it is {\it not} observable as implied in Eq.~\eqref{eq:dsang}.
Indeed, the coordinate transformation in Eq.~\eqref{eq:coord} induces the
gauge-transformation
\beeq
\tilde s^\alpha=s^\alpha+{\ttt_\beta\LL^\beta_s\over\rbar_z}
~\ttt^\alpha+{\pp_\beta\LL^\beta_s\over\rbar_z}~\pp^\alpha~.
\eneq
Consider again an extended source in the sky of the observer that appears
subtended by an infinitesimal angular size $(d\ttt,d\pp)$ as in 
Sec.~\ref{sec:stdlensing}. The angular size of the source in a FRW coordinate
is then
\bear
ds^\alpha&=&\ttt^\alpha
\left[d\ttt+\left(d\ttt{\partial\over\partial\ttt}+d\pp{\partial\over\partial
\pp}\right)\dtt-d\pp~\cos\ttt~\sin\ttt~\dpp\right]~\\
&&
+\pp^\alpha\left[\sin\ttt~d\pp\left(1+\cot\ttt~\dtt+{\partial\over\partial\pp}
\dpp\right)+d\ttt{\partial\over\partial\ttt}\left(\sin\ttt~\dpp\right)\right]
-n^\alpha\left[\dtt~d\ttt+\sin^2\ttt~\dpp~d\pp\right]~.\nonumber
\enar
Compared to Eq.~\eqref{eq:infds}, we notice the presence of the additional
term along the line-of-sight.
Furthermore, the source position in a FRW coordinate is indeed specified
in terms of its four-dimensional position:
\beeq
\eta_s=\bar\eta_z+\Delta\eta_s~,\qquad\qquad
x_s^\alpha=\bar x^\alpha_z+\drr~n^\alpha+\Delta x^\alpha_\perp
=\rbar_z s^\alpha+\drr~n^\alpha~,
\eneq
Therefore, the source size in a FRW coordinate that would appear
subtended by angular size~$(d\ttt,d\pp)$ at the fixed observed redshift~$z$
is then
\beeq
dx_s^\alpha=\rbar_z~ds^\alpha+d\left(\drr\right)n^\alpha+\drr~dn^\alpha
\equiv\rbar_z\left(dn^\alpha+\Delta s^\alpha\right)~,
\eneq
where the first term represents the source size in the absence of any
perturbations and the second term with~$\Delta s^\alpha$ is defined to
capture any deviations due to the perturbations
\bear
\label{eq:dsalpha}
\Delta s^\alpha&=&\left(ds^\alpha-dn^\alpha\right)
+{d\left(\drr\right)\over\rbar_z}n^\alpha
+{\drr\over\rbar_z}dn^\alpha
=
\ttt^\alpha\left[d\ttt\left({\partial\over\partial\ttt}\dtt+{\drr\over\rbar_z}
\right)+\sin\ttt~d\pp\left({1\over\sin\ttt}{\partial\over\partial\pp}\dtt
-\cos\ttt~\dpp\right)\right]~\nnn
&&
+\pp^\alpha\left[\sin\ttt~d\pp\left(\cot\ttt~\dtt+{\partial\over\partial\pp}
\dpp+{\drr\over\rbar_z}\right)
+d\ttt{\partial\over\partial\ttt}\left(\sin\ttt~\dpp\right)\right] \nnn
&&
-n^\alpha\left[d\ttt\left(\dtt-{1\over\rbar_z}{\partial\over\partial\ttt}\drr
\right)+\sin\ttt~d\pp\left(\sin\ttt~\dpp-{1\over\rbar_z\sin\ttt}{\partial\over
\partial\pp}\drr\right)\right]~.
\enar
We will not need to consider the variation in the time 
coordinate~$\Delta\eta_s$ at the linear order. Compared to 
Eq.~\eqref{eq:infds}, the radial distortion~$\drr$ also contributes to
the source size~$\Delta s^\alpha$ in a FRW coordinate. As we further proceed,
it will become clear that the radial component in proportion to~$n^\alpha$
will be projected out, so that ignoring this part in Eq.~\eqref{eq:infds}
causes no systematic errors at the linear order in perturbations.
However, note the presence of the radial distortion~$\drr$ in proportion
to~$\ttt^\alpha$ and~$\pp^\alpha$, compared to Eq.~\eqref{eq:infds}.

In Section~\ref{sec:observables}, we set up the tetrad basis 
at the observer position to relate the local observables
to the photon wavevector~$k^\mu$ in a FRW coordinate. By solving the
geodesic equation in Section~\ref{sec:src}, the photon wavevector is traced
back to the source position. To properly relate the source size~$dx_s^\mu$
in a FRW  coordinate to the physical size~$dx_s^i$ 
of the source in its rest-frame, we need 
to construct the tetrad basis in the source rest-frame. The expressions
for the tetrad basis vectors at the source position are almost identical to
those at the observer position in Eqs.~\eqref{eq:srcv} and~\eqref{eq:stetrad},
except the fact that all the quantities are evaluated at the source position
and the velocity vector~$\VV^\alpha$ represents the velocity of the source.
Finally, we need to define a two-dimensional plane in the rest-frame of
the source, perpendicular to the light propagation. A small area in this
plane such as a circle or an ellipse will appear distorted at the observer, 
and its observed shape compared to that in the source rest-frame will
determine the physical lensing observables such as the magnification
(convergence), the shear, and the rotation that we can measure

First, the photon propagation direction measured by the source would be 
expressed in a FRW coordinate as
\beeq
\label{eq:nsrc}
n_s^\mu=-{k^\mu\over \omega}+u^\mu \simeq
\frac1a(0,n^\alpha)+\frac1a\left[n^\beta(\VV-\BB)_\beta,~\Delta n^\alpha
\right]_s+\OO(2)~,
\eneq
where we defined the perturbation to the photon propagation at the source
position
\beeq
 \Delta n_s^\alpha\equiv \dea_s^\alpha+\VV_s^\alpha-\dhnu_s~n^\alpha~.
\eneq
By using Eqs.~\eqref{eq:inita}, \eqref{eq:dea}, \eqref{eq:dzfull}, 
we can explicitly derive the perturbation as
\bear
\Delta n_s^\alpha&=&\left(-V^\alpha_\perp+C^\alpha_\beta n^\beta
-\TCP n^\alpha-\epsilon^\alpha{}_{ij}n^i\Omega^j\right)_o
-\left(H\chi+\px-\TCP\right)n^\alpha
+\left(V^\alpha_\perp-2C^\alpha_\beta n^\beta
-\CCG^\alpha{}_{,\beta}n^\beta\right)\nnn
&&
+\int_0^{\rbar_z} d\rbar~{1\over\rbar}\left[-\hat\nabla^\alpha
\left(\ax-\px\right)+n^\beta\hat\nabla^\alpha\pv_\beta+
n^\beta n^\gamma\hat\nabla^\alpha C_{\beta\gamma}\right] ~.
\enar
Again, the photon wavevector $k^\mu$ at the source position is fully
determined (up to $\Omega^i_o$) given the observation at the observer position.
Once the source velocity is specified as an ``observer'' (in fact, an
``emitter''), 
the photon propagation direction~$n^\mu_s$ (or $\Delta n^\alpha_s$)
measured by this observer is then fully determined, as indicated above.

Next, this photon propagation direction~$n^\mu_s$ in a FRW coordinate needs to
be projected into the rest-frame of the source to define a two-dimensional
area perpendicular to the light propagation.
In the source rest-frame, the ``observed angle~$n_s^i$'' measured by
an ``observer'' at the source position is then
\bear
\label{eq:obssrc}
n^i_s&=&
g_{\mu\nu}e_i^\mu n^\nu_s\simeq g_{\alpha\beta}e_i^\alpha n^\beta+\OO(2)
=n^i+2\CC_{\alpha\beta}n^\beta \delta^{\alpha i}+\Delta n^\alpha\delta_\alpha^i
+n_\alpha \delta e^\alpha_j~\delta^{ji}~\\
&=&n^i+\left[V^i_\perp+\TCP n^i-C^i_\alpha n^\alpha
-\epsilon^i{}_{jk}n^j\Omega^k\right]^s_o
+\int_0^{\rbar_z} d\rbar~{1\over\rbar}\left[-\hat\nabla^i
\left(\ax-\px\right)+n^\alpha\hat\nabla^i\pv_\alpha+
n^\alpha n^\beta\hat\nabla^i C_{\alpha\beta}\right] ~, \nonumber
\enar
where the integration represents the perturbation contribution of the
photon propagation along the line-of-sight, the terms at the source
and the observer positions are due to the misalignment of their rest-frames
and FRW coordinates, and the expression is fully gauge-invariant.
This equation is indeed the correct lens equation, which generalizes
over Eqs.~\eqref{eq:basic} and~\eqref{eq:dsang}.

The observed angle~$n^i_s$ 
at the source (or the emission angle) in the rest-frame
is identical to the observed angle~$n^i$ at the observer position in the 
background due to the construction of our local tetrad basis.
Even in the absence of any perturbations, these two angles can be different,
simply by setting up the local coordinates differently. In perturbation theory,
all these differences are, however, shifted to the perturbations, while the
background remains unaffected, and indeed the difference in the observed angles
is proportional to the difference in rotation of the local tetrad bases
at the source and the observer positions:
\beeq
\label{eq:nrot}
n^i_s-n^i\ni -\epsilon^i{}_{jk}n^j\left(\Omega_s^k-\Omega_o^k\right)
=\ttt^i\left(\Omega^\pp_s-\Omega^\pp_o\right)
-\pp^i\left(\Omega^\ttt_s-\Omega^\ttt_o\right)~.
\eneq
At the moment, the rotations~$\Omega^i$ of the local tetrad bases at both
positions are undetermined.
By parameterizing the source angle with perturbations $(\Delta\ttt,\Delta\pp)$
in the source rest-frame as
\beeq
\label{eq:sang}
n^i_s\equiv(\ttt_s,\pp_s)~,\qquad \ttt_s\equiv\ttt+\Delta\ttt~,\qquad 
\pp_s\equiv\pp+\Delta\pp~,
\eneq
we can express the light propagation direction in the source rest-frame
\beeq
\label{eq:lens3}
n^i_s= n^i+\ttt^i\Delta\ttt+\pp^i\sin\ttt~\Delta\pp~,
\eneq
and set up two orthonormal bases~$\ttt^i_s$ and~$\pp^i_s$ 
that are perpendicular to the propagation direction and form an orthonormal
basis
\bear
\label{eq:twob}
\ttt^i_s&=&-n^i\Delta\ttt+\ttt^i+\pp^i\cos\ttt~\Delta\pp
\equiv \ttt^i+\Delta\ttt_s^i~,\\
\pp^i_s&=&-n^i\sin\ttt~\Delta\pp-\ttt^i\cos\ttt~\Delta\pp+\pp^i
\equiv\pp^i+\Delta\pp^i_s~,
\enar
where we defined the perturbation vectors~$\Delta\ttt^i_s$ 
and~$\Delta\pp^i_s$ for the notational simplicity and note that $(\ttt,\pp)$
in Eq.~\eqref{eq:sang} is the same observed angle at the observer position.
Given the explicit expression for~$n_s^i$ in Eq.~\eqref{eq:obssrc}, we
can compute the perturbations to the source angle~$(\ttt_s,\pp_s)$ as
\bear
\label{eq:deltas}
\Delta\ttt&=&\ttt_i\left[V^i_\perp-C^i_\alpha n^\alpha
-\epsilon^i{}_{jk}n^j\Omega^k\right]^s_o
-\ttt_i\int_0^{\rbar_z} d\rbar\left[\left(\ax-\px\right)^{,i}
-\pv_\alpha{}^{,i}n^\alpha-C_{\alpha\beta}{}^{,i}n^\alpha
n^\beta\right]~\\
&=&\ttt_i\left[V^i_\perp-C^i_\alpha n^\alpha
-\epsilon^i{}_{jk}n^j\Omega^k\right]^s_o
-\ttt_i\int_0^{\rbar_z} d\rbar\left(\ax-\px-\pvP-\TCP\right)^{,i}
-\int_0^{\rbar_z}{d\rbar\over\rbar}\ttt_i\left(\pv^i+2~C_\parallel^i\right)~,
\nonumber
\enar
and the expression for~$\sin\ttt~\Delta\pp$ is equivalent with $\ttt_i$
replaced by~$\pp_i$. These perturbations in angle are gauge-invariant, and
they contain the rotation of the local tetrad bases
\beeq
\Delta\ttt\ni-\epsilon_{ijk}\ttt^in^j\left(\Omega^k_s-\Omega^k_o\right)
=\Omega^\pp_s-\Omega^\pp_o~,\qquad
\qquad
\sin\ttt~\Delta\pp\ni-\epsilon_{ijk}\pp^in^j\left(\Omega^k_s-\Omega^k_o\right)
=-\Omega_s^\ttt+\Omega_o^\ttt~.
\eneq

Having obtained two orthonormal basis vectors, we are in a position to
compute the physical size in the source rest-frame that is expressed
as $dx^\mu_s$ in a FRW coordinate and would be measured at the
observer position with angular size~$(d\ttt,d\pp)$. In reality, the source
position would be measured as the observed angle~$(\ttt,\pp)$ at the
observed redshift~$z$, and the physical size of a standard ruler in the
source rest-frame would be subtended by its angular size~$(d\ttt,d\pp)$,
as opposed to our reverse construction of the standard ruler.
So by measuring the angular size with prior knowledge of the standard ruler,
we can construct the lensing observables, and compared to the standard
lensing formalism, the lensing observables obtained this way
are physically well-defined. The source size~$dx^\mu_s$ in a FRW coordinate
can be projected into the rest-frame along the two orthogonal directions
in the plane perpendicular to the light propagation direction:
\bear
\label{eq:size}
dL_{\ttt_s}&=&g_{\mu\nu}e_i^\mu dx_s^\nu\ttt_s^i
\simeq g_{\alpha\beta}e_i^\alpha\ttt^i dx_s^\beta+\OO(2)
=a_s\rbar_z\left[d\ttt+\ttt_\alpha\Delta s^\alpha
+2\CC_{\alpha\beta}\ttt^\alpha dn^\beta+dn_\alpha\left(\Delta\ttt^\alpha_s
+\delta e^\alpha_i\ttt^i\right)\right]~,\nnn
dL_{\pp_s}&=&g_{\mu\nu}e_i^\mu dx_s^\nu\pp_s^i
=a_s\rbar_z\left[\sin\ttt~ d\pp+\pp_\alpha\Delta s^\alpha
+2\CC_{\alpha\beta}\pp^\alpha dn^\beta+dn_\alpha\left(\Delta\pp_s^\alpha 
+\delta e^\alpha_i\pp^i\right)\right]~,
\enar
where we used $dx_s^\eta=d(\Delta\eta_s)=\OO(1)$. 
The perturbations in the light propagation are captured by~$\Delta s^\alpha$,
and their contributions to the physical sizes in the rest-frame are
from Eq.~\eqref{eq:dsalpha}
\bear
\ttt_\alpha\Delta s^\alpha
&=&d\ttt\left({\partial\over\partial\ttt}\dtt+{\drr\over\rbar_z}\right)
+\sin\ttt~d\phi\left({1\over\sin\ttt}{\partial\over\partial\pp}\dtt
-\cos\ttt~\dpp\right)~,\\
\pp_\alpha\Delta s^\alpha
&=&
\sin\ttt~d\phi\left(\cot\ttt~\dtt+{\partial\over\partial\pp}\dpp
+{\drr\over\rbar_z}\right)
+d\ttt~{\partial\over\partial\ttt}\left(\sin\ttt~\dpp\right)~.
\enar
The changes in the observed angles at the source position due to the distortion
of the local frame contribute to the physical sizes
\bear
dn_\alpha\left(\Delta\ttt_s^\alpha+\delta e^\alpha_i\ttt^i\right)
&=&d\ttt~\delta e^\ttt_\ttt+\sin\ttt ~d\pp
\left(\delta e^\pp_\ttt+\cos\ttt~\Delta\pp\right)~,\\
dn_\alpha\left(\Delta\pp_s^\alpha+\delta e^\alpha_i\pp^i\right)
&=&d\ttt\left(\delta e^\ttt_\pp-\cos\ttt~\Delta\pp\right)+
\sin\ttt ~d\pp~\delta e^\pp_\pp~, 
\enar
where we used the short-hand notation~$\delta e^\ttt_\pp\equiv\ttt_\alpha
\delta e^\alpha_j\pp^j$ for instance. Putting it altogether, the physical sizes
in Eq.~\eqref{eq:size} can be simplified as
\bear
{dL_{\ttt_s}\over a_s\rbar_z}&=&d\ttt\left(1+{\drr\over\rbar_z}+{\partial\over
\partial\ttt}\dtt+\CC_{\ttt\ttt}\right)
+\sin\ttt~ d\pp\left(
{1\over\sin\ttt}{\partial\over\partial\pp}\dtt-
\cos\ttt~\dpp+2~\CC_{\ttt\pp}+\delta e^\phi_\ttt+\cos\ttt~\Delta\pp\right)~,
\nnn
{dL_{\pp_s}\over a_s\rbar_z}&=&
d\ttt\left[{\partial\over\partial\ttt}\left(\sin\ttt
~\dpp\right)+2~\CC_{\ttt\pp}+\delta e^\ttt_\phi-\cos\ttt~\Delta\pp\right]
\nnn
&&\qquad
+\sin\ttt ~d\pp\left(
1+{\drr\over\rbar_z}+\cot\ttt~\dtt+{\partial\over\partial\pp}
\dpp+\CC_{\pp\pp}\right)~,~~~~~~~~~~~
\enar
where we used the same short-hand notation for~$\CC_{\alpha\beta}$
and noted that $\delta e^\ttt_\ttt=-\CC_{\ttt\ttt}$ and 
$\delta e^\pp_\pp=-\CC_{\pp\pp}$.
Noting that the angular diameter distance in the background is 
$\bar D_A=a_z\rbar_z$  and the scale factor~$a_s$ 
at the source is related to the perturbation~$\dz$
in the observed redshift
\beeq
a_s={1+\dz\over 1+z}=a_z\left(1+\dz\right)~,
\eneq
we can define the distortion matrix~$\hDD$ ({\it with hat}) as
\beeq
\label{eq:DD3}
\left(\begin{array}{c}dL_{\ttt_s} \\ dL_{\pp_s}\end{array}\right)
\equiv\bar D_A
\left(\begin{array}{cc}\hDD_{11}&\hDD_{12}\\\hDD_{21}&\hDD_{22}\end{array}
\right)\left(\begin{array}{c}d\ttt\\\sin\ttt~d\pp\end{array}\right)~,
\eneq
and the elements of the matrix can be read off as
\bear
\hDD_{11}&=&\left(1+{\partial\over\partial\ttt}\dtt\right)
+\dz+{\drr\over\rbar_z}+\CC_{\ttt\ttt}~,\\
\hDD_{22}&=&\left(1+\cot\ttt~\dtt+{\partial\over\partial\pp}\dpp\right)
+\dz+{\drr\over\rbar_z}+\CC_{\pp\pp}~,\\
\hDD_{12}&=&\left({1\over\sin\ttt}{\partial\over\partial\pp}\dtt-
\cos\ttt~\dpp\right)+2~\CC_{\ttt\pp}+\delta e^\phi_\ttt+\cos\ttt~\Delta\pp~,\\
\hDD_{21}&=&\left(\cos\ttt+\sin\ttt{\partial\over\partial\ttt}\right)\dpp
+2\CC_{\ttt\pp}+\delta e^\ttt_\phi-\cos\ttt~\Delta\pp~,
\enar
where the elements of the distortion matrix~$\DD$ ({\it without hat})
in Eq.~\eqref{eq:DD2}
are put in the parenthesis to facilitate the comparison.
There exist notable differences in Eq.~\eqref{eq:DD3} that are physically
unambiguous, when compared to the distortion matrix~$\DD$:
The source position in the sky
appears to the observer as the observed angle~$(\ttt,\pp)$ at the 
observed distance~$\bar D_A(z)$ (or at the observed redshift~$z$).
Given the knowledge of the physical sizes and the orientation
$(dL_{\ttt_s},dL_{\pp_s})$ of a standard ruler, their ``observed angular size''
$(ds_\ttt,ds_\pp)$ in the absence of perturbations would be
\beeq
\left(\begin{array}{c}dL_{\ttt_s} \\ dL_{\pp_s}\end{array}\right)
\equiv\bar D_A
\left(\begin{array}{c}ds_\ttt\\\sin\ttt~ds_\pp\end{array}\right)~,
\eneq
and hence the distortion matrix defined above is indeed the lensing
distortion matrix, providing the relation between
the angular sizes at the source and at the observer:
\beeq
\left(\begin{array}{c}ds_\ttt\\\sin\ttt~
ds_\pp\end{array}\right)=
\left(\begin{array}{cc}\hDD_{11}&\hDD_{12}\\\hDD_{21}&\hDD_{22}\end{array}
\right)\left(\begin{array}{c}d\ttt\\\sin\ttt~d\pp\end{array}\right)~.
\eneq
Note that their observed angular position is $(\ttt,\pp)$,
not $(\ttt+\dtt,\pp+\dpp)$, naturally resolving the ambiguity in 
Eq.~\eqref{eq:DD2}.

To ensure these expressions are correct and physically well-defined, 
we explicitly verify the gauge-invariance of the distortion matrix. 
Any physically well-defined quantities should be gauge-invariant 
at the linear order \cite{YODU17}. For a general coordinate transformation
in Eq.~\eqref{eq:coord}, 
we already derived in Sec.~\ref{sec:src} how the geometric
distortions $(\drr,\dtt,\dpp)$ of the source position transform.
Given that $\Delta\pp$ in Eq.~\eqref{eq:deltas} is gauge-invariant,
it is relatively straightforward to show that each component of
the distortion matrix~$\hDD$ is indeed gauge-invariant, 
by using how the remaining components transform
\bear
\CC_{\alpha\beta}&=&\left(\px+H\chi\right)\delta_{\alpha\beta}+\CCG_{\alpha,
\beta}+C_{[\beta,\alpha]}^{(v)}+C_{\alpha\beta}~~\mapsto~~
\tilde\CC_{\alpha\beta}=\CC_{\alpha\beta}
-\HH T\delta_{\alpha\beta}-\LL_{\alpha,\beta}-L^{(v)}_{[\beta,\alpha]}~,\nnn
-\delta e^\alpha_i&=&\left(\px+H\chi\right)\delta^\alpha_i
+\CCG^\alpha{}_{,i}+C^\alpha_i+\epsilon^\alpha{}_{ij}\Omega^j
~~\mapsto~~-\widetilde{\delta e}{}^\alpha_i=-\delta e^\alpha_i
-\HH T\delta^\alpha_i-\LL^\alpha{}_{,i}~.
\enar

Having verified the gauge-invariance of the distortion matrix, we will
decompose the distortion matrix and derive the lensing observables.
First, we consider the convergence of the distortion matrix.
Consider two physical separation vectors $dL_{d\ttt}^i$
and $dL_{d\pp}^i$ in the source rest-frame that would appear to the observer
subtended by $d\ttt$ and $d\pp$, respectively. Using the distortion matrix
in Eq.~\eqref{eq:DD3}, we can derive these two separation vectors in
the source rest-frame:
\bear
dL_{d\ttt}^i&=&
\bar D_A\left(\hDD_{11}\ttt_s^i+\hDD_{21}\pp_s^i\right)d\ttt
=\bar D_A\left(ds_\ttt^{d\ttt}\ttt_s^i+\sin\ttt ~ds_\pp^{d\ttt}
\pp_s^i\right)~,\\
dL_{d\pp}^i&=&
\bar D_A\left(\hDD_{12}\ttt_s^i+\hDD_{22}\pp_s^i\right)\sin\ttt~d\pp
=\bar D_A\left(ds_\ttt^{d\pp}\ttt_s^i+\sin\ttt~ds_\pp^{d\pp}\pp_s^i\right)~.
\enar
Note that we have two separation vectors and two angular vectors in the
source rest-frame. The physical area $dA$ spanned by these two separation
vectors in the source rest-frame is 
\beeq
dA_s=\epsilon^{ijk}n_s^i~dL_{d\ttt}^j ~dL_{d\pp}^k
=\bar D_A^2\det\hDD~d\Omega_o~,\qquad\qquad 
d\Omega_o=\sin\ttt d\ttt d\pp~,
\eneq
where the relation is non-perturbative, given the distortion matrix.
This relation is indeed the definition for the angular diameter 
distance~$\dA$:
\beeq
dA_s\equiv
\dA^2~d\Omega_o~,\qquad\qquad \dA\equiv\bar D_A\left(1+\ddD\right)~,\qquad
\qquad\det\hDD=\left(1+\ddD\right)^2~,
\eneq
where $\ddD$ is the fluctuation in the angular diameter 
distance.\footnote{The fluctuation~$\ddD$ in the angular diameter distance
is identical to the fluctuation in the luminosity distance, because
the luminosity distance is $\dL=(1+z)^2\dA$.}
By expressing the physical area in terms of the observed angles in the
absence of perturbations (or the source angles),
\beeq
dA_s=\epsilon^{ijk}n_s^i~dL_{\ttt_s}^j ~dL_{\pp_s}^k
=\bar D_A^2~d\Omega_s~,\qquad
\qquad d\Omega_s=\sin\ttt\left(ds_\ttt^{d\ttt}ds_\pp^{d\pp}
-ds_\ttt^{d\pp}ds_\pp^{d\ttt}\right)~,
\eneq
the determinant of the distortion matrix is indeed related 
to the observed magnification~$\mu$ as
\beeq
\det\hDD={d\Omega_s\over d\Omega_o}=\mu^{-1}~.
\eneq
At the linear order in perturbations, the determinant of the distortion matrix
is
\beeq
\label{eq:ddD}
\left(\det\hDD\right)^{1/2}\simeq
1+\dz+{\drr\over\rbar_z}-\kappa+\frac12\PP^{\alpha\beta}\CC_{\alpha\beta}
+\OO(2)=1+\ddD~,
\eneq
where the convergence~$\kappa$ in the standard lensing formalism
is given in Eq.~\eqref{eq:kappa} and we used
\beeq
\PP^{\alpha\beta}\CC_{\alpha\beta}=\CC_{\ttt\ttt}+\CC_{\pp\pp}=\CC^\alpha_
\alpha-\CC_\parallel=2\left(\px+H\chi\right)-\TCP-{1\over\rbar_z}
\hat\nabla_\alpha\CCG^\alpha~.
\eneq
The convergence~$\hat\kappa$ ({\it with hat})
 of the distortion matrix~$\hDD$ is then
\beeq
\label{eq:hkappa}
\hat\kappa\equiv1-\frac12\up{Tr}~\hDD\simeq1-\left(\det\hDD\right)^{1/2}
+\OO(2)=-\ddD~.
\eneq
Therefore, the diagonal component~$\hat\kappa$ ({\it with hat})
of the lensing observables is related to the physical magnification 
(or the luminosity distance), and it is indeed gauge-invariant,
as opposed to the gauge-dependent convergence~$\kappa$ ({\it without hat})
in Eq.~\eqref{eq:kappa}. The expression is independent of the 
rotations~$\Omega^i$ at the source and the observer positions, because
it only involves the ratio of the physical and the inferred areas.
The detailed properties of the gauge-invariant luminosity distance
fluctuation are studied in \cite{BIYO16,BIYO17}.

Next, we compute the gravitational lensing shear of the distortion 
matrix~$\hDD$. Two components of the lensing shear can be readily read off 
from the distortion matrix as
\bear
\label{eq:hgamma1}
2\hat\gamma_1&\equiv&\hDD_{22}-\hDD_{11}=\left(
\cot\ttt~\dtt+{\partial\over\partial\pp}\dpp-{\partial\over\partial\ttt}\dtt
\right)+\CC_{\pp\pp}-\CC_{\ttt\ttt}~,\\
\label{eq:hgamma2}
-2\hat\gamma_2&\equiv&\hDD_{12}+\hDD_{21}=\left({1\over\sin\ttt}
{\partial\over\partial\pp}\dtt+\sin\ttt{\partial\over\partial\ttt}\dpp\right)
+\delta e^\phi_\ttt+\delta e^\ttt_\phi+4~\CC_{\ttt\pp}~,
\enar
where the shear components~$\gamma_1$ and~$\gamma_2$ in Eqs.~\eqref{eq:gamma1}
and~\eqref{eq:gamma2} are shown in the parenthesis. The additional
contributions in the shear components~$\hat\gamma_1$ and~$\hat\gamma_2$
arise due to the change of a FRW frame to the source rest-frame:
\bear
2\hat\gamma_1&\ni&\CC_{\pp\pp}-\CC_{\ttt\ttt}=\left(\pp^\alpha\pp^\beta
-\ttt^\alpha\ttt^\beta\right)\left(C_{\alpha\beta}+\CCG_{\alpha,\beta}\right)
~,\\
-2\hat\gamma_2&\ni&\delta e^\phi_\ttt+\delta e^\ttt_\phi+4~\CC_{\ttt\pp}
=\left(\ttt^\alpha\pp^\beta+\pp^\alpha\ttt^\beta\right)
\left(C_{\alpha\beta}+\CCG_{\alpha,\beta}\right)~,
\enar
which exactly cancel the gauge-dependent terms in~$\gamma_1$ and~$\gamma_2$
and add the tensor contribution~$C_{\alpha\beta}$ at the source position
(or the FNC term at the source \cite{SCJE12b}).
In the same way in Sec.~\ref{sec:stdlensing}, we can construct the
spin~$\pm2$ shear components 
${}_{\pm2}\hat\gamma=\hat\gamma_1\pm i\hat\gamma_2$ and derive the shear matrix
\bear
\label{eq:hgamma}
\hat\gamma_{\alpha\beta}&=&-\left(C_{\alpha\beta o}+C_{\alpha\beta}\right)
-\int_0^{\rbar_z}d\rbar\left({\partial\over\partial 
x^\beta}\right)\left(\pv_\alpha+2C_{\alpha\gamma}n^\gamma\right)~\\
&&
+\int_0^{\rbar_z}d\rbar\left({\rbar_z-\rbar\over\rbar_z\rbar}\right)
\left[\rbar^2\left({\partial^2\over\partial x^\alpha\partial x^\beta}\right)
\left(\ax-\px-\pvP-\TCP\right)\right]~.\nonumber
\enar
Compared to Eq.~\eqref{eq:gammaij}, the shear matrix~$\hat\gamma_{\alpha\beta}$
({\it with hat}) is gauge-invariant, 
and there exist additional tensor contribution at the
source position, or the FNC term \cite{SCJE12b}. This contribution arises
because the physical length is defined in the source rest-frame, not in
the FRW frame. Indeed, its presence is necessary to prevent the 
infrared-divergences of the shear matrix (see Sec.~\ref{ssec:GW}).

Finally, we derive the last remaining lensing observable of the distortion
matrix~$\hDD$, or the rotation ({\it with hat})
\bear
\label{eq:homega}
2\hat\omega&=&\hDD_{21}-\hDD_{12}=\left(\sin\ttt{\partial\over\partial\ttt}
\dpp-{1\over\sin\ttt}{\partial\over\partial\pp}\dtt+2\cos\ttt~\dpp\right)
+\delta e^\ttt_\pp-\delta e^\pp_\ttt-2~\cos\ttt~\Delta\pp \nnn
&=&2\left(\Omega^n_o-\Omega_s^n\right)-2~\cos\ttt~\Delta\pp
-\int_0^{\rbar_z}d\rbar~\bdi{n}\cdot\nabla\times
\left(\pv^\alpha+2C^\alpha_\parallel\right)~, 
\enar
where the rotation~$\omega$ ({\it without hat})
from~$\DD$ in Eq.~\eqref{eq:rot} is shown
in the parenthesis. The additional contribution to the rotation~$\hat\omega$
from the frame change is
\beeq
2\hat\omega\ni\delta e^\ttt_\pp-\delta e^\pp_\ttt-2\cos\ttt~\Delta\pp
=-\left(\CCG_{\ttt,\pp}-\CCG_{\pp,\ttt}\right)-2\Omega^n_s-2\cos\ttt~\Delta\pp
~,
\eneq
which cancels the gauge-dependent part in~$\omega$ ({\it without hat}).
Compared to Eq.~\eqref{eq:rot}, the rotation~$\hat\omega$ has extra
terms that involves the rotation at the source position, i.e., 
$\Omega^n_s$ and $\Delta\pp$. Any physical rotation should be measured
against the local tetrad bases, and hence it naturally involves the 
difference $(\Omega^n_o-\Omega^n_s)$ in orientation. 

In short, the standard lensing formalism lacks the specification of the source
rest-frame, in which the physical size and shape are defined. By setting
up a local tetrad basis at the source position, we fixed the standard
weak lensing formalism and explicitly verified the gauge-invariance
of the lensing observables.

\subsection{Physical rotation of the images vs rotation of the tetrad basis}
\label{ssec:rotation}
We have computed the rotation~$\hat\omega$ as one of the lensing observables
in Sec.~\ref{ssec:geo} by comparing the orientation of the physical 
lengths~$L_{\ttt_s}$ and~$L_{\pp_s}$ 
along two orthogonal directions in the source
rest-frame to the orientation they appear in the observer rest-frame.
However, as apparent in Eq.~\eqref{eq:homega}, the rotation~$\hat\omega$
depends on the orientations~$\Omega_s$ and~$\Omega_o$
of the local rest-frames of the source and
the observer around the light propagation direction. Technically, if we
change~$\Omega_o$, the observed angle~$(\ttt,\pp)$ changes or rotates.
But in reality, the fact that we assign some numbers
to the observed light propagation direction $(\ttt,\pp)$ means that
we already fixed the orientation~$\Omega_o$ of our local coordinate.
However, the orientation~$\Omega_s$ of the source remains unspecified,
and the lensing observable~$\hat\omega$ depends on this unspecified
orientation. In other words, by rotating the orientation of the local 
coordinate in the source rest-frame, we can set the rotation to zero.
The question then naturally arises, ``is the rotation physical and 
measurable?''

Let's revisit how the lensing observables are defined physically. The lensing
convergence~$\hat\kappa$
 (or the determinant at the linear order) can be defined as
the ratio of the physical area in the source rest-frame
to the area inferred from the observed angular size. So it is independent
of the orientations at the source and the observer positions. The lensing shear
can be defined as the observed ellipticity of a circular object 
in the source rest-frame. Hence the shear is again
independent of the orientation of the source, while
it depends on the orientation of the observer coordinate.
Though Eq.~\eqref{eq:hgamma} is independent of~$\Omega_o$ at the linear
order, the shear matrix 
rotates in general as we rotate the local coordinate, and
two shear components~$\hat\gamma_1$ and~$\hat\gamma_2$ represent the
shear amplitude and orientation (the latter of which depends on the orientation
of the observer coordinate). To define the rotation in a physical way,
we need to synchronize the orientations at the source and the observer
positions, from which any deviation can be assigned to rotation.
In a curved spacetime, this can be achieved 
by parallel transporting the local tetrad basis along the photon propagation
direction. Indeed, this is the only physically meaningful way,
as the parallel transport is path-dependent. 

One subtlety associated with this procedure is that when the local tetrad
vectors~$e^\mu_a$ are parallel transported to the source position, the 
transported
timelike vector~$\ee_0^\mu$ (with {\it check}) at the source position
is different from the source velocity~$u^\mu_s$. To construct the
parallel transported basis~$\hat e^\mu_a$ in the source rest-frame, we need
to Lorentz boost the transported basis~$\ee^\mu_a$ with the source velocity
vector~$u^\mu_s$. Indeed, one can show \cite{MIYO18} that any 
vectors~$A^i_\perp$ and~$B^i_\perp$
defined in a plane perpendicular to a photon propagation
direction~$n^i$ of a null vector in Eq.~\eqref{eq:rest} satisfy
\beeq
\label{eq:inv}
A^i_\perp B^i_\perp=\hat A^i_\perp \hat B^i_\perp~,
\eneq
where those with hat represent the components after the Lorentz boost
(see, also, \cite{LECH06}).
In our case, the photon propagation direction~$\breve{n}^i$ constructed 
from~$n^\mu_s$ with the transported timelike vector~$\ee_0^\mu$ 
changes to~$\hat n^i$ due to the Lorentz boost, and indeed we already
derived the photon propagation direction $\hat n^i=n^i_s$
in Eq.~\eqref{eq:obssrc}.

For our purposes, the plane at the observer is defined in terms of two
orthonormal vectors $\ttt^\mu$-$\pp^\mu$ (or $\ttt^i$-$\pp^i$ in the
rest-frame), and the transported plane
spanned by $\eett_s^\mu$-$\eepp_s^\mu$ defines an 
``oriented'' plane perpendicular
to the photon propagation direction~$n^\mu_s$ but with the transported
velocity~$\ee_0^\mu$. Equation~\eqref{eq:inv} implies that the physical
size and the shape $\hat A^i_\perp$ defined in a plane perpendicular to the 
photon propagation direction~$n^\mu_s$ with the correct source
velocity~$u^\mu_s$
is equivalent to those defined in a plane perpendicular to~$n^\mu_s$
with~$\ee_0^\mu$. Therefore, the source size~$dx^\mu$ that appears
subtended by the angular size~$(d\ttt,d\pp)$ can be projected into any of 
the two planes for computing the lensing observables. This argument
is indeed crucial to the Jacobi mapping method in Sec.~\ref{ssec:jacobi}
that is often implicitly assumed (see, e.g., \cite{BONVI08} for the
discussion).

\begin{figure}
\centerline{\psfig{file=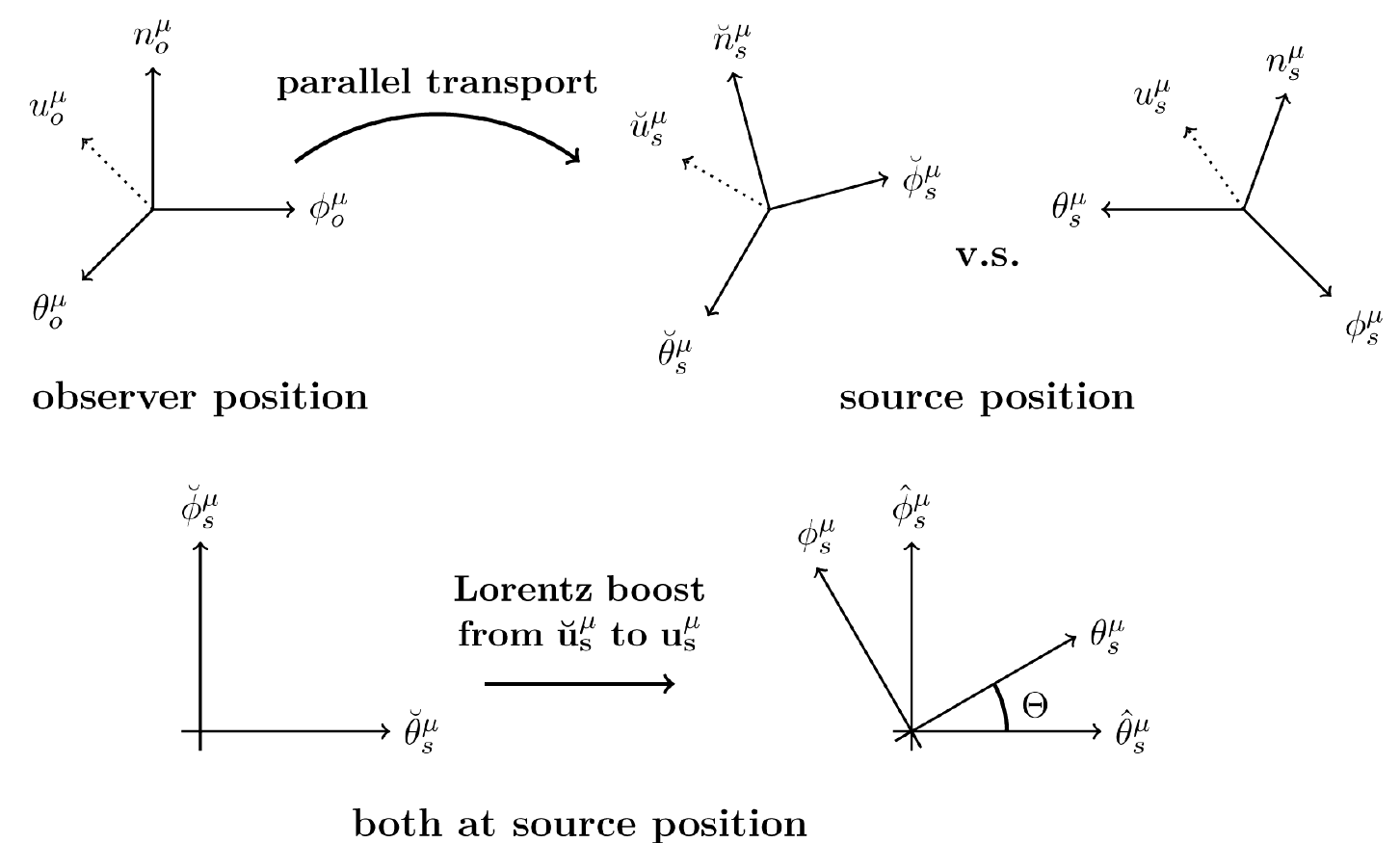, width=4.0in}}
\caption{Schematic view of the basis vectors. At the observer position,
the rest frame is determined by the observer four velocity~$u^\mu_o$, and
the two spatial directions are set by $\ttt^\mu_o$-$\pp^\mu_o$ perpendicular
to the light propagation direction~$n^\mu_o$. When these basis vectors
are parallel transported to the source position, these four vectors
(with {\it check}) are misaligned with the orthonormal basis vectors
in the source rest-frame set by~$u^\mu_s$. In particular, the two-dimensional
plane spanned by $\eett^\mu_s$-$\eepp^\mu_s$ needs to be Lorentz boosted
to the source rest-frame to form the correct and oriented plane denoted
by $\hat\ttt^\mu_s$-$\hat\pp^\mu_s$. This plane is identical to the plane
spanned by $\ttt_s^\mu$-$\pp^\mu_s$, but two sets of the basis vectors
make an angle~$\Theta$.}
\label{figure}
\end{figure}

Given the photon propagation direction~$n^\mu_s$ with the correct source
velocity~$u^\mu_s$, we computed the photon propagation direction~$n^i_s$
in Eq.~\eqref{eq:obssrc} and the plane $\ttt^i_s$-$\pp^i_s$ perpendicular
to~$n^i_s$. To compute the physical rotation, first we need to parallel 
transport the local tetrad basis vectors~$\ttt^\mu$ and~$\pp^\mu$ to the
source position, then the transported vectors~$\eett^\mu_s$ and~$\eepp^\mu_s$
need to be Lorentz boosted to the source rest-frame with~$u^\mu_s$.
Finally, we need to subtract the angle~$\Theta$ from the rotation~$\hat\omega$
in Eq.~\eqref{eq:homega} that the Lorentz boosted vectors
$\hat\ttt^\mu_s$ and $\hat\pp^\mu_s$ make against~$\ttt^\mu_s$ and~$\pp^\mu_s$,
respectively (note that the plane spanned by~$\hat\ttt^\mu_s$-$\hat\pp^\mu_s$
is identical to that by~$\ttt^\mu_s$-$\pp^\mu_s$, but their individual
directions are not aligned). This procedure is illustrated in 
Figure~\ref{figure}.

Fortunately, the second step is trivial at the linear order in perturbations,
as the Lorentz boost by~$\VV^\alpha_s$ of the source is already at the
linear order. The spatial tetrad vectors after the Lorentz boost are
\beeq
\hat e^i_\alpha=\Lambda^i{}_b \ee^b_\alpha=\Lambda^i{}_0\ee^0_\alpha
+\Lambda^i{}_j\ee^j_\alpha=\OO(2)+\ee^i_\alpha ~,\qquad\qquad
\hat e^\alpha_i=\ee^\alpha_i+\OO(2)~,
\eneq
identical in their components at the linear order. Therefore, the
directional vectors of~$\hat\ttt^\mu_s$ and~$\hat\pp^\mu_s$ in the source
rest-frame given the tetrad basis~$e_i^\mu$ are
\beeq
\hat\ttt^i_s=g_{\mu\nu}\hat e_\ttt^\mu e_i^\nu\simeq g_{\alpha\beta}
\ee_\ttt^\alpha e_i^\beta+\OO(2)\equiv\ttt^i+\Delta\hat\ttt^i_s~,\qquad\qquad
\hat\pp^i_s\simeq g_{\alpha\beta}\ee_\pp^\alpha e_i^\beta+\OO(2)
\equiv\pp^i+\Delta\hat\pp^i_s~,
\eneq
where a particular attention needs to be paid to the difference in notation
of $\Delta\ttt^i_s$ and $\Delta\pp^i_s$ (without hat) in
Eq.~\eqref{eq:twob}. So the angle~$\Theta$ between two directional vectors
is
\bear
\cos\Theta&=&\hat\ttt_s^i\ttt_s^i=\hat\pp_s^i\pp_s^i\simeq1+\OO(2)~,\qquad
\qquad 0=\Delta\hat\ttt_s^i+\Delta\ttt^i_s=\Delta\hat\pp_s^i+\Delta\pp^i_s~,
\\
\sin\Theta&=&\hat\pp_s^i\ttt_s^i=\hat\ttt_s^i\pp_s^i\simeq\Theta+\OO(2)~,
\qquad\qquad \Theta=\pp^i\Delta\ttt^i_s+\ttt^i\Delta\hat\pp_s^i
=\pp^i\Delta\hat\ttt^i_s+\ttt^i\Delta\pp_s^i~.~~~~
\enar
The rotation of the local coordinate in the source rest-frame by the 
angle~$\Theta$ will affect the distortion matrix~$\hDD$ in Eq.~\eqref{eq:DD3}
as $R(\Theta)\hDD$, where~$R$ is the rotation matrix. This needs to be
contrasted by the usual rotation of coordinates, where the distortion matrix
changes as $R(\Theta)\hDD R^t(\Theta)$. The latter arises when we rotate
the local coordinates both at the source and the observer positions by
the same angle~$\Theta$, while the former in our case rotates only the
local coordinate at the source position. Given this change, only
the rotation among the lensing observables is affected as
\beeq
\hat\omega~\mapsto~\hat\omega+\Theta~,
\eneq
at the linear order in perturbations, while the convergence~$\hat\kappa$
and the shear~$\hat\gamma_{\alpha\beta}$ remain unaffected.
Therefore, our task boils down to computing 
the transported vectors~$\eett^\mu_s$ and~$\eepp^\mu_s$ to derive
the angle~$\Theta$.

The condition for the parallel transport of the tetrad basis vectors is
\beeq
0={D\over d\Lambda}\ee_a^\mu={d\over d\Lambda}\ee_a^\mu
+\Gamma^\mu_{\rho\sigma}\ee_a^\rho k^\sigma~~\mapsto~~
0={d\over d\lambda}\ee_a^\mu+\Gamma^\mu_{\rho\sigma}\ee_a^\rho\hat k^\sigma~,
\eneq
where we used the conformal transformation relation in 
Eq.~\eqref{eq:affine}. At the background, we can readily check that
two spatial tetrad vectors perpendicular to the light propagation are
trivially transported as
\beeq
\ee_\ttt^\mu=\left(0,~\frac1{a_s}\ttt^i\delta_i^\alpha\right)~,\qquad\qquad
\ee_\pp^\mu=\left(0,~\frac1{a_s}\pp^i\delta_i^\alpha\right)~,
\eneq
while the propagation direction vector and the four velocity 
are parallel transported in a  non-trivial way, even at the background, i.e., 
the simple parametrizations 
\beeq
e_n^\mu=\left(0,~\frac1an^i\delta_i^\alpha\right)~,\qquad\qquad
e_0^\mu=\left(\frac1a,~0\right)~,
\eneq
do not satisfy the parallel transport condition. For our purposes, we will
focus on the two spatial tetrad vectors $\eett^\mu$ and~$\eepp^\mu$.
Integrating the parallel transport
condition, we can derive the  perturbation~$\delta \ee^\mu_\ttt$
of the spatial tetrad vectors parametrized in Eq.~\eqref{eq:tetdef} as
\bear
\delta \ee_\ttt^\alpha&=&(\delta e_\ttt^\alpha)_o-\int_0^{\rbar_z}d\rbar
\bigg[\HH\delta \ee_\ttt^\eta~n^\alpha-\HH\dnu~\ttt^\alpha-
\frac12\left(\BB_\beta{}^{|\alpha}-\BB^\alpha{}_{|\beta}\right)\ttt^\beta
\nnn
&&
\qquad\qquad
-\CC^{\alpha\prime}_\ttt+\left(2\CC^\alpha_{(\beta|\gamma)}
-\CC_{\beta\gamma}{}^{|\alpha}\right)\ttt^\beta n^\gamma\bigg]~,
\enar
where the initial condition $(\delta e^\alpha_\ttt)_o$
is set by Eq.~\eqref{eq:stetrad}.
Therefore, the directional vectors~$\hat\ttt^i_s$ and~$\hat\pp^i_s$
that are parallel transported and Lorentz boosted are
\beeq
\label{eq:sang2}
\hat\ttt^i_s=\ttt^i+\Delta\hat\ttt_s^i=g_{\alpha\beta}\ee_\ttt^\alpha
e_i^\beta=\ttt^i+2\CC^i_\ttt+\delta \ee_\ttt^i+\delta e^\ttt_i~,\qquad\qquad
\hat\pp^i_s=\pp^i+\Delta\hat\pp^i_s=\pp^i+2\CC^i_\pp+\delta\ee_\pp^i
+\delta e^\pp_i~,
\eneq
where the last terms $\delta e_i^{\ttt,\pp}$
in both expressions are the perturbations of the
spatial tetrad vectors at the source position we used for computing the
lensing observables, not the parallel transported tetrad vectors
$\ee^i_{\ttt,\pp}$. So, the angle between two basis vectors is then
\bear
\label{eq:mis}
\Theta&=&\pp^i\Delta\ttt_s^i+\ttt^i\Delta\hat\pp^i_s=\cos\ttt~\Delta\phi+
2\CC_{\ttt\pp}+\delta e^\pp_\ttt+\delta \ee^\ttt_\pp \nnn
&=&\cos\ttt~\Delta\pp+
\left(\CC_{\ttt\pp}+\delta e^\pp_\ttt+C_{[\beta,\alpha]}\ttt^\alpha\pp^\beta
\right) \nnn
&&\qquad\qquad
+\left(\CC_{\ttt\pp}+\delta e^\ttt_\pp-C_{[\beta,\alpha]}
\ttt^\alpha\pp^\beta\right)_o 
+\int_0^{\rbar_z}d\rbar\left[\pv_{[\beta,\alpha]}+2C_{\parallel[\beta,\alpha]}
\right]\ttt^\alpha\pp^\beta \nnn
&=&\cos\ttt~\Delta\pp+\Omega^n_s-\Omega^n_o
+\int_0^{\rbar_z}d\rbar~\frac12\bdi{n}\cdot\nabla\times
\left(\bm{\pv}^\alpha+2\bm{C_\parallel}^\alpha\right) ~,
\enar
where we simplified it with an integration by part and the boundary terms are
\beeq
\CC_{\ttt\pp}+\delta e^\pp_\ttt=C^{(v)}_{[i,j]}\ttt^i\pp^j+\Omega^n~,
\qquad\qquad
\CC_{\ttt\pp}+\delta e^\ttt_\pp=C^{(v)}_{[j,i]}\ttt^i\pp^j-\Omega^n~.
\eneq
The rotation angle~$\Theta$ is the rotation of the local tetrad
basis vectors ($\ttt^\mu$-$\pp^\mu$),
as we parallel transport them to the source position, and  it is
indeed the rotation angle~$\hat\omega$ in Eq.~\eqref{eq:homega}. However,
we should emphasize that the rotation of the tetrad basis is a ``rotation''
against an arbitrary global FRW coordinate and there is no way to compare the
orientations at two different spacetime points without parallel transporting
the basis. In this regard, the so-called Skrotsky effect is not physical.
Therefore, by setting the local coordinate at the source position to that
of the observer parallel transported and Lorentz boosted, the misalignment
angle~$\Theta$ is zero, and the orientation of the local coordinate at
the source position is completely determined as
\beeq
\Omega_s^n=\Omega^n_o-\cos\ttt~\Delta\phi-
\int_0^{\rbar_z}d\rbar~\frac12\bdi{n}\cdot\nabla\times
\left(\bm{\pv}^\alpha+2\bm{C_\parallel}^\alpha\right) ~.
\eneq
Putting this in Eq.~\eqref{eq:homega}, we find that the physical
lensing rotation is 
\beeq
\hat\omega=0~,
\eneq
with no further arbitrariness left.
We want to emphasize that this is the only physically meaningful way
to define the rotation and the physical rotation~$\hat\omega$
vanishes {\it only} at the linear order. 
In other words, the lensing rotation is {\it not} a coordinate artifact
(but the Skrotsky effect is), 
and it is non-zero beyond the linear order in perturbations.

\subsection{E-B decomposition and its relation to the rotation}
\label{ssec:EB}
The traceless symmetric shear tensor~$\gamma_{\alpha\beta}$ (or $\gamma_1$
and $\gamma_2$) was decomposed in terms of the helicity 
eigenstates~$_{\pm2}\gamma$ of spin $s=\pm2$. However, as introduced
in \cite{STEBB96}, the shear tensor can be decomposed in terms of
two potentials~$\Phi_E$ and~$\Phi_B$ (e.g., see
\cite{KABAET98,CRNAET02,HISE03b,HISE03a}). Following the approach
developed in \cite{HISE03b}, the lens equation can be generically written as
\beeq
s^i=n^i-\hat\nabla^i\Phi_E+\epsilon^i{}_j\hat\nabla^j\Phi_B
=n^i+\ttt^i\left(-{\partial\over\partial\ttt}\Phi_E+{1\over\sin\ttt}
{\partial\over\partial\pp}\Phi_B\right)-\pp^i\left({1\over\sin\ttt}{\partial
\over\partial\pp}\Phi_E+{\partial\over\partial\ttt}\Phi_B\right)~,
\eneq
where $\epsilon_{ij}$ is the Levi-Civita in two-dimensional angular basis,
the E-mode potential~$\Phi_E$ represents the usual lensing potential, and
the B-mode potential~$\Phi_B$
represents the pseudo vector of the deflection angle.\footnote{Here
our notation convention differs in the following quantities with minus
sign from those quantities defined in \cite{HISE03b}: 
\beeq
\Phi_E\RA-\Phi~,\quad \Phi_B\RA-\Omega~,\quad \kappa\RA-\kappa~,\quad
\gamma_1\RA-\gamma_Q~,\quad \gamma_2\RA-\gamma_U~,\quad\omega\RA-\omega~.
\eneq
The shear matrix in \cite{KABAET98,CRNAET02} is indeed the ellipticity
matrix, and hence their shear matrix corresponds to $2\gamma_{\alpha\beta}$ 
in our notation, giving rise to a factor~$2$ difference in the E-B
decomposition.}
Given the expression for the angular position of the source, the distortion
matrix in Eq.~\eqref{eq:DD2} can be derived as in Sec.~\ref{ssec:relens}:
\bear
\DD_{11}&=&1-{\partial^2\over\partial\ttt^2}\Phi_E+{1\over\sin\ttt}
\left({\partial^2\over\partial\ttt\partial\pp}-\cot\ttt{\partial\over\partial
\pp}\right)\Phi_B~,\\
\DD_{22}&=&1-\left(\cot\ttt{\partial\over\partial\ttt}+{1\over\sin^2\ttt}
{\partial^2\over\partial\pp^2}\right)\Phi_E+{1\over\sin\ttt}\left(\cot\ttt
{\partial\over\partial\pp}-{\partial^2\over\partial\ttt\partial\pp}\right)
\Phi_B~,\\
\DD_{12}&=&{1\over\sin\ttt}\left(\cot\ttt{\partial\over\partial\pp}-
{\partial^2\over\partial\ttt\partial\pp}\right)\Phi_E+\left(\cot\ttt
{\partial\over\partial\ttt}+{1\over\sin^2\ttt}{\partial^2\over\partial\pp^2}
\right)\Phi_B~,\\
\DD_{21}&=&{1\over\sin\ttt}\left(\cot\ttt{\partial\over\partial\pp}-
{\partial^2\over\partial\ttt\partial\pp}\right)\Phi_E-{\partial^2
\over\partial\ttt^2}\Phi_B~,
\enar
and the lensing observables can be directly read off from the distortion 
matrix as:
\bear
\label{eq:kEB}
2\kappa&=&\left({\partial^2\over\partial\ttt^2}
+\cot\ttt{\partial\over\partial\ttt}+{1\over\sin^2\ttt}
{\partial^2\over\partial\pp^2}\right)\Phi_E=\hat\nabla^2\Phi_E~,\\
2\gamma_1&=&\left({\partial^2\over\partial\ttt^2}-
\cot\ttt{\partial\over\partial\ttt}-{1\over\sin^2\ttt}
{\partial^2\over\partial\pp^2}\right)\Phi_E
-{2\over\sin\ttt}\left({\partial^2\over\partial\ttt\partial\pp}
-\cot\ttt{\partial\over\partial\pp}\right)\Phi_B~,\\
2\gamma_2&=&{2\over\sin\ttt}\left({\partial^2\over\partial\ttt\partial\pp}
-\cot\ttt{\partial\over\partial\pp}\right)\Phi_E+\left({\partial^2
\over\partial\ttt^2}-\cot\ttt
{\partial\over\partial\ttt}-{1\over\sin^2\ttt}{\partial^2\over\partial\pp^2}
\right)\Phi_B~,\\
\label{eq:Brot}
2\omega&=&-\left({\partial^2\over\partial\ttt^2}
+\cot\ttt{\partial\over\partial\ttt}
+{1\over\sin^2\ttt}{\partial^2\over\partial\pp^2}\right)\Phi_B=-\hat\nabla^2
\Phi_B~.
\enar
These relations of the lensing observables to the E- and B-mode potentials
can be compactly represented as \cite{KABAET98,CRNAET02,HISE03b,HISE03a}
\bear
\label{eq:gEB}
&&\gamma_{ij}=\left(\hat\nabla_i\hat\nabla_j-\frac12\delta_{ij}^{2D}\hat
\nabla^2\right)\Phi_E+\frac12\left(\epsilon_{kj}\hat\nabla_i\hat\nabla_k
+\epsilon_{ki}\hat\nabla_k\hat\nabla_j\right)\Phi_B~,\\
&&{\hat\nabla^4} \Phi_E = 2~\hat\nabla_i\hat\nabla_j\gamma_{ij}~,\qquad\qquad
\qquad
{\hat\nabla^4} \Phi_B = 2~\epsilon_{ij}\hat\nabla_i\hat\nabla_k
\gamma_{jk}~,
\enar
where the notation for the two potentials $\gamma_E$ and~$\gamma_B$ 
in \cite{KABAET98} is $\gamma_E =\nabla^2\Phi_E/2$ and
$\gamma_B =\nabla^2 \Phi_B/2$. 
The spin~$\pm2$ shear components $_{\pm2}\gamma$
are also related to the two potentials as
\bear
\label{eq:tEB}
_{\pm2}\gamma&=&\gamma_1\pm i\gamma_2=\left[\frac12\left(
{\partial^2\over\partial\ttt^2}-
\cot\ttt{\partial\over\partial\ttt}-{1\over\sin^2\ttt}
{\partial^2\over\partial\pp^2}\right)\pm 
{i\over\sin\ttt}\left({\partial^2\over\partial\ttt\partial\pp}
-\cot\ttt{\partial\over\partial\pp}\right)\right]
\left(\Phi_E\pm i\Phi_B\right)~ \nnn
&=&m^i_{\mp}m^j_{\mp}\hat\nabla_i\hat\nabla_j\left(\Phi_E\pm i\Phi_B\right)~.
\enar
The two potentials are uniquely defined up to the transformation 
\beeq
\Phi_E\RA\Phi_E+\Psi_E~,\qquad\qquad \Phi_B\RA\Phi_B+\Psi_B~,
\qquad\qquad \hat\nabla^4\Psi_E=\hat\nabla^4\Psi_B=0~
\eneq
that supplies the same lensing observables under the conditions:
\bear
&&
\left({\partial^2\over\partial\ttt^2}-
\cot\ttt{\partial\over\partial\ttt}-{1\over\sin^2\ttt}
{\partial^2\over\partial\pp^2}\right)\Psi_E=
{2\over\sin\ttt}\left({\partial^2\over\partial\ttt\partial\pp}
-\cot\ttt{\partial\over\partial\pp}\right)\Psi_B~,\\
&&{2\over\sin\ttt}\left({\partial^2\over\partial\ttt\partial\pp}
-\cot\ttt{\partial\over\partial\pp}\right)\Psi_E=-\left({\partial^2
\over\partial\ttt^2}-\cot\ttt
{\partial\over\partial\ttt}-{1\over\sin^2\ttt}{\partial^2\over\partial\pp^2}
\right)\Psi_B~.
\enar

The lens equation written in terms of unit angular vector has two 
degrees of freedom, and these two degrees of freedom
are captured by two independent
potentials~$\Phi_E$ and~$\Phi_B$. As derived above,
all the lensing observables are therefore expressed in terms of~$\Phi_E$
and~$\Phi_B$: The convergence~$\kappa$ is exclusively described by~$\Phi_E$, 
and the rotation~$\omega$ is also exclusively by~$\Phi_B$, 
while the shear matrix~$\gamma_{ij}$ receives the contributions of 
both~$\Phi_E$ and~$\Phi_B$. To compute the two potentials, we need to
use the lens equation and the lensing observables.
From the lens equation~\eqref{eq:basic} in the standard lensing formalism, 
the two potentials can be read off as
\beeq
\Phi_E=\Phi=\int_0^{\rbar_s}d\rbar\left({\rbar_s-\rbar\over\rbar_s\rbar}
\right)~2\psi~,\qquad\qquad \Phi_B=0~,
\eneq
where the E-mode potential~$\Phi_E$
is the projected lensing potential, sourced
by the density fluctuations, while the B-mode potential is absent $\Phi_B=0$
to all orders.

In Sec.~\ref{ssec:relens} we generalized the standard weak lensing formalism
by solving the geodesic equation and accounting for the contributions from
the scalar, vector, and tensor perturbations, and we derived the
lens equation~\eqref{eq:dsang} and the lensing observables in 
Eqs.~\eqref{eq:kappa}, \eqref{eq:gamma1}, \eqref{eq:gamma2}, 
and~\eqref{eq:rot}. The two potentials in the generalized lensing formalism
can then be derived as 
\bear
\Phi_E&=&2\hat\nabla^{-2}\kappa=\left(\VP-\frac12\TCP\right)_o
-{n_\alpha\over\rbar_z}\bigg[\left(\delta x^\alpha
+\CCG^\alpha\right)_o-\CCG^\alpha_s\bigg] +{1\over\rbar_z}\hat\nabla^{-2}
\left(\hat\nabla_\alpha\CCG^\alpha\right) \\
&&
+\int_0^{\rbar_z}d\rbar\left({\rbar_z-\rbar\over\rbar_z\rbar}\right)
\left(\ax-\px-\pvP-\TCP\right)+\int_0^{\rbar_z}{d\rbar\over\rbar}\left(n_\alpha
+\hat\nabla^{-2}\hat\nabla_\alpha\right)\bigg(\pv^\alpha+2C^\alpha_\parallel
\bigg)~,\nnn
\Phi_B&=&-2\hat\nabla^{-2}\omega=\Omega_o^n+\hat\nabla^{-2}\left[\bdi{n}\cdot
\nabla\times\CCG^\alpha\right]
+\int_0^{\rbar_z}d\rbar~\hat\nabla^{-2}\left[
\bdi{n}\cdot\nabla\times\left(\pv^\alpha+2C^\alpha_\parallel\right)\right]~,
\enar
where $\hat\nabla^{-2}$ is the inverse (angular) Laplacian operator and
 we used the useful relations (and their inverse relations)
\beeq
\hat\nabla^2 n^i=-2n^i~,\qquad \qquad \hat\nabla^2\left(n^in^j\right)
=-6n^in^j+2\delta^{ij}~.
\eneq
Compared to the
standard formalism, the E-mode potential~$\Phi_E$ 
is generalized to include the vector
and the tensor contributions to the projected lensing potential with
extra terms at the observer and the source positions. The B-mode 
potential~$\Phi_B$ is also excited in the presence of the vector and
the tensor perturbations, giving rise to the lensing rotation 
$2\omega=-\hat\nabla^2\Phi_B$ in Eq.~\eqref{eq:rot}. However, as emphasized
in Sec.~\ref{ssec:relens}, all the lensing observables in this simple
generalization of the standard formalism are gauge-dependent, and consequently
the two potentials derived above are also gauge-dependent:
\beeq
\widetilde{\Phi}_E=\Phi_E-{1\over\rbar_z}\left(n_\alpha
+\hat\nabla^{-2}\hat\nabla_\alpha\right)\LL^\alpha~,\qquad\qquad
\widetilde{\Phi}_B=\Phi_B-\hat\nabla^{-2}\left[\bdi{n}\cdot\nabla\times\LL
^\alpha\right]~.
\eneq

To resolve this issue, we projected the image in a FRW coordinate
into the source rest-frame to derive the relation of the observed angular
size to the physical size of a standard ruler in Sec.~\ref{ssec:geo}.
The lens equations~\eqref{eq:obssrc} and~\eqref{eq:lens3} are now expressed
in the local coordinates of the observer and the source frames, and their
relation is gauge-invariant. Consequently, the lensing observables
in Eqs.~\eqref{eq:hkappa}, \eqref{eq:hgamma1}, \eqref{eq:hgamma2}, 
and~\eqref{eq:homega} are gauge-invariant, as they are derived from the
distortion matrix~$\hDD$ in Eq.~\eqref{eq:DD3} constructed out of the physical
size of a standard ruler in the source rest-frame. However, it is evident
that we need to bring additional information to build the physically 
well-defined distortion matrix~$\hDD$, namely, the source position
(or the distance to the source), and this additional information is
{\it not} captured by two potentials in the lens equation. Simply put,
while the distortion matrix~$\DD$ is just a function of~$\Phi_E$ and~$\Phi_B$,
the distortion matrix~$\hDD$ in the gauge-invariant formalism
is not. For instance, in the presence of
the vector and the tensor perturbations, the shear matrix~$\hat\gamma_{ij}$
in Eq.~\eqref{eq:hgamma} is non-vanishing, or the B-mode is non-zero
as shown in Eqs.~\eqref{eq:gEB}$-$\eqref{eq:tEB}. However, as we showed
in Sec.~\ref{ssec:rotation}, 
the physical lensing rotation~$\hat\omega$ at the linear order is 
{\it zero}, or vanishing B-mode in Eq.~\eqref{eq:Brot},
which shows the inconsistency of the E-B decomposition.

In summary, while the E-B decomposition is a useful tool, the physical
lensing observables ($\hat\kappa$, $\hat\gamma_{ij}$, $\hat\omega$, or $\hDD$)
are not fully described by~$\Phi_E$ and~$\Phi_B$ alone, and hence the
relation in Eqs.~\eqref{eq:kEB}$-$\eqref{eq:Brot} breaks down for the physical
lensing observables.

\section{Comparison to Previous Work}
\label{sec:comparison}
Here we compare our gauge-invariant lensing formalism to previous work
in literature. While there exists an extensive work in lensing, 
there are relatively
few papers that treat the weak lensing observables in a fully relativistic
framework. In most cases, previous work adopts the conformal Newtonian
gauge and computes only the scalar contributions. Few papers considered
the vector and the tensor contributions, and even fewer papers checked
the gauge-invariance of the lensing observables. In this section,
we first compare our results 
in the presence of tensor modes, highlighting the need for the
tensor contributions at the source and the observer positions.
Then we discuss the Cosmic Ruler papers \cite{SCJE12a,SCJE12b} that
computed the lensing observables with a different approach and provided
comprehensive accounts of the lensing observables.
While in literature there exist several work
\cite{DOKOET05,KRHI10,SULI14,COHU02,BEBOVE10,BEBOET12} that computed
the lensing observables beyond the linear order,
we will focus on our comparison to those only at the linear order
in perturbations.

\subsection{Contributions of the primordial gravity waves to the lensing
observables}
\label{ssec:GW}
The primordial gravity waves in the early Universe provide the most natural
way to excite the tensor modes on large scales at the linear order in
perturbations. The presence of the tensor perturbations $C_{\alpha\beta}$
affects the lensing observables in many subtle ways. The photon path is
deflected (in Sec.~\ref{sec:src}),
as it propagates along the line-of-sight direction; The rest-frames 
of the source and the observer are affected (in Sec.~\ref{sec:GIlensing}),
changing the relation between the photon wavevector in the FRW frame and
the observables in the observer rest-frame.
These tensor contributions to gravitational lensing have been considered
in the past \cite{DOROST03,SCJE12a,SCJE12b,ADDUTA16}. Here we briefly
compare our calculations to the previous work.

Dodelson, Rozo, \& Stebbins \cite{DOROST03} considered 
the tensor contributions to the lensing observables. Focusing only on
the tensor modes, they solve
the geodesic equation, and derive the light deflection due to the tensor
modes. Upon identifying their notation 
\beeq
\eta_0-\eta~~\mapsto~~\rbar_z=\bar\eta_o-\bar\eta_z~,
\qquad\qquad \bdv{H}~~\mapsto~~2C_{\alpha\beta}~,
\eneq
the source position~$\bdv{r}$ in their Equation~(3) can be arranged to 
match the angular distortion, when only the
transverse directions are considered:
\beeq
\bm{\ttt}\cdot\bdv{r}=-\rbar_z~C_{\alpha\beta o}n^\alpha \ttt^\beta
-\int_0^{\rbar_z}d\rbar~2C_{\alpha\beta}n^\alpha \ttt^\beta
+\int_0^{\rbar_z}d\rbar
\left({\rbar_z-\rbar\over\rbar}\right){\partial\over\partial\ttt}\TCP~.
\eneq
Compared to our angular distortion in Eq.~\eqref{eq:theta2},
\beeq
\rbar_z\dtt=\rbar_z~C_{\alpha\beta o}n^\alpha\ttt^\beta
-\int_0^{\rbar_z}d\rbar~{\rbar_z\over\rbar}~2C_{\alpha\beta}n^\alpha\ttt^\beta
+\int_0^{\rbar_z}d\rbar\left({\rbar_z-\rbar\over\rbar}\right)
{\partial\over\partial\ttt}\TCP~,
\eneq
we find some errors in the angular distortion in their Equation~(3).
Furthermore, since the source and the observer positions are slightly
different from those in the background, their time coordinate in fact
corresponds to the distance in the background and additional perturbations
\beeq
\eta_o-\eta~~\mapsto~~\rbar_z+\delta\eta_o+\Delta\eta_s~.
\eneq
While $\delta\eta_o$ vanishes when only the tensor perturbations are
considered, the time lapse at the source position is non-vanishing:
\beeq
\Delta\eta_s={\dz\over\HH}={1\over\HH}\int_0^{\rbar_z}d\rbar~\TCP'~.
\eneq

Furthermore, they computed the rotation~$\omega$ and its power spectrum
as a measure of the inflationary primordial gravity waves. The rotation
was derived by using the relation of the E-B decomposition~$\Phi_E$
and~$\Phi_B$. Their rotation 
\beeq
\omega=\int_0^{\rbar_z}d\rbar~\bdi{n}\cdot\nabla\times C^\alpha_\parallel~,
\eneq
is exactly the rotation~$\omega$ ({\it without hat}) in Eq.~\eqref{eq:rot},
except the rotation~$\Omega^i_o=0$ and the 
minus sign due to the different sign convention. However, as we
showed, this rotation~$\omega$ is a coordinate artifact due to a
rotation of the tetrad basis along the
light propagation, when compared against a global coordinate.
Interestingly, by imposing the consistency condition among the lensing
observables, they introduced the tensor contribution at the source position
(so called the metric shear), which is what we need to fix the
gauge-dependent shear matrix~$\gamma_{\alpha\beta}$ when
we transform from the FRW frame to the source rest-frame.

Recently, Adamek, Durrer \& Tansella \cite{ADDUTA16} considered the tensor
contributions~$h_{ij}$ from the primordial gravitational waves
to the lensing observables, and they used the relativistic cosmological
simulations \cite{ADDUKU14,ADDAET16} to compare the gravitational wave signals 
in the lensing observables to the tensor contributions generated by 
the nonlinear evolution of large scale structure in the late time.
They solved the geodesic equation and derived
the deflection angle~$\alpha_a$, which  corresponds
to our angular distortion 
\beeq
\alpha_a~~\mapsto~~(\dtt,~\sin\ttt~\dpp)~,\qquad\qquad
h_{ij}\mapsto2C_{\alpha\beta}~.
\eneq
Their Equation~(2.5) is equivalent to our Eq.~\eqref{eq:theta}, but
without the tensor contribution at the observer position. 
The absence of the tensor contribution
causes the infrared divergences 
in the power spectrum, which they regulate by introducing a counter term. 
These tensor contributions at the source and the observer positions
are not constrained by the geodesic equation, as they arise in defining
the rest-frames. However, their significance
is apparent in both cases \cite{DOROST03,ADDUTA16}. 

In application to CMB polarization, Dai \cite{DAI14} computes 
the rotation of CMB polarization and argues that it is maximally correlated 
with the lensing rotation, affecting the CMB polarization power spectrum. 
However, the lensing rotation by the vector and the tensor vanishes
at the linear order, and the rotation in \cite{DAI14} is a coordinate
artifact. Furthermore, since
polarization is parallel transported, the rotation of polarization is
zero to all orders, and its correlation to the lensing rotation beyond
the linear order is therefore zero.

\subsection{Cosmic Ruler approach: lensing by a standard ruler}
\label{ssec:ruler}
It is well known from the early days in lensing literature
that the geodesic equation and its infinitesimal deviation vectors
provide the essential ingredients for describing the lensing phenomena
(see, e.g., \cite{KRSA66,GUNN67}). However, as emphasized throughout
this work, the photon wavevector obtained by solving the geodesic equation
needs to be related to the observables in the observer rest-frame and
the physical size and shapes in the source rest-frame. This requires
setting up the local tetrad bases at the observer and the source positions.
In this regard, a comprehensive and pioneering work has been done
\cite{SCJE12a,SCJE12b} under the name of the ``Cosmic Ruler,'' with which
we briefly compare our calculations.

The cosmic ruler approach assumes a standard ruler in the rest-frame of 
the source and computes its relation to the observables.
Imagine a stick of a known length~$L$ (or the standard ruler) and 
measure the light from this stick. In particular, we measure the angular
positions~$n^i_1$ and~$n^i_2$ and the observed redshifts~$z_1$ and~$z_2$
of the two end points of the stick:
\beeq
x^\mu_{1,2}=\bar x^\mu_{z_1,z_2}+\Delta x^\mu_{s_1,s_2}~,\qquad \qquad
x^\mu_2-x^\mu_1=\delta\bar x^\mu+\Delta x^\mu_s~,
\eneq
where two separation vectors are defined as
\beeq
\delta\bar x^\mu\equiv\bar x^\mu_2-\bar x^\mu_1~,\qquad \qquad
\Delta x^\mu_s\equiv\Delta x^\mu_{s_2}-\Delta x^\mu_{s_1}~.
\eneq
In the limit the size of the ruler goes to zero ({\it small ruler}), 
we consider a case, in which two end points of the ruler are 
at the same observed redshift $(z_1=z_2)$,
but with slightly different angular positions $(n^i_2=n^i_1+dn^i$; 
{\it transverse ruler}).
We will refer to this limit as the ``small transverse ruler limit for 
lensing,''
which is relevant for our purposes.\footnote{It is noted that the cosmic
ruler approach in \cite{SCJE12a,SCJE12b} includes the longitudinal ruler,
in addition to the transverse ruler we consider here.}
In this limit, these separation vectors correspond to 
\beeq
x_2^\alpha-x_1^\alpha~\mapsto~dx_s^\alpha~,\qquad\qquad
\delta \bar x^\alpha~\mapsto~\rbar_z dn^\alpha~,\qquad\qquad
\Delta x^\alpha~\mapsto~\rbar_z \Delta s^\alpha~.
\eneq

Given these positions in a FRW coordinate, the (known) length of the standard 
ruler can be computed by projecting two end points to the rest-frame of the
source as
\beeq
L^2=\HH_{\mu\nu}\left(x_1^\mu-x_2^\mu\right) \left(x_1^\nu - x_2^\nu\right)~,
\eneq
where the projection tensor for the source position~$x_1^\mu$ is
\beeq
\label{eq:proj}
\HH_{\mu\nu}(x_1)=g_{\mu\nu} + u_\mu u_\nu\simeq 
a^2 \left(\begin{array}{cc}0 & 
-\VV_\alpha \\ - \VV_\alpha & g_{\alpha\beta}\end{array}\right)+\OO(2)~.
\eneq
This projection tensor ensures that when contracted, any four vectors
are projected in the rest-frame of the source. 
Since the projection and any other operations in this case
involve two end points with two different redshifts,
there always exists ambiguity regarding the choice of the operation point, 
i.e., either at $x_1^\mu$ or $x_2^\mu$. The observed
size~$L_z$ of the standard ruler will be inferred based on the observed
angular size and the redshift (again we use~$z_1$) as
\beeq
L_z^2\equiv a_z^2\delta_{\alpha\beta}\delta\bar x^\alpha\delta\bar x^\beta
\equiv a_z^2\left(\delta
\bar x_\parallel^2+\delta\bar x_\perp^2\right)=\bar D_A^2(z)\left[\left({\delta
\bar x_\parallel\over\rbar_z}\right)^2+\left({\delta\bar x_\perp\over\rbar_z}
\right)^2\right]~,
\eneq
where the orientation of the ruler is further decomposed \cite{SCJE12a}
along the line-of-sight direction~$\delta\bar x_\parallel$ and the transverse
direction~$\delta\bar x_\perp$ (by using $n_1$). Using Eq.~\eqref{eq:proj},
we can compute the size of the standard ruler at the linear order 
in perturbations and relate it to the observed size~$L_z$ as
\bear
L^2&=&L^2_z(1+2~\dz)+2a_z^2\left(-\VV_\alpha\delta\bar\eta~\delta\bar x^\alpha
+\CC_{\alpha\beta}\delta\bar x^\alpha\delta\bar x^\beta
+\delta_{\alpha\beta}\delta\bar x^\alpha\Delta x_s^\beta\right)\\
&=&L_z^2(1+2~\dz)+2a_z^2\CC_{\alpha\beta}\delta\bar x^\alpha\delta\bar x^\beta
+2a_z^2\left(\VV_\parallel\delta\bar x_\parallel^2
+\VV_\perp\delta\bar x_\perp\delta\bar x_\parallel\right)
+2a_z^2\delta_{\alpha\beta}\delta\bar x^\alpha\left(\delta\bar 
x_\parallel\partial_{\rbar}+\delta\bar x_\perp\partial_\perp\right)\Delta
x^\beta_{s_1}~, \nonumber
\enar
where $\delta\bar\eta=-\delta \bar x_\parallel$ is assumed \cite{SCJE12a}
and the deviation vector~$\Delta x^\beta_s$ is expanded as
\beeq
\Delta x_s^\beta=\Delta x_{s_2}^\beta-\Delta x_{s_1}^\beta
=\rbar_z\Delta n^\gamma{\partial\over\partial x^\gamma}\Delta x_{s_1}^\beta~.
\eneq

Having derived the relation between the physical size~$L$ and the observed
size~$L_z$ and measured six observables (two redshifts and two angular
positions), we can determine six physical quantities associated with
the distortion of the standard ruler as \cite{SCJE12a}
\beeq
1-\frac{L}{L_z}\equiv\mathcal{C}\frac{(\delta\bar x_\parallel)^2}{L_c^2} 
+ \mathcal{B}_\alpha \frac{\delta\bar x_\parallel \delta\bar x_\perp^\alpha}
{L_c^2}+ \mathcal{A}_{\alpha\beta} 
\frac{\delta\bar x_\perp^\alpha \delta\bar x_\perp^\beta}{L_c^2}~,
\qquad\qquad \delta\bar x^\alpha=n^\alpha\delta\bar x_\parallel
+\delta\bar x_\perp^\alpha~,
\eneq
where we defined the inferred comoving size 
$L_c^2\equiv\delta\bar x_\parallel^2+\delta\bar x_\perp^2$.
The scalar~$\mathcal{C}$ and the two-component transverse vector~$\BB_{\alpha}$
are related to the radial distortion. For our purposes of deriving the
lensing observables, we take the small transverse
ruler limit ($\delta\bar x_\parallel=0$), and the
symmetric matrix~$\mathcal{A}_{\alpha\beta}$ 
plays the equivalent role of the distortion matrix~$\hDD$.
The symmetric matrix can be further decomposed in terms of
the trace~$\mathcal{M}$ and the spin-2 shear~$\gamma_1$ \&~$\gamma_2$  as
\beeq
\AA_{\alpha\beta}=\left(\begin{array}{ccc}\mathcal{M}/2+\gamma_1&\gamma_2 &0\\
\gamma_2&\mathcal{M}/2-\gamma_1&0\\0&0&0\end{array}\right)~,
\eneq
where the components of the matrix is written in a Cartesian coordinate
aligned with $(\ttt,\pp, n)$
and the magnification scalar~$\mathcal{M}$ and the shear matrix~$\gamma_{ij}$
are
\bear
\mathcal{M}&=&- 2~\dz- (\CC^\alpha_\alpha-\CC_\parallel)
+ 2~\kappa - \frac{2~\drr}{\rbar} ~,\\
\gamma_{ij}&=&{}_{\pm2}\gamma ~m_\pm^i m_\pm^j=
-\left(\mathcal{P}^k_i\mathcal{P}^l_j-\frac12
\mathcal{P}_{ij}\mathcal{P}^{kl}\right)\CC_{kl}
-\frac12\left(\partial_{\perp i}
\Delta x_{\perp j}+\partial_{\perp j}\Delta x_{\perp i}\right)
-\mathcal{P}_{ij}\kappa~.~~~
\enar
Identifying their notation
\beeq
\Delta x_\parallel~\mapsto~\drr~,\qquad\qquad
\Delta x_\perp^i~\mapsto~\rbar_z\left(\dtt~\ttt^i+\sin\ttt~\dpp~\pp^i\right)~,
\eneq
we can derive the relation of the ruler observables to our lensing observables
as
\beeq
\kappa=-\frac12\partial_{\perp i}\Delta x^i_\perp~\mapsto~\kappa~,\qquad
\mathcal{M}~\mapsto~2\hat\kappa=-2~\ddD~,\qquad
\gamma_{ij}~\mapsto~\hat\gamma_{\alpha\beta}~.
\eneq

In their calculations \cite{SCJE12a}, the photon wavevector~$\hat k^\mu$
is related to the observed angle and redshift with an implicit assumption
$\dhnu_o=0$ in our calculation.
Furthermore, they assume that the observer position is
$x^\mu_o=\bar x^\mu_o$, i.e., the coordinate lapse $\delta \eta_o$ and 
the spatial shift $\delta x^\alpha_o$ are ignored. At the linear order
in perturbations, the magnification~$\mathcal{M}$ and the 
shear matrix~$\gamma_{ij}$
are independent of the spatial shift~$\delta x^\alpha_o$, but the
magnification depends on the coordinate lapse~$\delta\eta_o$. Its absence
is the source for the infrared divergence in the variance of the luminosity
distance (see \cite{BIYO16}).\footnote{However, see their arXiv version
(1204.3625v3) of the Cosmic Ruler paper \cite{SCJE12a}, 
where they included the coordinate
lapse~$\delta\eta_o$ in the expressions, exactly to prevent the
infrared divergence of the theoretical predictions. The spatial shift~$\delta
x^\alpha_o$ is not included, but again its contribution drops out in the
lensing observables.} 
The tetrad basis at the source position
is essential to our approach in relating the photon wavevector to the
physical size and the shape in the source rest-frame. This part was
replaced in the cosmic ruler by the projection tensor~$\HH_{\mu\nu}$
in Eq.~\eqref{eq:proj}. In fact, the projection operation has to be
performed not only to the rest-frame of the source, but also to a plane
perpendicular to the photon propagation. So, using the photon 
direction~$n^\mu$ in Eq.~\eqref{eq:nsrc} at the source position,
the correct projection tensor can be derived as
\beeq
\HH_{\mu\nu}=g_{\mu\nu}+u_\mu u_\nu-n_\mu n_\nu=e_\mu^\ttt e_\nu^\ttt
+e_\mu^\pp e_\nu^\pp~,\qquad\qquad e^\mu_\ttt=e_i^\mu\ttt_s^i~,
\qquad e^\mu_\pp=e_i^\mu\pp_s^i~.
\eneq
However, at the linear order, the additional term from~$n_\mu n_\nu$
is always multiplied by the transverse deviation $\delta\bar x^\mu_\perp$, and
hence these contributions vanish as $\bar n^\alpha_s \propto n^\alpha$.

Furthermore, in the cosmic ruler, the orientation of the standard ruler
is not specified, and hence the rotation is absent 
in~$\mathcal{A}_{\alpha\beta}$ by construction.
While the rotation~$\hat\omega$ at the linear order indeed vanishes,
this is not true beyond the linear order. 
A straightforward extension of the cosmic ruler to include the lensing
rotation would be to consider
two (distinguishable) standard rulers in the source rest-frame, e.g.,
a red stick and a blue stick with known sizes and orientations in the
source rest-frame. Despite these points for improvement, the cosmic
ruler approach \cite{SCJE12a,SCJE12b} provides a very comprehensive
account of the lensing observables with an explicit check of
gauge-invariance and the consideration
of the source rest-frame, in which the standard ruler is defined.
Moreover, their calculations of the magnification and the shear matrix 
are fully consistent with our calculations, and their work has been further
extended \cite{SCJE12b,SCPAZA14}
to include the intrinsic shear in a gauge-invariant way, which
is beyond the scope of our current work.

\subsection{Jacobi mapping approach}
\label{ssec:jacobi}
The Jacobi mapping approach provides a different way to formulate the
lensing observables (see, e.g., 
\cite{SESCEH94,LECH06,BONVI08,BEBOVE10,BEBOET12}). Consider an
infinitesimal field~$\xi^\mu$ (or the Jacobi field)
that connects two photon paths parametrized by the same affine 
parameter.\footnote{Only in this subsection, we use~$\xi^\mu$ to denote
the Jacobi field. This should not be confused with an infinitesimal
coordinate transformation in Eq.~\eqref{eq:coord}.}
The Jacobi field can be expressed in terms of the local tetrad basis
parallel transported along the photon path as
\beeq
\xi^\mu_\lambda\equiv\xi^0\ee_0^\mu+\xi^nn^\mu_\lambda+\xi^A\ee_A^\mu~,
\qquad\qquad 0=k_\mu \ee^\mu_A~,
\eneq
where $\ee^\mu_A$ ($A=\ttt,\pp$) forms the two-dimensional plane
perpendicular to the light propagation and $\ee_0^\mu$ is the parallel
transported velocity~$u^\mu_\lambda$. 
This basis is often referred to as the Sachs basis 
\cite{SACHS61} and is more convenient than any other tetrad basis~$\ee_a^\mu$
that are parallel transported, because its spatial basis vectors are
aligned with the light propagation direction~$n^\mu_\lambda$ and two
orthonormal directions~$\ee_A^\mu$. 
Since the photon path is described by the null vector~$k^\mu$, the 
infinitesimal Jacobi field can be further simplified as
\beeq
\xi^\mu_\lambda={\xi^n\over k\cdot u}k^\mu+\xi^A\ee_A^\mu
\equiv \xi^kk^\mu+\xi^A\ee_A^\mu~,
\eneq
where we used Eq.~\eqref{eq:direction} to obtain the relation 
$\xi^0=-\xi^n$. The geodesic deviation equation for the Jacobi field is
\beeq
{D^2\over d\Lambda^2}
\xi^\mu\equiv k^\rho\nabla_\rho\left(k^\nu\nabla_\nu\xi^\mu
\right)=-R^\mu{}_{\nu\rho\sigma}k^\nu\xi^\rho k^\sigma~,
\eneq
where $R^\mu{}_{\nu\rho\sigma}$ is the Riemann tensor.
By projecting the Jacobi field into the Sachs basis, the governing 
equation for the Jacobi field can be derived \cite{YOSC16} as
\beeq
{d^2\over d\Lambda^2}\xi^A=-\mathfrak{R}^A_B\xi^B~,\qquad\qquad
\mathfrak{R}^A_B\equiv
\left(R^\mu{}_{\nu\rho\sigma}k^\nu k^\sigma\right)\ee^A_\mu\ee_B^\rho~.
\eneq
The solution of the Jacobi field can be obtained by doubly integrating
the source term on the right-hand side, and it can be expressed in terms
of the linear map, or the Jacobi map~$\mathfrak{D}_{AB}$ as
\beeq
\xi^A(\Lambda)\equiv\mathfrak{D}^A{}_B(\Lambda)\dot \xi_o^B~,\qquad\qquad
{d^2\over d\Lambda^2}~\mathfrak{D}_{AB}=-\mathfrak{R}_{AC}\mathfrak{D}_{CB}~.
\eneq

Since the Jacobi field~$\xi^A$ is the physical separation
in a plane perpendicular to the photon propagation direction in the
rest-frame of an ``observer'' with~$u^\mu_\lambda$, the Jacobi field
$\xi^A$ at the source position can be related to
the physical length and the shape in the source rest-frame, once we
correct the difference between the parallel transported 
velocity~$u^\mu_\lambda$ and the source velocity~$u^\mu_s$.
However, this correction is trivial as described in Eq.~\eqref{eq:inv}.
Therefore, the distortion matrix that relates the physical size and the
shape in the source rest-frame to the observed angles in the observer
rest-frame can be readily constructed by using the Jacobi map
\cite{GRYO18}.
In particular, the anti-symmetric part of the Jacobi map will be related
to the lensing rotation. 
Given the governing differential equation for the Jacobi map, the solution
is in general {\it not} symmetric (hence non-vanishing rotation), though
the source term~$\mathfrak{R}_{AB}$ is symmetric. However, the background
solution is trivial $\mathfrak{D}_{AB}\propto\delta_{AB}$, and hence the
linear-order solution is symmetric with vanishing rotation,
in full agreement with our result.
The Jacobi mapping approach provides a non-perturbative description
of the lensing observables. However, since its formalism is based on the
parallel transported tetrad basis, its specification is often absent
and hence the relation to the observables is left unspecified. In contrast,
our geometric approach provides this missing link, and in particular our
explicit calculation of the parallel transport of the tetrad basis illuminates
the physical origin of the vanishing rotation, which appears mysterious
in the Jacobi mapping approach.

\section{Discussion and Summary}
\label{sec:discussion}

We have presented a gauge-invariant formalism of cosmological weak lensing,
accounting for all the relativistic effects associated with the lensing
observables at the linear order in perturbations. Without choosing a
gauge condition, we have solved the geodesic equation for the light
propagation and constructed the tetrad bases for the rest-frames
of the observer and the source. The last step is important in 
establishing the relation of the photon wavevectors
to the observed angle and redshift in the observer rest-frame
and the physical size and shape in the source rest-frame.
We have demonstrated that the standard weak lensing formalism can be naturally
generalized to account for all the relativistic effects associated with
the light propagation by using the solution of the  geodesic equation.
However, without specifying the observer and the source rest-frames, the
standard lensing formalism is shown to be deficient and gauge-dependent.
Using the tetrad bases at the observer and the source positions, 
we have improved the standard lensing formalism and
derived the lensing observables such as 
the gravitational lensing convergence, the lensing shear, and the rotation
in a gauge-invariant way.
With full generality, the derivations are lengthier than when a gauge
condition is adopted, but by choosing a gauge
condition, one loses a way to check the gauge-invariance of the lensing 
observables. Indeed, very little attention has been paid to this aspect in
lensing literature, and 
some of the gauge issues in the standard weak lensing in 
Sec.~\ref{sec:stdlensing} could have been
 readily spotted by comparing numerical results
in another gauge conditions, for instance, the synchronous gauge.
We emphasize that it is important to derive equations with the general metric
condition and explicitly verify their gauge-invariance.
This procedure provides a great sanity check of nonlinear perturbation
calculations \cite{YODU17}.

The key point in developing the gauge-invariant formalism of weak lensing
and improving upon the standard lensing formalism is to identify
the rest-frames, in which observable and physical
quantities are defined. The tetrad basis vectors $e_a^\mu$ form an orthonormal
basis with the Minkowski metric and connect the local rest-frame to
the FRW frame, providing the exact ingredient for our purpose. 
In cosmology, there exists a privileged timelike direction set by the observer
four velocity $e_0^\mu=u^\mu$, which defines the rest-frame of the observer.
Furthermore, the spatial directions~$e_i^\mu$ in this rest-frame are
used to measure any directional quantities in cosmology; for example,
the observed propagation direction of
the photon wavevector~$k^\mu$ in the FRW frame is measured in the rest
frame by contracting against the spatial tetrads as in Eq.~\eqref{eq:rest}.
Any measurements in the observer rest-frame are indeed expressed in terms
of diffeomorphism scalars constructed by contracting against~$e_a^\mu$,
and this condition automatically
guarantees that any observables measured in the rest-frame
are gauge-invariant at the linear order
in perturbations \cite{YODU17}. For this reason,
it is difficult to overemphasize 
the significant role of the tetrad bases in cosmology that is unfortunately
often neglected in literature.
It is indeed more advantageous to go beyond the tetrad 
bases at the observer and the source positions and consider a tetrad field
that consists of the observer families in describing physical observables,
as it will lead to a fully nonlinear formalism in cosmology without
coordinates \cite{MIYO18}.

With four tetrad vectors, there exist 16 degrees of freedom, 10 out of
which are constrained by the metric~$g_{\mu\nu}=\eta_{ab}e^a_\mu e^b_\nu$
at each point. By setting the privileged direction $e_0^\mu=u^\mu$,
we fix three degrees of freedom associated with the boosts, but three
remain unconstrained. This remaining freedom is unconstrained by the 
metric tensor, and it represents the rotation
of the spatial tetrad vectors. While this anti-symmetric part of the
spatial tetrad in Eqs.~\eqref{eq:pij} and~\eqref{eq:aij} is often
neglected in literature, it is necessary to include the anti-symmetric
part to ensure that
the spatial tetrad vectors transform as four vectors in a FRW coordinate.
Fortunately though, the missing part at the linear order 
results in the systematic errors, only in the vector perturbations and 
the rotation of the spatial tetrad vectors that are often subdominant.
In the same spirit, the rest-frame of the source is as important as the
rest-frame of the observer. For our purpose, the tetrad basis at the source 
position needs to be used to establish the rest-frame of the source, 
in which the physical size and shape of the source galaxy are defined.
In conjunction with the tetrad basis at the observer position,
this part is crucial for deriving the gauge-invariant expressions for
the lensing
observables. It is well-known \cite{DOROST03,ADDUTA16} that the tensor
perturbations yield the infrared divergences in the lensing shear
and this pathology was fixed by introducing a counter term. However,
we have shown that this correction term 
naturally arises when we transform the FRW frame to the
source rest-frame (this correction term is also known as the FNC 
term \cite{SCJE12b},\footnote{The Fermi normal coordinate (FNC) is a coordinate
system, in which the metric at the origin is the Minkowski. This coincides
our tetrad basis at the origin. However, the FNC is indeed a coordinate
that describes the neighborhood around the origin, which is more than what
we need to define the rest-frame and its observables. In fact, the information
about the nearby region is fully contained in the tetrad field.}
or the metric shear \cite{DOROST03,YANATA13}).
Such correction terms due to the frame change
exist not only in tensor perturbations, but also
in scalar and vector perturbations. 

In addition to the frame change, there exists another ingredient often
missing in the perturbation calculations ---
the observer position is different from the position
$\bar x^\mu_o=(\bar\eta_o,0)$ in a homogeneous universe, as the observer
drifts away from the background path in the presence of perturbations.
This deviation is characterized by the coordinate lapse~$\delta\eta_o$ in
Eq.~\eqref{eq:lapse} and the coordinate shift~$\delta x^\alpha_o$ in
Eq.~\eqref{eq:shift}. These perturbations are uniquely determined,
and their presence is necessary for deriving the gauge-invariant expressions.
These deviations vanish in the comoving gauge, but
they cannot be set zero with other gauge conditions. 
At the linear order, however, the coordinate shift~$\delta x^\alpha_o$ 
drops out
in all the lensing observables, but the coordinate lapse~$\delta\eta_o$
contributes to the lensing convergence (or the luminosity distance). 
Their absence in
the calculation of the variance of the luminosity distance is shown
\cite{BIYO16,BIYO17} to be the cause of the infrared divergences.

In the presence of the vector and the tensor perturbations, the tetrad
basis vectors ``rotate'' even at the linear order in perturbations
as they are parallel transported along the photon path, when they are
seen from a global FRW coordinate aligned with the local tetrad basis
at the observer position. This rotation of the tetrad basis, often
known as the Skrotsky effect \cite{SKROT57}, translates
into the lensing rotation, and its potential
measurements may be considered as a probe
of the primordial gravitational waves, though its constraining power is
expected to be low \cite{DOROST03}. However,
we have shown that the Skrotsky effect
is an artifact of using a global FRW coordinate for observables,
and the only physically meaningful way to compare two points in curved space
is to parallel transport the basis vectors.  We stress that even in the
presence of the vector and tensor perturbations 
 the lensing rotation vanishes (but only) at the linear order
in perturbations, when compared to the basis parallel transported 
to the source position. However, this point should not be confused with the 
statement that the vanishing rotation implies the vanishing lensing B-mode;
Instead, the vector and the tensor perturbations contribute to the lensing 
shear (or non-vanishing B-mode),
while the physical rotation is zero, as described in Sec.~\ref{ssec:EB}.

Indeed,
one way to measure the lensing rotation even with a single source was already
 discussed \cite{KRDYET91,BUDYET04} in the past by using
the polarization measurements. Though the polarization vector is 
perpendicular to the photon propagation as the lensing images,
it belongs to the tangent space of the central geodesic, while lensing
images are extended in space, albeit infinitesimally small.
Therefore, polarization is parallel transported (to all orders 
in perturbations), and hence its measurement
can be used to infer the base direction to synchronize the local coordinates
at the observer and the source positions, in the absence of any significant 
magnetic field along the path.\footnote{In principle, multi-frequency
observations can decode the Faraday rotation due to the magnetic fields.}
Under the assumption that
the morphology of the source is aligned with its polarization, one can
infer the lensing rotation from a single system, when combined with
the polarization measurements. 

In summary, the physical lensing observables can be found in 
Eq.~\eqref{eq:hkappa} for the lensing convergence, in Eq.~\eqref{eq:hgamma}
for the lensing shear, and in Eq.~\eqref{eq:homega} for the rotation.
Compared to the standard lensing formalism, there exist additional
relativistic effects in all the lensing observables.
The physical rotation is zero~$(\hat\omega=0)$ in Eq.~\eqref{eq:homega},
because the spatial orientation~$\Omega^n_s$ of the source frame indeed
cancels the remaining terms (see Sec.~\ref{ssec:rotation}). The lensing
convergence~$\hat\kappa$ 
we measure is indeed the luminosity distance, not the usual
(coordinate) convergence~$\kappa$ in Eq.~\eqref{eq:kappa}.
Compared to the standard lensing formalism, the additional velocity
contributions in the luminosity distance, sometimes referred to as the
Doppler lensing, are significant and already measured in current surveys
(see, e.g., \cite{BAANET14,BIYO17}), while the contributions of
the gravitational potential or the primordial gravity waves are small, 
demanding special techniques to be measured in the upcoming surveys
(see, e.g., \cite{YOHAET12}).
Looking to the future, we believe that 
the relativistic effects in large scale structure will provide a great
opportunity to probe the nature of gravity and understand the physical
mechanism of the perturbation generation in the early Universe.
Our gauge-invariant lensing formalism will be essential in providing
correct theoretical predictions for the upcoming surveys
(see, e.g., \cite{GHDUSE18}).

\acknowledgments
We acknowledge useful discussions with Donghui Jeong, Fabian Schmidt,
Eduardo Rozo, and Scott Dodelson. We thank Uro\v s Seljak for 
useful discussions about the lensing rotation and clarifying the E-B
decomposition of the lensing potential in Sec.~\ref{ssec:EB}.
We acknowledge support by the Swiss National Science Foundation.
J.Y. and E.M. are further supported by
a Consolidator Grant of the European Research Council (ERC-2015-CoG grant
680886).

\appendix

\section{Metric Convention and Gauge Transformation}
\label{appendix}
Here we present our notation convention used in this paper. A concise
summary is given in Table~\ref{table}. To model the background universe,
we adopt a spatially flat Robertson-Walker metric and choose a Cartesian
coordinate:
\beeq
ds^2=g_{\mu\nu}~dx^\mu dx^\nu=-a^2(\eta)d\eta^2
+a^2(\eta)\delta_{\alpha\beta}dx^\alpha dx^\beta~,
\eneq
where $\eta$ is the conformal time and $a(\eta)$ is the scale factor.
Small perturbations are introduced to capture the deviation from the background
in the real universe. Our notation convention for the metric tensor is
\beeq
\label{eq:perturb}
\delta g_{\eta\eta}\equiv-2~a^2\mathcal{A}~,\qquad\qquad
\delta g_{\eta\alpha}\equiv-a^2\BB_\alpha~,\qquad\qquad
\delta g_{\alpha\beta}\equiv2~a^2\CC_{\alpha\beta}~.
\eneq
According to the rotational properties, 
these metric perturbations are further decomposed 
into scalar $\alpha,\beta,\varphi,\gamma$, 
transverse vector $B_\alpha,C_\alpha$ and transverse traceless symmetric
tensors $C_{\alpha\beta}$:
\beeq
\label{eq:decom}
\mathcal{A}=\alpha~,\qquad \BB_{\alpha}=
\beta_{,\alpha}+B_\alpha~,\qquad\qquad
\CC_{\alpha\beta}=\varphi~\delta_{\alpha\beta}+\gamma_{,\alpha\beta}
+{1\over2}\left(C_{\alpha,\beta}+C_{\beta,\alpha}\right)+C_{\alpha\beta}~,
\eneq
where the commas represent the spatial derivative.\footnote{The separation
of the scalar, the vector, and the tensor perturbations is straightforward,
based on their spatial indices.} 

General relativity is diffeomorphism invariant, allowing for any 
coordinate systems to describe the physical systems. We consider the most
general coordinate transformation
\beeq
\label{eq:coord}
\tilde x^\mu=x^\mu+\xi^\mu~,\qquad\qquad
\xi^\mu=(T,\mathcal{L}^\alpha)~,\qquad\qquad
\mathcal{L}^\alpha\equiv L^{,\alpha}+L^\alpha~,
\eneq
where two coordinates describe the same physical point and
the infinitesimal transformation~$\xi^\mu$ is further decomposed
in terms of scalar~$T,L$ and transverse vector~$L^\alpha$.
This coordinate transformation involves the change in the correspondence
to the background universe, accompanying the gauge transformation for
the metric perturbations \cite{BARDE80}. Since physical observables
are expressed in terms of diffeomorphism invariant scalars, it is
important to check if our expressions for physical observables
are indeed gauge-invariant at the linear order \cite{YODU17}.
Under the coordinate transformation in Eq.~\eqref{eq:coord},
the scalar perturbations gauge transform as
\beeq
\tilde\alpha=\alpha-T'-\HH T~,\qquad  \tilde\beta=\beta-T+L'~,\qquad
\tilde\varphi=\varphi-\HH T~, \qquad \tilde\gamma=\gamma-L~,
\eneq
and the vector perturbations transform as
\beeq
\tilde B_\alpha=B_\alpha+L'_\alpha~,\qquad\qquad
\tilde C_\alpha=C_\alpha-L_\alpha~,
\eneq
and the tensor perturbations remain unaffected, where the prime
indicates the derivative with respect to~$\eta$ 
and the conformal Hubble parameter is $\HH=a'/a$. 

At the linear order
in perturbations, the spatial shift~$\LL^\alpha$ is absent in any
physical quantities, and only the temporal shift~$T$ represents the
real physical choices of the time slicing. For this reason, the pure
gauge modes can be combined \cite{YOO14a} as
\beeq
\CCG^\alpha\equiv \gamma^{,\alpha}+C^\alpha~,\qquad\qquad
\tilde\CCG^\alpha=\CCG^\alpha-\LL^\alpha~.
\eneq
Moreover, the 
presence of the scalar spatial shift~$L$ signals that the physical
quantities cannot depend on~$\beta$ directly, instead they depend on the
combination~$\chi$ that is absent of~$L$:
\beeq
\chi\equiv a\left(\beta+\gamma'\right)~,\qquad\qquad
\tilde\chi=\chi-aT~.
\eneq
This combination introduced in \cite{BARDE88}
is indeed the scalar shear of the normal observer.\footnote{A normal 
observer~$n^\mu$
($n_\alpha\equiv0$) is a timelike four velocity 
($-1=n^\mu n_\mu=n^\eta n_\eta$) associated with the given coordinate system
(often used in the ADM formalism \cite{ADM}). Its flow can be covariantly
decomposed as
\beeq
n_{\alpha;\beta}=\frac13\theta\delta_{\alpha\beta}+\sigma_{\alpha\beta}~,
\qquad\qquad \theta=3H(1-\alpha)+3\dot\phi+{\Delta\over a^2}\chi~,
\qquad\qquad
\sigma_{\alpha\beta}=\chi_{,\alpha\beta}+a\pv_{(\alpha|\beta)}~,
\eneq
where $\theta$ is the expansion and $\sigma_{\alpha\beta}$ is the shear
of the flow. The rotation and the acceleration of the normal flow vanish
at the linear order.}
According the gauge-transformation properties, it is natural to
construct and work with the gauge-invariant variables \cite{BARDE80}.
The gauge-invariant variables are
\beeq
\label{eq:gigi}
\ax=\alpha-\frac1a~\chi'~,\qquad\qquad\px=\varphi-H\chi~,\qquad\qquad
\pv_\alpha=B_\alpha+C'_\alpha~,
\eneq
and these gauge-invariant variables correspond to the Bardeen variables:
\beeq
\ax~~\mapsto~~\Phi_A~,\qquad\qquad \px~~\mapsto~~\Phi_H~,\qquad \qquad
\pv_\alpha~~\mapsto~~\Psi Q^{(1)}_\alpha~,
\eneq
where $Q_\alpha^{(1)}$ is the vector harmonics in \cite{BARDE80}.

Timelike four vectors~$u^\mu$ are important in establishing the rest-frames
of the observer and the source. Given the timelike condition,
the four velocity vector can be parametrized as
\beeq
u^\mu={1\over a}\left(1-\AA,~\VV^\alpha\right)~,\qquad\qquad
\VV^\alpha\equiv-U^{,\alpha}+U^\alpha~,
\eneq
where we again decomposed the spatial velocity into the scalar~$U$
and the transverse vector~$U^\alpha$, representing the degree of freedom
associated with the flow. Under the coordinate transformation,
the spatial velocity components gauge-transform as
\beeq
\tilde U=U-L'~,\qquad\qquad \tilde U_\alpha=U_\alpha+L'_\alpha~.
\eneq
According to their gauge-transformation properties, we introduce 
a scalar~$v$ perturbation,
independent of the spatial gauge mode
\beeq
v\equiv U+\beta~,\qquad\qquad \tilde v=v-T~,
\eneq
and the gauge-invariant variables
\beeq
\vx=v-{1\over a}\chi~,\qquad\qquad 
v_\alpha=U_\alpha-B_\alpha ~,
\eneq
corresponding to the Bardeen variables:
\beeq
\vx~~\mapsto~~ v_s^{(0)}~,\qquad\qquad v_\alpha~~\mapsto~~ v_cQ_\alpha^{(1)}~.
\eneq
With our interest in the observer four velocity, we introduce a
gauge-invariant variable~$V^\alpha$ by combining the scalar and the vector
gauge-invariant variables
\beeq
V_\alpha\equiv -v_{\chi,\alpha}+v_\alpha~.
\eneq

Given the observed angle and frequency in the observer rest-frame,
the photon wavevector~$k^\mu$ in Eq.~\eqref{eq:photon} can be constructed
in a FRW coordinate. Under the coordinate transformation in 
Eq.~\eqref{eq:coord}, this photon wavevector should transform as a
four vector, and this transformation property
constrains how the conformally transformed 
wavevector~$\hat k^\mu$ in Eq.~\eqref{eq:affine} should transform.
Noting that the normalization condition in Eq.~\eqref{eq:normalization}
is imposed at the same physical point~$p$, we can derive the
gauge-transformation properties of the perturbations $(\dnu,\dea^\alpha)$
to the photon wavevector~$\hat k^\mu$
\beeq
\widetilde{\dnu}=\dnu+2\HH T-\HH_pT_p+{d\over d\lambda}~T~,  \qquad\qquad
\widetilde{\dea}{}^\alpha=\dea^\alpha+\left(2\HH T-\HH_pT_p\right)n^\alpha
-{d\over d\lambda}\LL^\alpha~,
\eneq
where the normalization point~$p$ in the main text is the observer position,
but here we left unspecified (it can be the source position or  any point).
When imposed at the observer position
($\dhnu_o=0$), the initial conditions $(\dnu_o,\dea^\alpha_o)$
in Eqs.~\eqref{eq:init} and~\eqref{eq:inita}
indeed match the gauge-transformation properties derived above.
The contributions at the normalization point~$p$ was neglected in Eq.~(2.21)
in \cite{YOO14a}. Based on the gauge-transformation properties,
the gauge-invariant variables can be constructed as
\beeq
\dnu_\chi=\dnu+2H\chi+{d\over d\lambda}\left({\chi\over a}\right)-H_p\chi_p~,  
\qquad\qquad
\dea^\alpha_\chi=\dea^\alpha+2H\chi ~n^\alpha-{d\over d\lambda}~\CCG^\alpha
-H_p\chi_pn^\alpha~.
\eneq

\bibliography{ms.bbl}

\end{document}